
(TeXPS: dvi->PostScript Driver dvitps, Version 3.11 of September 5,
1990\n)print flush
(TeXPS: release number dvitps of 3.14\n)print flush
save
/@bop0
{
PsPrintEnabled { (Pass0: Page ) print == flush } {pop} ifelse
}	def
/@eop0 {
pop
}	def
/@bop1
{
PsPrintEnabled { (Pass1: Page ) print == flush } {pop} ifelse
save
DviTranslateVector-X DviTranslateVector-Y translate
DviOrientationMatrix concat
Page-H-Offset Page-V-Offset translate
3 1 roll
translate
0 0 moveto
Draft-Flag { @Draft } if
/DriverDelta 0 def
} def
/@eop1 {
pop
showpage
restore
} def
/@end {
(Done!\n) print flush
} def
/@ReportVmStatus {
(\n@ReportVmStatus: ) print
print (: ) print
vmstatus
(max: ) print 20 string cvs print
(, used: ) print 20 string cvs print
(, level: ) print 20 string cvs print
(\n) print flush
} def
/@ReportVmStatusNew {
(VM: ) print
exch print
(, printer: ) print
/Vmfree vmstatus pop exch pop def Vmfree (       ) cvs print
(, driver: ) print dup (      ) cvs print
(, delta: ) print
Vmfree sub (      ) cvs print
(\n) print flush
} def
/@Draft {
gsave
Page-H-Offset neg Page-V-Offset neg translate
-45 rotate
-150 0 moveto
/Helvetica-Bold findfont
[120.0 0 0 -120.0 0 0 ] makefont setfont
(DRAFT) show
grestore
gsave
Page-H-Offset neg Page-V-Offset neg translate
300 -100 moveto
/Helvetica-Bold findfont
[60.0 0 0   -60.0 0 0 ] makefont setfont
Date-and-Time		show
(   ::   )	      	show
Dvi-File-Name		show
grestore
} def
/a { rmoveto } def
/DriverDelta 0 def
/b { exch p dup /DriverDelta exch def 0 rmoveto } def
/c { p DriverDelta 4 sub dup /DriverDelta exch def 0 rmoveto } def
/d { p DriverDelta 3 sub dup /DriverDelta exch def 0 rmoveto } def
/e { p DriverDelta 2 sub dup /DriverDelta exch def 0 rmoveto } def
/f { p DriverDelta 1 sub dup /DriverDelta exch def 0 rmoveto } def
/g { p DriverDelta 0 rmoveto } def
/h { p DriverDelta 1 add dup /DriverDelta exch def 0 rmoveto } def
/i { p DriverDelta 2 add dup /DriverDelta exch def 0 rmoveto } def
/j { p DriverDelta 3 add dup /DriverDelta exch def 0 rmoveto } def
/k { p DriverDelta 4 add dup /DriverDelta exch def 0 rmoveto } def
/l { p -4 0 rmoveto } def
/m { p -3 0 rmoveto } def
/n { p -2 0 rmoveto } def
/o { p -1 0 rmoveto } def
/q { p 1 0 rmoveto } def
/r { p 2 0 rmoveto } def
/s { p 3 0 rmoveto } def
/t { p 4 0 rmoveto } def
/p { show } def
/w { 0 exch rmoveto } def
/x { 0 rmoveto } def
/y { 3 -1 roll p rmoveto } def
/u-string ( ) def
/u { u-string exch 0 exch put
u-string show
} def
/v { u-string exch 0 exch put
currentpoint
u-string show
moveto
} def
/z
{   /dy exch def
/dx exch def
currentpoint
currentpoint
transform round exch round exch itransform
newpath
moveto
dx 0 rlineto
0 dy rlineto
dx neg 0 rlineto
closepath
fill
moveto
}
def
/z
{   /dy exch def
/dx exch def
currentpoint
0.2 0.2 rmoveto
currentpoint
newpath
moveto
dx 0 rlineto
0 dy rlineto
dx neg 0 rlineto
closepath
fill
moveto
}
def
letter
/Dvi-File-Name (my_second_paper_ms.dvi) def
(Dvi file name: ") print Dvi-File-Name print (".\n) print
/Draft-Flag false def
/#copies 1 def
/NumCharsInPixelFonts 256 def
/HostName (lulu) def
(This PostScript file was produced on host \") print HostName print (\".\n)
print
/PsPrintEnabled true def
/Page-H-Offset   0.000000 def
/Page-V-Offset   0.000000 def
/ExecPsConv {0.30 mul} def
/Date-and-Time (Wed Jul 13 14:13 1994) def
/DviTranslateVector-X   72.000 def
/DviTranslateVector-Y  720.000 def
/DviOrientationMatrix [    0.240    0.000    0.000
	   -0.240 0.0 0.0 ] def
/@newfont
{
/newname exch def
newname 7 dict def
newname load begin
/FontType 3 def
/FontMatrix [1 0 0 -1 0 0] def
/FontBBox [0 0 1 1] def
/BitMaps NumCharsInPixelFonts array def
/BuildChar {CharBuilder} def
/Encoding NumCharsInPixelFonts array def
0 1 NumCharsInPixelFonts 1 sub {Encoding exch /.undef put} for
end
newname newname load definefont pop
} def
/ch-image {ch-data 0 get} def
/ch-width {ch-data 1 get} def
/ch-height {ch-data 2 get} def
/ch-xoff  {ch-data 3 get} def
/ch-yoff  {ch-data 4 get} def
/ch-tfmw  {ch-data 5 get} def
/CharBuilder
{
/ch-code exch def
/font-dict exch def
/ch-data font-dict /BitMaps get ch-code get def
ch-data null eq not
{
ch-tfmw   0
ch-xoff neg
ch-height ch-yoff sub neg
ch-width  ch-xoff neg add
ch-yoff
setcachedevice
0 0 transform round exch round exch itransform translate
ch-width ch-height true
[1 0  0 1 ch-xoff ch-height ch-yoff sub] {ch-image} imagemask
}
if
} def
/@dc
{
/ch-code exch def
/ch-data exch def
currentfont /BitMaps get
ch-code ch-data put
currentfont /Encoding get
ch-code (   ) cvs   
cvn
ch-code exch put
} def
/@sf /setfont load def

28 @bop0
/@F7 @newfont
@F7 @sf
[<
FFFFFE>
	 23 1 0 12 24] 123 @dc
[<
FFFFC07FFFC03FFFC030004018006008002004002002002001000001800000C00000600000
3000001800001C00000E000007000007800003C00003C00003E02003E0F801E0F801E0F801
E0F003E08003E04003C04003C02007801007000C1C0003F000>
	 19 33 -2 32 24] 50 @dc
[<
03F0000E0E001803003000806000C0600040C00060C00060C00060C00060C000E06000E060
01C03007C0100F80083F80067F0003FC0003F8000FF8001FC4003F02003E01007801807000
C06000C06000C06000C02000C0200180100180080300060E0001F800>
	 19 34 -2 32 24] 56 @dc
[<
FFFE00000FC000000780000007800000078000000780000007800000078000000780000007
8000000780000007802000078020000780200007802000078060000780E00007FFE0000780
E0000780600007802000078020000780200007802020078000200780002007800020078000
600780004007800040078000C0078001C00F8007C0FFFFFFC0>
	 27 34 -2 33 32] 70 @dc
[<
FFC00E000E000E000E000E000E000E000E000E000E000E000E000E000E000E000E000E001E
00FE000E00000000000000000000000000000000001C001E003E001E001C00>
	 10 34 -1 33 14] 105 @dc
[<
03FE000E03803800E0600030600030C00018C00018C000184000186000303800F00FFFE00F
FFC01FFE0018000018000010000010000019F0000F1C000E0E001C07001C07003C07803C07
803C07803C07801C07001C07000E0E18071E1801F198000070>
	 21 33 -1 21 24] 103 @dc
[<
70F8F8F870>
	 5 5 -4 4 14] 46 @dc
[<
FFFFFFFFFFFF>
	 48 1 0 12 49] 124 @dc
[<
8FC0D030E018C008C00C800C800C801C003C01F80FF03FE07F80F000E008C008C008C01860
1830780F88>
	 14 21 -2 20 19] 115 @dc
[<
FFE7FF0E00700E00700E00700E00700E00700E00700E00700E00700E00700E00700E00700E
00700E00700E00700E00700F00700F00700E80E00E60C00E1F800E00000E00000E00000E00
000E00000E00000E00000E00000E00000E00000E00001E0000FE00000E0000>
	 24 35 -1 34 27] 104 @dc
[<
01FC000707000E03801C01C03800E07800F0700070F00078F00078F00078F00078F00078F0
0078F000787000707000703800E01800C00C018007070001FC00>
	 21 21 -1 20 24] 111 @dc
[<
00600600006006000060060000F00F0000F00F0000F00D0001C81C8001C81C8001C8188003
84384003843840038430400702702007027020070260200E01E0100E01E0100E01C0181C01
C0181E01E03CFF8FF8FF>
	 32 21 -1 20 35] 119 @dc
[<
01F0030807080E040E040E040E040E040E040E000E000E000E000E000E000E000E000E000E
000E00FFF83E001E000E000600060006000200020002000200>
	 14 31 -1 30 19] 116 @dc
[<
00FC000703000E00801C0040380020780020700000F00000F00000F00000F00000F00000FF
FFE0F000E07000E07801E03801C01C01C00C038007070001FC00>
	 19 21 -1 20 22] 101 @dc
[<
00100000380000380000380000740000740000E20000E20000E20001C10001C10003808003
80800380800700400700400E00200E00200E00301E0078FFC1FE>
	 23 21 -1 20 26] 118 @dc
[<
FFE00E000E000E000E000E000E000E000E000E000E000E000E000E000E000E000E000E000E
000E000E000E000E000E000E000E000E000E000E000E000E000E001E00FE000E00>
	 11 35 -1 34 14] 108 @dc
[<
01F8000706000C01001C0080380040780040700000F00000F00000F00000F00000F00000F0
0000F000007000007800003803001C07800C078007030001FE00>
	 18 21 -2 20 22] 99 @dc
[<
3C0000430000F18000F08000F0400000400000200000200000200000100000100000380000
380000380000740000740000E20000E20000E20001C10001C1000380800380800380800700
400700400E00200E00200E00301E0078FFC1FE>
	 23 31 -1 20 26] 121 @dc
[<
FFE0000E00000E00000E00000E00000E00000E00000E00000E00000E00000E3F000E41C00E
80E00F00700E00380E003C0E003C0E001E0E001E0E001E0E001E0E001E0E001E0E001E0E00
1C0E003C0E00380F00700E8060FE61C00E1F00>
	 23 31 -1 20 27] 112 @dc
[<
01F0FE070CF00C02E01801E03800E07800E07000E0F000E0F000E0F000E0F000E0F000E0F0
00E0F000E07000E07800E03800E01C01E00C02E00704E001F8E00000E00000E00000E00000
E00000E00000E00000E00000E00000E00000E00000E00001E0000FE00000E0>
	 23 35 -2 34 27] 100 @dc
[<
0FC1E03C2390781708F00F08F00708F00708F007087007007807003C07001E070007C70000
FF000007000007000007001807003C0E003C0C001838000FE000>
	 21 21 -2 20 24] 97 @dc
[<
FFE7FF0E00700E00700E00700E00700E00700E00700E00700E00700E00700E00700E00700E
00700E00700E00700E00700F00700F00701E80E0FE60C00E1F80>
	 24 21 -1 20 27] 110 @dc
[<
7FF80780070007000700070007000700070007000700070007000700070007000700070007
000700FFF8070007000700070007000700070007000700030F038F018F00C6003C>
	 16 35 0 34 15] 102 @dc
[<
FFFFFC000F800F00078007C0078003E0078001F0078001F0078000F8078000F8078000F807
8000F8078000F8078000F8078000F0078001F0078001E0078003C00780078007FFFE000780
3E0007800F80078007C0078003E0078001E0078001F0078001F0078001F0078001F0078001
F0078001E0078003E0078003C0078007800F800E00FFFFF800>
	 29 34 -2 33 35] 66 @dc
[<
4020101008080404040474FCFCF870>
	 6 15 -4 34 14] 39 @dc
[<
00020000800000030001800000070001C00000070001C00000070001C000000F8003E00000
0F8003E000000F8003E000001E40079000001E40079000001E40079000003C200F0800003C
200F0800003C200F0800007C101E04000078101E04000078101E040000F8183E060000F008
3C020000F0083C020000F0083C020001E00478010001E00478010001E00478010003C002F0
008003C002F0008003C002F00080078001E00040078001E00040078001E000400F0003C000
200F0003C000200F0003C000701F8007E000F8FFF03FFC03FE>
	 47 35 -1 33 50] 87 @dc
[<
FFE000201F0000600E000060040000E0040001E0040001E0040003E0040003E0040007A004
000F2004000F2004001E2004003E2004003C20040078200400F8200400F0200401E0200401
E0200403C0200407802004078020040F0020041F0020041E0020043C0020047C0020047800
2004F0002004F0002005E0002007C0007007C000F8FF8007FF>
	 32 34 -2 33 37] 78 @dc
[<
FFE0FFE0>
	 11 2 -1 11 16] 45 @dc
[<
083F000C41C00C80600F00700E00380E003C0E001C0E001E0E001E0E001E0E001E0E001E0E
001E0E001E0E001C0E003C0E00380F00300E80600E61C00E1F000E00000E00000E00000E00
000E00000E00000E00000E00000E00000E00000E00001E0000FE00000E0000>
	 23 35 -1 34 27] 98 @dc
[<
FFE3FF8FFE0E003800E00E003800E00E003800E00E003800E00E003800E00E003800E00E00
3800E00E003800E00E003800E00E003800E00E003800E00E003800E00E003800E00E003800
E00E003800E00F003C00E00F003C00E01E807201C0FE60E183800E1FC07F00>
	 39 21 -1 20 41] 109 @dc
[<
00FC7F0382780601700E00F00E00F00E00700E00700E00700E00700E00700E00700E00700E
00700E00700E00700E00700E00700E00701E00F0FE07F00E0070>
	 24 21 -1 20 27] 117 @dc
[<
FFF00F000E000E000E000E000E000E000E000E000E000E000E000E000E000F000F060F0F1E
8FFE460E3C>
	 16 21 -1 20 19] 114 @dc
[<
4020101008080404040474FCFCF870>
	 6 15 -4 4 14] 44 @dc
/@F3 @newfont
@F3 @sf
[<
003F000001C0E000070010000E000C001C0002003C000100380001007800008078000040F8
000040F0000040F0000020F0000020F0000000F0000000F0000000F8000000F8000000F800
0000F8000000F80000007C0000007C0000083C0000083E0000081E0000081E0000180F0000
180700001C0380001C01C0003C00E0003C0070005C003C008C000F03060001FC02>
	 31 36 -6 34 35] 67 @dc
[<
007F000003C1E000070078000E001C001C000E003C00070078000780780003C0780001E0F8
0001F0F80000F0F00000F8F00000F8F000007CF000007CF800007CF800003EF800003EF800
003EF800003EF800003E7C00003E7C00003E3C00003E3C00003E1E00003E1E00003C0F0000
3C0700003C0780007803C0007801E000F0007000E0003801C0000E07000001FC00>
	 31 36 -6 34 38] 79 @dc
[<
FFFFFC0007C00F80078003C003C001E003C000F003C000F803C0007C03C0007C03C0003C01
E0003E01E0003E01E0003E01E0003C01E0007C01E0007800F000F800F001F000FFFFC000F0
07C000F001F000F000780078003C0078003E0078001F0078001F0078001F0078000F003C00
0F003C000F003C001F003C001E003C003C003E007803FFFFE0>
	 32 34 -2 33 35] 66 @dc
[<
FFFFFFE007C003E0078000F003C0003003C0001003C0001803C0000803C0000803C0000401
E0000401E0000401E0080201E0080001E0080001E0080000F00C0000F01C0000FFFC0000F0
1C0000F0040000F00400007806000078020000780202007800020078000200780002003C00
02003C0002003C0002003C0006003C000E003E001F03FFFFFF>
	 32 34 -2 33 33] 69 @dc
@F7 @sf
[<
00200040008001000300060004000C000C00180018003000300030007000600060006000E0
00E000E000E000E000E000E000E000E000E000E000E000E000E00060006000600070003000
30003000180018000C000C000400060003000100008000400020>
	 11 50 -4 36 19] 40 @dc
[<
81FC00C60700C80180F000C0E000C0C00060C0006080007080007080007080007000007000
00F00000F00001E00007E0003FC003FF800FFF001FFE003FF0007F0000780000F00000F000
00E00020E00020E00020E00060E000606000607000E03001E01802600C0C6003F020>
	 20 36 -3 34 27] 83 @dc
[<
FFFE07C0038003800380038003800380038003800380038003800380038003800380038003
800380038003800380038003800380038003800380F3800F8003800080>
	 15 33 -4 32 24] 49 @dc
[<
0FC000103000201800700C007806007807003003000003800003800001C00001C00001C003
E1E00619E00C05E01805E03803E07003E07001E0F001E0F001E0F001E0F001E0F001E0F001
C0F001C0F001C07003807003803803801807000C0600060C0001F000>
	 19 34 -2 32 24] 57 @dc
[<
03F0000C1C00100F002007804007804003C0F003C0F803E0F803E07003E02003E00003E000
03C00003C0000780000780000F00001C0003F000003800000E00000F000007000007800007
803807C07807C07803C07807C04007C02007801007000C1E0003F800>
	 19 34 -2 32 24] 51 @dc
[<
800040002000100018000C000400060006000300030001800180018001C000C000C000C000
E000E000E000E000E000E000E000E000E000E000E000E000E000E000C000C000C001C00180
01800180030003000600060004000C0018001000200040008000>
	 11 50 -3 36 19] 41 @dc
[<
03FFFF00000FC0000007800000078000000780000007800000078000000780000007800000
07800000078000000780000007800000078000000780000007800000078000000780000007
80000007800000078000000780000007800080078004800780048007800480078004C00780
0C40078008400780084007800860078018780780787FFFFFF8>
	 30 34 -2 33 35] 84 @dc
[<
0400003C000C00004200060000C100030001810003000380800180030080018003004000C0
07004000600700400060070040003007004000180700400018070040000C07004000060300
40000603008000030380800003018100000180C1000000C042000780C03C00184060000018
20300000301030000070101800006010180000E0080C0000E008060000E008060000E00803
0000E008018000E008018000E00800C000E0080060006013FF6000701C00B0003010007000
18200018001840000C000780000C00>
	 34 40 -3 36 41] 37 @dc
[<
FFE3FE0E00F80E00F00E01E00E01C00E03C00E07800E07000E0E000E1E000F1C000EF8000E
38000E10000E08000E04000E02000E01800E01C00E01F00E03FC0E00000E00000E00000E00
000E00000E00000E00000E00000E00000E00000E00001E0000FE00000E0000>
	 23 35 -1 34 26] 107 @dc
[<
01FFF0001F00000E00000E00000E00000E00000E00000E00000E00FFFFF8800E00400E0020
0E00200E00100E00100E00080E00040E00040E00020E00020E00010E00008E00008E00004E
00004E00002E00001E00001E00000E00000E00000600000200>
	 21 33 -1 32 24] 52 @dc

28 @eop0

0 0 28 @bop1 910 45 a @F7 @sf({)17 b(28)g({)-1040 1158 y(Fig.)46 b(8.|)24 b
(sho)o(ws)i(the)e(v)o(elo)q(cit)o(y)o 24 x(elli)o(psoid)g(at)h(the)f(p)q
(osition)h(of)g(Baade's)g(Windo)o(w)f(of)h(a)g(N-b)q(o)q(dy)-1949 96 y
(sim)o(ulated)19 b(bar,)i(the)f(ligh)o(t)g(distribution)g(of)g(whic)o(h)g
(resem)o(bles)f(the)h @F3 @sf(COBE)g @F7 @sf(map)g(\(Sellw)o(o)q(o)q(d)h
(1993\).)-1949 97 y(The)16 b(data)h(p)q(oin)o(ts)g(corresp)q(ond)g(to)f
(particles)f(of)i(the)f(N-b)q(o)q(dy)h(bar,)f(98
(found)f(to)h(ha)o(v)o(e)-1951 96 y(prograde)g(sense)g(of)f(rotation.)23 b
(The)16 b(striking)g(similarit)o(y)o 15 x(with)h(Baade's)f(Windo)o(w)g
(metal)g(ric)o(h)f(sample)-1950 96 y(\(Figure)k(1\))i(and)f(the)g
(prograde)h(stars)g(\(Figure)e(4\))i(implie)o(s)e(that)h(it)g(is)g(p)q
(ossible)g(for)g(the)g(metal)f(ric)o(h)-1951 97 y(Bulge)d(to)g(gro)o(w)h
(out)g(of)f(the)g(disk)g(and)h(form)f(a)h(nearly)e(self-gra)o(vitating)h
(bar.)-1460 1109 y
28 @eop1

27 @bop0
@F7 @sf
[<
01800003C00003C00003C00003C00003C00003C00003C00003C00001C00001C00001C00001
C00000C00000C00000E0000060000060000060000020000030000010000008000008000004
00800200800200800100C001004000807FFFC07FFFC07FFFE0600000400000>
	 19 35 -3 33 24] 55 @dc
[<
01F000071C000C06001C07003803803803807803C07001C07001C07001C0F001E0F001E0F0
01E0F001E0F001E0F001E0F001E0F001E0F001E0F001E0F001E0F001E0F001E0F001E07001
C07001C07001C07001C03803803803801803000C0600071C0001F000>
	 19 34 -2 32 24] 48 @dc
[<
03F0000C1C001006002007004003804003C08001C0E001C0F001E0F001E07001E00001E000
01E00001E00001E00001C00001C0100380180380140700130E0010F8001000001000001000
0010000010000010000013E0001FF8001FFE001FFF001E0700100080>
	 19 34 -2 32 24] 53 @dc
/@F6 @newfont
@F6 @sf
[<
07C000001870000030180000600E000060070000E0030000E0038000E001C000E001C000E0
01C0007000E0007000E0007000E0003800E0003800E0001C00E0000E01C0000783C00003FF
FF8001FFFF80007FFF80>
	 25 21 -2 20 28] 27 @dc
@F7 @sf
[<
3E006180F180F0C060E000E000E000E000E000E000E000E000E000E000E000E000E000E000
E000E000E000E000E000E000E000E000E000E001E00FE001E0000000000000000000000000
0000000001C003E003E003E001C0>
	 11 44 2 33 15] 106 @dc
[<
FF03FE3E01F00C00E00401E00201C0010380018700008E00004E00003C0000380000780000
780000E40001C2000383000381000700800E00C01F01F0FF83FE>
	 23 21 -1 20 26] 120 @dc
@F6 @sf
[<
1C0064006200E200E200E200710070007000700038003800380038001C001C001C001C000E
000E000E000E000700070007000700038003800380038001C001C001C01FC001E0>
	 11 35 -3 34 15] 108 @dc
/@F10 @newfont
@F10 @sf
[<
3C004600C100818080C080C0C060C060C060C0606060704068C06780300030003000300018
001800180018007C00>
	 11 23 -2 22 15] 98 @dc
[<
387046C8C1C480C480C480C0C060C060C06060606070307018F00710>
	 14 14 -2 13 19] 97 @dc
[<
3000300018001800180018000C000C000C008C0046304638451838F0>
	 13 14 -1 13 16] 114 @dc
@F7 @sf
[<
FFFFFFFEFFFFFFFE0000000000000000000000000000000000000000000000000000000000
000000FFFFFFFEFFFFFFFE>
	 31 12 -3 17 38] 61 @dc
/@F5 @newfont
@F5 @sf
[<
FFFFFFFCFFFFFFFC>
	 30 2 -4 13 39] 0 @dc
@F10 @sf
[<
3E006180C0C0C06080308030C018C018C0186018201810100C2007C0>
	 13 14 -2 13 17] 111 @dc
@F7 @sf
[<
FF800FFF3E0001F80C0000F00C0000F0040001E0040001E0040003E0020003C0020003C003
0007C0010007800100078000FFFF0000800F0000800F0000401E0000401E0000401E000020
3C0000203C0000203C0000107800001078000010F8000008F0000008F000000DF0000005E0
000005E0000003C0000003C0000003C000000180000001800000018000>
	 32 35 -2 34 37] 65 @dc
[<
7FE1FF80070038000700380007003800070038000700380007003800070038000700380007
00380007003800070038000700380007003800070038000700380007003800070038000700
380007007800FFFFF800070000000700000007000000070000000700000007000000070000
000700300007007800038078000180380000C0100000702000001FC000>
	 25 35 0 34 27] 12 @dc
[<
FFFF807007807801803800801C00800E00C00F004007004003804001C00001E00000E00000
7000403800403C00401C00600E002007003007803803803FFFC0>
	 18 21 -1 20 22] 122 @dc
@F6 @sf
[<
00C000000000C000000000E000000000E000000000F000000000F800000001F800000001E4
00000001E200000001E200000001E100000001E180000001E080000003E040000003E04000
0003C020000003C010000003C010000003C008000003C00C000003C004000007C002000007
800200000780010000078000800007800080000780004000078000600007800020000F8000
10000F000010000F000008000F00000C000F80000F00FFF8007FC0>
	 34 35 -3 33 28] 86 @dc
@F10 @sf
[<
70D0C8C8C8C06060606030303030181818180C0C0C0C3E>
	 7 23 -2 22 11] 108 @dc
@F7 @sf
[<
C00000C00000C000006000006000006000003000003000003000001800001800001800000C
00000C00000C00000600000600000600000300000300000300000180000180000180000180
0000C00000C00000C000006000006000006000003000003000003000001800001800001800
000C00000C00000C00000600000600000600000300000300000300000180000180000080>
	 17 49 -3 36 24] 47 @dc
@F7 @sf
[<
7FE3FF80070078000700700007007000070070000700700007007000070070000700700007
00700007007000070070000700700007007000070070000700700007007000070070000700
700007007000FFFFFFC0070070000700700007007000070070000700700007007000070070
0007007000070070000380F0780180F87800C07C7800706E30001F83E0>
	 29 35 0 34 28] 11 @dc
@F6 @sf
[<
4020101008080404040474FCFCF870>
	 6 15 -4 4 14] 59 @dc
[<
0F000030C000307000603800601C00E01C00E01E00E00F00E00F00E00F0070078070078070
07807007803807803803803803803C07001E07001D8E001C78001C00000E00000E00000E00
000E00000700000700000700000700000380000380000380003F800003C000>
	 17 35 -3 34 21] 98 @dc
@F7 @sf
[<
01F000070C000C06001C03001803803801C03801C07001E07001E07001E0F001E0F001E0F0
01E0F001E0F001E0F001C0F801C0F80380F40300F40600F30C00F0F8007000007000007800
003800003800001801801C03C00E03C00601C003008001C100007E00>
	 19 34 -2 32 24] 54 @dc
@F7 @sf
[<
7FF3FF80070038000700380007003800070038000700380007003800070038000700380007
00380007003800070038000700380007003800070038000700380007003800070038000700
380007003800FFFFF800070038000700380007003800070038000700380007003800070038
000700380007003800038078000180780000C0780000703800001FD800>
	 25 35 0 34 27] 13 @dc
@F7 @sf
[<
07C000187000203800401C00F01E00F80E00F80F00F80F00700F00000F00000F00000F0000
0F00000F00000F00000F00000F00000F00000F00000F00000F00000F00000F00000F00000F
00000F00000F00000F00000F00000F00000F00000F00000F00001F0003FFF0>
	 20 35 -2 33 25] 74 @dc
@F6 @sf
[<
FFFFFFE00007C003F00003C000700003C000300003C000180003C000080001E0000C0001E0
00040001E000020001E000020000F000020000F004010000F004000000F004000000780600
0000780600000078060000007FFE0000003C070000003C030000003C010000003C01000000
1E008000001E008080001E000080001E000080000F000080000F000080000F0000C0000F00
00C000078000C000078000C000078003C0007FFFFFC0>
	 34 34 -2 33 36] 69 @dc
@F10 @sf
[<
3F000041C000E0E000F07000F07000607000003800003800003800003800001C00001C0000
1C00001C00000E00000E00000E00000E00000700000700000700000700007FE0>
	 19 23 -2 22 19] 74 @dc
/@F11 @newfont
@F11 @sf
[<
1F0020C040608030E038E0384038003800380030207030E02F80200020002000200024003F
C03FE02030>
	 13 21 -2 20 18] 53 @dc
@F6 @sf
[<
C00000C00000C000006000006000006000003000003000003000001800001800001800000C
00000C00000C00000600000600000600000300000300000300000180000180000180000180
0000C00000C00000C000006000006000006000003000003000003000001800001800001800
000C00000C00000C00000600000600000600000300000300000300000180000180000080>
	 17 49 -3 36 24] 61 @dc
@F11 @sf
[<
FFF07FF03FF0101808080C0806000300018000C000600070007000384038E038C038803040
7030E00F80>
	 13 21 -2 20 18] 50 @dc
@F7 @sf
[<
000FFE0000E00000E00000E00000E00000E00000E00000E00000E00000E001F0E0070CE00C
02E01C01E03801E07800E07000E0F000E0F000E0F000E0F000E0F000E0F000E0F000E07800
E07800E03801E01C01600E026007046001F820>
	 23 31 -2 20 26] 113 @dc
[<
FEFEC0C0C0C0C0C0C0C0C0C0C0C0C0C0C0C0C0C0C0C0C0C0C0C0C0C0C0C0C0C0C0C0C0C0C0
C0C0C0C0C0C0C0C0C0C0FEFE>
	 7 49 -5 36 14] 91 @dc
@F7 @sf
[<
FFFC3FFF0FC003F0078001E0078001E0078001E0078001E0078001E0078001E0078001E007
8001E0078001E0078001E0078001E0078001E0078001E0078001E0078001E007FFFFE00780
01E0078001E0078001E0078001E0078001E0078001E0078001E0078001E0078001E0078001
E0078001E0078001E0078001E0078001E00FC003F0FFFC3FFF>
	 32 34 -2 33 37] 72 @dc
[<
FEFE0606060606060606060606060606060606060606060606060606060606060606060606
06060606060606060606FEFE>
	 7 49 -1 36 14] 93 @dc
[<
0000007C00FFFC01E2000FC003C100078007C08007800FC08007800F808007800F80000780
0F800007800F800007800F000007800F000007800F000007800F000007800E000007801E00
0007801C00000780380000078070000007FFE0000007803C000007800E0000078007800007
8007C000078003C000078003E000078003E000078003E000078003E000078003E000078003
C000078007C000078007800007800E00000F803C0000FFFFE00000>
	 33 35 -2 33 36] 82 @dc

27 @eop0

0 0 27 @bop1 910 45 a @F7 @sf({)17 b(27)g({)-1040 208 y(Fig.)28 b(4.|)18 b
(sho)o(ws)h(the)f(v)o(elo)q(cit)o(y)f(distribution)h(at)h(Baade's)f(Windo)o
(w)h(in)f(gra)o(y)g(scale)g(as)h(the)g(result)f(of)-1950 97 y(20,000)k
(sim)o(ulated)d(orbits.)33 b(The)20 b(distribution)g(for)g(the)g(prograde)h
(\(righ)o(t)f(column\))g(and)g(retrograde)-1949 96 y(orbits)e(\(left)g
(column\))f(should)i(b)q(e)g(compared)f(with)g(the)g(distribution)g(of)g
(the)g(metal)g(ric)o(h)f(and)i(metal)-1951 96 y(p)q(o)q(or)27 b(sample)d
(in)h(Figure)f(1)i(resp)q(ectiv)o(e)o(ly)l(.)o 47 x(The)f(elli)o(pses)f
(sho)o(w)i(the)e(1.5)i @F6 @sf(\033)g @F7 @sf(and)g(2.5)f @F6 @sf(\033)i
@F7 @sf(lev)o(e)o(ls)d(of)-1950 97 y(the)d(v)o(elo)q(cit)o(y)o 20 x
(ellipsoids.)35 b(The)21 b(bar's)g(ma)s(jor)g(axis)g(is)g(assumed)h(to)f
(b)q(e)g(p)q(oin)o(ting)h(to)o(w)o(ard)f(longitude)-1950 96 y @F6 @sf(l)p
7 w @F10 @sf(bar)20 -7 y @F7 @sf(=)d @F5 @sf(\000)p @F7 @sf(45)p -18 w
@F10 @sf(o)3 18 y @F7 @sf(.)28 b(As)19 b(stars)g(with)f(a)i(\014xed)e
(sense)g(of)h(rotation)h(cannot)f(ha)o(v)o(e)f(zero)g(angular)i(momen)o
(tum,)o -1951 96 a(there)j(is)h(an)h(empt)o(y)e(region)h(near)g @F6 @sf(V)p
7 w @F10 @sf(l)29 -7 y @F7 @sf(=)k(0.)45 b(This)24 b(empt)o(y)f(region,)i
(ho)o(w)o(ev)o(er)e(will)g(b)q(e)h(\014lled)f(b)o(y)h(a)-1950 96 y(t)o
(ypical)14 b(20)j(km/s)e(observ)m(ational)h(v)o(elo)q(cit)o(y)e(errors.)
21 b(F)l(urther,)15 b(metal)g(ric)o(h/p)q(o)q(or)h(stars)g(ma)o(y)f(not)h
(all)f(on)-1949 97 y(prograde/retrogra)q(de)i(orbits.)-583 183 y(Fig.)23 b
(5.|)17 b(sho)o(ws)h(the)f @F6 @sf(V)p 7 w @F10 @sf(l)14 -7 y @F5 @sf(\000)
11 b @F6 @sf(V)p 7 w @F10 @sf(r)20 -7 y @F7 @sf(v)o(elo)q(cit)o(y)k
(ellipsoids)h(at)i(minor)e(axis)h(and)h(o\013-axis)g(\014elds)f(of)g(the)g
(Bulge)-1951 96 y(\()p @F6 @sf(l)q(;)8 b(b)p @F7 @sf(\))18 b(with)i(the)f
(bar)g(at)h(four)g(p)q(ossible)f(angles)h @F6 @sf(l)p 7 w @F10 @sf(bar)2
-7 y @F7 @sf(.)30 b(The)20 b(solid)f(and)h(dotted)f(ellipses)f(sho)o(w)i
(the)f(1)g @F6 @sf(\033)-1948 96 y @F7 @sf(con)o(tours)12 b(for)g
(prograde)h(and)g(retrograde)f(orbits)g(resp)q(ectiv)o(el)o(y)l(.)18 b
(The)12 b @F6 @sf(V)p 7 w @F10 @sf(r)15 -7 y @F7 @sf(o\013sets)h(b)q(et)o
(w)o(een)e(the)h(cen)o(troids)-1951 97 y(of)18 b(the)f(prograde)i(and)f
(retrograde)g(ellipses)e(at)i(o\013-axis)g(\014elds)f(are)h(due)f(to)h
(rotation,)g(not)g(triaxialit)o(y)l(.)o -1951 96 a(Notice)h(the)g(lo)o(w)h
(latitude)f(\014elds)h(sho)o(w)h(more)e(ob)o(vious)h(v)o(ertex)f
(deviations)g(than)i(the)f(high)g(latitude)-1951 96 y(\014elds,)15 b(b)q
(ecause)i(regular)f(prograde/retrogra)q(de)h(orbits)g(are)f(concen)o
(trated)g(to)g(the)g(plane.)-1711 183 y(Fig.)50 b(6.|)26 b(sho)o(ws)h(the)e
(sim)o(ulated)g(surface)h(densit)o(y)f(of)h(the)g(prograde)h(b)q(o)o(xy)f
(orbits)g(in)g(con)o(tour)-1950 97 y(map)20 b(\(solid)f(lines\))g(and)h
(gra)o(y)g(scale)f(for)h(comparison)g(with)g @F3 @sf(COBE)g @F7 @sf
(bulge.)31 b(The)20 b(dashed)g(con)o(tours,)-1950 96 y(re\015ections)27 b
(of)h(the)f(righ)o(t)h(side)f(con)o(tours,)k(are)c(added)h(to)g(sho)o(w)h
(the)e(asymmetry)g(of)h(a)g(triaxial)-1951 96 y(distribution.)21 b(The)16 b
(bar's)h(long)f(axis)g(is)g(assumed)h(to)g(b)q(e)f(p)q(oin)o(ting)g(to)o
(w)o(ard)h @F6 @sf(l)p 7 w @F10 @sf(bar)16 -7 y @F7 @sf(=)d @F5 @sf(\000)p
@F7 @sf(45)p -18 w @F10 @sf(o)3 18 y @F7 @sf(.)-1689 186 y(Fig.)22 b(7.|)
17 b(sho)o(ws)h(the)e(Jacobi's)h(in)o(tegral)f @F6 @sf(E)p 7 w @F10 @sf(J)
22 -7 y @F7 @sf(in)g(units)h(of)g(10)p -18 w @F11 @sf(5)3 18 y @F7 @sf(\()p
(km)p @F6 @sf(=)p @F7 @sf(s)p(\))p -18 w @F11 @sf(2)18 18 y @F7 @sf(of)g
(eac)o(h)f(star)i(computed)e(with)-1950 96 y(equation)g(\(6\))g(vs)g(its)f
(metallici)o(t)o(y)f([F)l(e)p @F6 @sf(=)p @F7 @sf(H])g(from)i(Ric)o(h)f
(\(1988\).)22 b(A)16 b(t)o(ypical)e(error)i(bar)g(is)g(plotted)f(based)-1949
96 y(on)23 b(uncertain)o(ties)f(in)g(metallici)o(t)o(y)l(,)o 23 x(v)o(elo)q
(cit)o(y)f(and)i(line-of-sigh)o(t)g(distance.)41 b(W)l(e)22 b(\014nd)h(a)h
(marginally)-1951 96 y(larger)16 b @F6 @sf(E)p 7 w @F10 @sf(J)22 -7 y
@F7 @sf(for)h(the)f(metal)f(p)q(o)q(or)j(stars)f(than)g(for)f(the)g(metal)g
(ric)o(h)f(stars.)-1358 160 y
27 @eop1

26 @bop0
@F10 @sf
[<
07800C6018101810180818080C040C040C04860446064606260E1C04>
	 15 14 -1 13 17] 118 @dc
@F7 @sf
[<
0007F000003C0C0800E0031801C000B8038000B8070000780F0000781E0000781E0000783C
0000783C0000787C00007878000078780000F8F8001FFFF8000000F8000000F8000000F800
0000F8000000F8000000F800000078000008780000087C0000083C0000183C0000181E0000
181E0000380F00003807000078038000F801C001B800E00218003C0C180007F008>
	 32 36 -3 34 38] 71 @dc
@F6 @sf
[<
FFFFC00001F0000000F0000000F0000000F0000000F0000000780000007800000078000000
780000003C0000003C0000003C0000003C0000001E0000001E0000001E0000001E0000000F
0000000F0000000F0000000F0000000780008007800480078004400780044003C0044003C0
042003C0062003C0063001E0061801E0061E01E00E1FFFFFFE>
	 31 34 -2 33 29] 84 @dc
@F7 @sf
[<
0600C000000600C000000600C0000003006000000300600000030060000003006000000180
30000001803000000180300000018030000000C018000000C018000000C018000000C01800
0000600C0000FFFFFFFFC0FFFFFFFFC0003006000000300600000030060000003006000000
18030000001803000000180300000018030000FFFFFFFFC0FFFFFFFFC0000C018000000600
C000000600C000000600C000000600C0000003006000000300600000030060000003006000
000300600000018030000001803000000180300000018030000000C018000000C018000000
C01800>
	 34 45 -3 34 41] 35 @dc
@F6 @sf
[<
3C1F00423080E17040F0E020F0E01060E01000700800700000700000700000380000380000
3800003800201C00201C18201C3C101A3C081A1C06310C01E0F0>
	 22 21 -2 20 28] 120 @dc
@F6 @sf
[<
3E000041800080C000E06000F03000F03800601C00001C00000E00000E0003CE000C3E001C
0F001C07001C07001C07001C03801C03801C03801C03800E01C00E01C00E01C00701C08700
E08380E08380E04380E04380702300701E0030>
	 20 31 -2 20 24] 121 @dc
@F6 @sf
[<
0F01C018C620302620701E10700E10F00E10F00708F00700F00700F0070078038078038078
03803803803C01C01C01C00E01C00601C00302E001C4E00078E00000E00000700000700000
7000007000003800003800003800003800001C00001C00001C0001FC00001E>
	 23 35 -2 34 25] 100 @dc
@F10 @sf
[<
3F00C180C0C0E0606060006003E00FC01E003800187018700C2007C0>
	 12 14 -2 13 16] 115 @dc
@F6 @sf
[<
000078000000FE000000FF000001FF000001C3800001C0800001C0C00001C0400000C02000
FEC02007C1C0200E40F0001C40B80038408E007840870070210380F01E03C0F00001E0E000
00E0E00000F0E0000078E0000078E000003CF000003CF000001CF000001EF000001EF00000
1E7800000F7800000F7800000F3C00000F3C00000F1E00000F0E00000F0F00000F0700000E
0380000E01C0001E00E0001C0070003800380038000E0060000381C000007F00>
	 32 45 -3 34 39] 81 @dc
@F6 @sf
[<
70F8F8F870>
	 5 5 -4 4 14] 58 @dc
@F6 @sf
[<
0000001800000078000001E00000078000001E00000078000003E000000F8000003C000000
F0000003C000000F0000003C000000F0000000F00000003C0000000F00000003C0000000F0
0000003C0000000F80000003E0000000780000001E0000000780000001E000000078000000
18>
	 29 28 -4 25 38] 60 @dc
[<
FFFFFF0007C01F0003C0078003C0018003C0008003C000C001E0004001E0004001E0002001
E0002000F0001000F0001000F0000000F0000000780000007800000078000000780000003C
0000003C0000003C0000003C0000001E0000001E0000001E0000001E0000000F0000000F00
00000F0000000F000000078000000780000007C000007FFE00>
	 28 34 -2 33 33] 76 @dc
@F10 @sf
[<
81C043E03C301008080804000200018000400020101018780FCC0784>
	 14 14 -1 13 16] 122 @dc
@F5 @sf
[<
70F8F8F870>
	 5 5 -4 14 14] 1 @dc

26 @eop0

0 0 26 @bop1 910 45 a @F7 @sf({)17 b(26)g({)-1040 256 y(Fig.)36 b(1.|)21 b
(The)g(upp)q(er)g(ro)o(w)h(of)f(plots)h(sho)o(w)f(the)g(distribution)g(of)g
(stars)h(in)f(the)g @F6 @sf(V)p 7 w @F10 @sf(l)17 -7 y @F5 @sf(\000)14 b
@F6 @sf(V)p 7 w @F10 @sf(r)25 -7 y @F7 @sf(plane)21 b(for)-1950 96 y(the)f
(whole)f(sample)h(at)g(Baade's)g(Windo)o(w,)g(the)f(metal)g(p)q(o)q(o)q(r)i
(sample,)f(and)g(the)g(metal)f(ric)o(h)f(sample)-1950 96 y(resp)q(ectiv)o
(el)o(y)l(.)j(The)16 b(lo)o(w)o(er)g(t)o(w)o(o)g(ro)o(ws)h(plot)g(the)f
(sample)g(distributions)g(in)g @F6 @sf(V)p 7 w @F10 @sf(l)14 -7 y @F5 @sf
(\000)11 b @F6 @sf(V)p 7 w @F10 @sf(b)19 -7 y @F7 @sf(plane)17 b(and)g
@F6 @sf(V)p 7 w @F10 @sf(r)14 -7 y @F5 @sf(\000)11 b @F6 @sf(V)p 7 w
@F10 @sf(b)-1947 90 y @F7 @sf(plane.)21 b(F)l(or)16 b(eac)o(h)f @F6 @sf(V)p
7 w @F10 @sf(l)12 -7 y @F5 @sf(\000)9 b @F6 @sf(V)p 7 w @F10 @sf(r)19 -7 y
@F7 @sf(diagram,)16 b(the)g(cross)g(near)f(the)h(cen)o(ter)e(and)i(the)g
(dashed)g(lines)f(indicate)f(the)-1950 96 y(origin,)19 b(the)f(v)o(ertex)g
(deviation)g @F6 @sf(l)p 7 w @F10 @sf(v)22 -7 y @F7 @sf(of)h(the)f(\014tted)h
(ellipsoid)e(and)j(the)e(cen)o(troid)g(of)h(the)f(ellipsoid.)28 b(The)-1950
96 y(v)o(ertex)14 b(deviations)i(for)g(the)f(metal)g(ric)o(h)g(and)h(the)f
(metal)g(p)q(o)q(or)j(sample)d(are)h(signatures)g(for)g(triaxialit)o(y)o
-1951 96 a(of)d(the)g(Bulge.)19 b(The)13 b(lac)o(k)f(of)i(clear)e(v)o
(ertex)f(deviations)i(in)f(the)h @F6 @sf(V)p 7 w @F10 @sf(l)7 -7 y @F5 @sf
(\000)t @F6 @sf(V)p 7 w @F10 @sf(b)16 -7 y @F7 @sf(and)g @F6 @sf(V)p 7 w
@F10 @sf(r)8 -7 y @F5 @sf(\000)t @F6 @sf(V)p 7 w @F10 @sf(b)16 -7 y @F7 @sf
(planes)g(is)f(consisten)o(t)-1950 97 y(with)18 b(the)f(Bulge)h(b)q(eing)g
(symmetric)o 17 x(with)g(resp)q(ect)f(to)h(the)g(Galactic)f(plane.)27 b
(The)17 b(disp)q(ersion)i(tensor)-1950 96 y(is)d(giv)o(en)f(in)h(T)l(able)g
(1.)-405 275 y(Fig.)35 b(2.|)21 b(The)g(solid)f(curv)o(e)g(sho)o(ws)i(the)f
(cum)o(ulativ)n(e)e(distribution)i(of)g(the)g @F6 @sf(T)27 b @F7 @sf
(statistic)21 b(de\014ned)f(in)-1950 96 y(equation)14 b(\(3\).)21 b(The)
14 b(star)h(sym)o(b)q(ol)f(marks)g(the)g(statistic)g(and)h(its)f
(signi\014cance)g(to)g(dra)o(w)h(our)g(metal)e(ric)o(h)-1951 97 y(and)k
(metal)e(p)q(o)q(or)j(samples)e(from)g(the)g(same)h(paren)o(t)f
(distribution.)-1265 278 y(Fig.)23 b(3.|)16 b(sho)o(ws)i(the)e(p)q
(ossible)h(orbits)g(of)g(an)h(actual)f(star)g(\(Arp)f(#)h(1102\))h(in)o
(tegrated)e(from)h(Baade's)-1950 96 y(Windo)o(w.)k(The)15 b(orbits)g(are)g
(pro)s(jected)f(on)h(the)g(bar's)g @F6 @sf(x)8 b @F5 @sf(\000)g @F6 @sf(y)
17 b @F7 @sf(plane.)j(The)15 b(left)f(panels)h(sho)o(w)g(retrograde)-1949
96 y(orbits)j(with)f(initial)f(line)h(of)g(sigh)o(t)h(distance)f @F6 @sf
(d)p 7 w @F10 @sf(l)p(os)3 -7 y @F7 @sf(=7.5)h(kp)q(c)f(while)g(the)g
(righ)o(t)g(panels)h(sho)o(ws)g(prograde)-1949 97 y(orbits)i(with)f @F6 @sf
(d)p 7 w @F10 @sf(l)p(os)3 -7 y @F7 @sf(=8.5)h(kp)q(c.)31 b(The)19 b(lo)o
(w)o(er)g(panels)g(are)h(launc)o(hed)f(with)g(the)h(observ)o(ed)f(v)o(elo)q
(cit)n(y)f(o\013set)-1949 96 y(b)o(y)g(20)h(km/s.)28 b(F)l(or)19 b(the)f
(prograde)h(orbits,)g(the)f(axis)h(ratio)g @F6 @sf(Q)e @F7 @sf(=)g(0)p
@F6 @sf(:)p @F7 @sf(75)j(and)f(the)f(angular)h(momen)o(tum)-1951 96 y(8)f
@F6 @sf(<)g(L)p 7 w @F10 @sf(z)22 -7 y @F6 @sf(<)g @F7 @sf(103)i(kp)q(c)p
@F5 @sf(\001)p @F7 @sf(km/s,)e(while)g(for)h(the)f(retrograde)h(orbits,)g
@F6 @sf(Q)f @F7 @sf(=)g(1)p @F6 @sf(:)p @F7 @sf(07)h(and)g @F5 @sf(\000)p
@F7 @sf(106)g @F6 @sf(<)f(L)p 7 w @F10 @sf(z)22 -7 y @F6 @sf(<)f @F5 @sf
(\000)p @F7 @sf(71)-1949 96 y(kp)q(c)p @F5 @sf(\001)p @F7 @sf(km/s.)j
(Notice)12 b(the)h(axis)g(ratio)g(of)h(the)f(p)q(oten)o(tial)g(0)p @F6 @sf
(:)p @F7 @sf(88)h(is)f(in)g(the)g(range)g(b)q(ounded)h(b)o(y)f(the)g
(prograde)-1949 97 y(orbits)k(and)g(the)f(retrograde)h(orbits.)23 b(These)
16 b(orbits)h(sho)o(w)g(that)g(the)g(crucial)e(data)i(for)g(classifying)f
(the)-1950 96 y(orbit)g(of)h(a)f(star)h(is)f(the)g(line)f(of)i(sigh)o(t)f
(distance)g(instead)g(of)h(the)f(absolute)h(prop)q(er)f(motion.)-1721 207 y
26 @eop1

25 @bop0
@F7 @sf
[<
FFFFFEF8007E78001E3C000E3C00061E00021E00020F000307800307800103C00103C00101
E00101F00000F000007800007800003C00003C00001E00001F00000F000007808007808003
C08003C0C001E0C000F04000F060007870007878003C7E003E7FFFFE>
	 24 34 -3 33 30] 90 @dc
@F3 @sf
[<
003F800001E07040038008C0060004C00E0003E01C0001E0380001E0780001E0780001E0F8
0001E0F00000F0F00000F0F00000F0F00001F0F0003FFFF0000000F8000000F8000000F800
0000F8000000F80000007C0000007C0000043C0000043E0000041E0000041E00000C0F0000
0C0780000E0380000E01C0001E00E0001E0070002E001C0046000701830000FE01>
	 32 36 -6 34 38] 71 @dc
[<
1F83C07867A0F01710F00F08F00F08F00708F007087807003807001C03800F038003E38000
7F800003800003800001C00601800F03800F030006060003FC00>
	 21 21 -3 20 24] 97 @dc
[<
FFE00F000E0007000700070007000700070003800380038003800380038001C001C001C001
C001C001C000E000E000E000E000E000E000700070007000700070007003F80078>
	 13 35 -1 34 14] 108 @dc
[<
07E0001C1800380600700100700100F00080F00000E00000F00000F00000F00000F00000F0
00007000007800003800001C01800C03C00603C00381C000FF00>
	 18 21 -4 20 22] 99 @dc
[<
07801C4038203820381038103810381038101C001C001C001C001C001C000E000E000E000E
000E00FFF81F000F0007000300030001000180008000800080>
	 13 31 -4 30 19] 116 @dc
[<
FFE00F000E0007000700070007000700070003800380038003800380038001C001C001C001
C00FC001C00000000000000000000000000000000000E001F001F000F00060>
	 12 34 -1 33 14] 105 @dc
[<
07E3F81C13C0380B803805C03803C03801C03801C03801C03801C01C00E01C00E01C00E01C
00E01C00E01C00E00E00700E00700E00700E00F0FE07F00E0070>
	 21 21 -4 20 27] 117 @dc
[<
03FC00001C0780003000C00060006000C0003000C0001800C0001800600018006000180030
0018001C00780007FFF00007FFE00007FF800006000000040000000400000004000000027E
000002C300000181C0000381E0000780E0000780F0000780F0000780F00003C0700003C070
0001C0700000E0E1800070D180001F888000000780>
	 25 33 0 21 24] 103 @dc
[<
03F0000E0C00180300300080700080F00040E00000E00000F00000F00000F00000F00000FF
FFE07000E07800E03800E01C00E00C00E00701C003838000FE00>
	 19 21 -3 20 22] 101 @dc
[<
87C0D830E008600C6006400640064006000E007E07FC0FF81FE01C0018021801180118030C
03060701F9>
	 16 21 -2 20 19] 115 @dc
@F7 @sf
[<
FFFFFFE00F8003E0078000E007800070078000300780003007800010078000100780001007
8000080780000807802008078020000780200007802000078060000780E00007FFE0000780
E0000780600007802000078020000780200007802020078000200780002007800020078000
600780004007800040078000C0078001C00F8007C0FFFFFFC0>
	 29 34 -2 33 33] 69 @dc
[<
FFFFF8000F801E0007800700078003C0078001E0078000E0078000F0078000780780007807
80007C0780003C0780003C0780003E0780003E0780003E0780003E0780003E0780003E0780
003E0780003E0780003E0780003C0780003C0780003C0780007C0780007807800078078000
F0078000E0078001C0078003C0078007000F801E00FFFFF000>
	 31 34 -2 33 37] 68 @dc
[<
FFFC07FF800FC000FC00078000F800078000F000078001E000078001E000078003C0000780
07C000078007800007800F000007800F000007801E000007803C000007C03C000007A07800
000790F000000788F000000789E000000787E000000783C000000781800000078080000007
80400000078020000007801000000780080000078004000007800200000780010000078000
8000078000400007800060000FC000F800FFFC03FF00>
	 33 34 -2 33 38] 75 @dc
[<
FFFC00000FC000000780000007800000078000000780000007800000078000000780000007
80000007800000078000000780000007800000078000000780000007FFF00007803C000780
0F0007800780078007C0078003C0078003E0078003E0078003E0078003E0078003E0078003
E0078003C0078007C00780078007800F000F803C00FFFFF000>
	 27 34 -2 33 33] 80 @dc
[<
70F8F8F870000000000000000000000070F8F8F870>
	 5 21 -4 20 14] 58 @dc
[<
FFE0203FFF1F007003F00E007001E004007001E00400F801E00400F801E00400F801E00401
E401E00401E401E00401E401E00403C201E00403C201E004078101E004078101E004078101
E0040F0081E0040F0081E0040F0081E0041E0041E0041E0041E0043C0021E0043C0021E004
3C0021E004780011E004780011E004780011E004F00009E004F00009E004F00009E005E000
05E005E00005E007C00003E00FC00003F0FFC00003FF>
	 40 34 -2 33 45] 77 @dc
[<
0007E00000381C0000E0020001C0010003800080070000400E0000401E0000201C0000203C
0000103C0000107C0000107800001078000000F8000000F8000000F8000000F8000000F800
0000F8000000F8000000F800000078000010780000107C0000103C0000303C0000301C0000
301E0000700E000070070000F0038000F001C0017000E00630003818300007E010>
	 28 36 -3 34 35] 67 @dc
[<
0003F000001C0800003006000060010000E0008001C0008003C0004003C000400380004007
80002007800020078000200780002007800020078000200780002007800020078000200780
00200780002007800020078000200780002007800020078000200780002007800020078000
2007800020078000200780002007800020078000700FC000F8FFFC07FF>
	 32 35 -2 33 37] 85 @dc
[<
FFFFFF000F803F0007800F0007800300078003000780010007800180078001800780008007
80008007800080078000800780000007800000078000000780000007800000078000000780
00000780000007800000078000000780000007800000078000000780000007800000078000
00078000000780000007800000078000000FC00000FFFE0000>
	 25 34 -2 33 30] 76 @dc
@F11 @sf
[<
FE03FE3800F01800E00800E00800E00401C00401C007FFC002038002038003078001070001
0700008E00008E00008E00005C00005C00005C00003800003800003800001000>
	 23 23 -1 22 26] 65 @dc
@F7 @sf
[<
FFE007FFC01F8001FC00070000F800030000F000010001F000008001E00000C003E0000040
07C000002007800000300F800000101F000000081E0000000C3E000000047C000000027800
000003F800000001F000000001E000000003E000000007C000000007A00000000FB0000000
1F100000001E080000003E0C0000007C040000007802000000F803000000F001000001E000
800003E000C00003C000E00007E001F8007FF807FF00>
	 34 34 -1 33 37] 88 @dc

25 @eop0

0 0 25 @bop1 910 45 a @F7 @sf({)17 b(25)g({)-1040 147 y(de)j(Zeeu)o(w,)h
(T.)f(1993,)j(in)d @F3 @sf(Galactic)g(Bulges)p @F7 @sf(,)h(Eds.)35 b(H.)
20 b(Dejonghe)h(and)g(H.J.)e(Habing)i(\(Klu)o(w)o(er)-1701 96 y(Academic)
15 b(Publ.:)21 b(Netherlands\),)15 b(p)h(191.)-954 117 y(Zhao,)g(H.S.,)f
(Ric)o(h,)g(R.M.,)o 15 x(Applegate,)g(J.H.,)g(Biell)o(o,)g(J.)h(1994,)h
(ApJ,)f(in)g(press.)-1513 117 y(Zhao,)g(H.S.)f(1994)q(,)i(Ph.)k(D.)16 b
(thesis,)g(Colum)o(bia)g(Univ)o(ersi)o(t)o(y)l(.)o -1132 2057 a 780 -2 z
88 79 a(This)g(man)o(uscript)g(w)o(as)h(prepared)f(with)g(the)g(AAS)f(L)
-15 -9 y @F11 @sf(A)-7 9 y @F7 @sf(T)-8 11 y(E)-5 -11 y(X)h(macros)h
(v3.0.)-1471 136 y
25 @eop1

24 @bop0
@F7 @sf
[<
03F003F0000E0C0E0C003C031C04003800B802007800700300F000E00100F001D00100F003
880000F007880000F007040000700E020000701E020000301C010000183C00800008380080
000470004000027000400001F000200001E000300000E000380001C0007C0001E003FF8001
D000000001C800000003880000000384000000038200000003820000000381000000038100
0000038100000003810000000181000000018100000000C20000000062000000003C000000
>
	 33 37 -2 35 38] 38 @dc
[<
003FFF00000003E000000001E000000001E000000001E000000001E000000001E000000001
E000000001E000000001E000000001E000000001E000000001E000000001E000000003E000
000003D000000007D800000007880000000F840000001F040000001E020000003E01000000
3C010000007C00800000F800C00000F000400001F000200001E000200003E000100007C000
180007800008000F80000C001F80001F00FFF0007FC0>
	 34 34 -1 33 37] 89 @dc
/@F4 @newfont
@F4 @sf
[<
0F80306070186004E002E002E000E000E000E000F000F000FFE0F018780438023C021C020E
02038400F8>
	 15 21 -6 20 22] 101 @dc
[<
1E003100708070407020702038103800380038001C001C001C001C000E000E000E000E0007
000700FFF80700038003800380038001C001C001C001C000C0>
	 13 31 -4 30 16] 116 @dc
[<
0F0780308C40305C40703C20701C20F01C20F00E10F00E00F00E00F00E0078070078070078
07003807003C03801C03800E03800E03800705C00185C000F8C0>
	 20 21 -5 20 25] 97 @dc
[<
38006400E200E200E200E200710070007000700038003800380038001C001C001C001C000E
000E000E000E000700070007000700038003800380038001C001C001C01FC001E0>
	 11 35 -4 34 12] 108 @dc
[<
E0F0F8F870>
	 5 5 -6 4 15] 46 @dc
@F7 @sf
[<
0000C000000000C000000000C000000001E000000001E000000003F000000003D000000003
D0000000078800000007880000000F8C0000000F040000000F040000001F020000001E0200
00001E020000003C010000003C010000007C0180000078008000007800800000F000400000
F000400000F000400001E000200001E000200003E000300003C000100003C0001000078000
080007800008000780000C000F00000C001F80001F00FFF0007FC0>
	 34 35 -1 33 37] 86 @dc
[<
FFFC0FC0078007800780078007800780078007800780078007800780078007800780078007
8007800780078007800780078007800780078007800780078007800FC0FFFC>
	 14 34 -2 33 18] 73 @dc
@F7 @sf
[<
381C7C3EFC7EFC7EB85C8040804080408040402040202010201010080804>
	 15 15 -6 34 24] 92 @dc
[<
4020201010081008080408040402040204020402743AFC7EFC7EF87C7038>
	 15 15 -2 34 24] 34 @dc

24 @eop0

0 0 24 @bop1 910 45 a @F7 @sf({)17 b(24)g({)-1040 147 y(Lynden-Bell,)e(D.)h
(1994,)h(priv)m(ate)f(comm)o(unication)-972 117 y(Merri\014eld,)o 15 x(M.)g
(R.)f(&)h(Kuijk)o(en,)f(K.)h(1993,)h(submitted)f(to)g(ApJ.)-1216 117 y
(Minniti,)e(D.)i(1993,)i(Ph.D.)e(thesis,)f(Univ)o(ersit)n(y)g(of)h
(Arizona.)-1144 117 y(Nak)m(ada,)h(Y.)e @F4 @sf(et)j(al.)23 b @F7 @sf
(1991,)17 b(Nature)f(353,)h(No.)k(6340,)d(p.140.)-1132 117 y(Pfenniger,)d
(D.)h(&)h(F)l(riedli)o(,)e(D.)h(1991,)h(A&A,)e(252,)i(No.)k(1,)16 b(75.)
-1170 117 y(Raha,)j(N.,)e(Sellw)o(o)q(o)q(d,)h(J.)f(A.,)g(James,)h(R.)f
(A.)g(&)h(Kahn,)h(F.)e(D.)h(1991,)h(Nature,)f(v.)25 b(352,)19 b(No.)27 b
(6334,)-1764 96 y(p.103.)-273 117 y(Ric)o(h,)15 b(R.M.,)o 15 x(1988,)j
(AJ,)d(95,)i(828.)-648 117 y(Ric)o(h,)e(R.M.,)o 15 x(1990,)j(ApJ,)d(362,)i
(604.)-699 117 y(Sadler,)f(E.M.,)e(T)l(erndrup,)i(D.M.,)f(&)h(Ric)o(h,)f
(R.M.)g(1994,)i(in)f(preparation.)-1414 117 y(Searle,)f(L.)h(&)h(Zinn,)e
(R.)h(1978,)h(ApJ,)f(225,)h(357.)-900 117 y(Sellw)o(o)q(o)q(d,)f(J.,)g
(1993,)i('Bac)o(k)d(to)i(the)f(Galaxy')g(the)h(3rd)f(Maryland)h
(conference,)e(ed.)22 b(Stephen)16 b(S.)g(Holt)-1798 97 y(and)h(F)l
(rances)f(V)l(erter.)-566 117 y(Sharples,)g(R.,)f(W)l(alk)o(er,A.)o(,)g(&)h
(Cropp)q(er,)g(M.)g(1990,)h(MNRAS,)e(246,)i(54.)-1367 117 y(Spaenhauer,)f
(A.)g(,)g(Jones,)g(B.F.)f(,)g(&)i(Whitford,)e(A.E.)h(1992,)h(AJ,)e(103,)i
(297.)-1459 117 y(T)o(yson,)f(N.D.,)f(&)h(Ric)o(h,)f(R.M.,)o 15 x(1991,)j
(ApJ,)d(326,)i(547.)-1044 117 y(Udry)l(,)e(S.)h(&)g(Pfenniger,)g(D.,)f
(1988,)j(A&A,)c(198,)j(135.)-1015 117 y(de)f(V)l(aucouleurs,)h(G.)f(1964,)i
(IA)o(U)d(Symp)q(osium)i(20:)23 b(The)16 b(Galaxy)h(and)g(the)g
(Magellanic)f(Clouds,)h(ed.)-1794 96 y(F.J.)f(Kerr)f(and)i(A.W.)e(Ro)q
(dgers)i(\(Sydney:)k(Australian)16 b(Academ)o(y)f(of)h(Science\),)f(195.)
-1781 117 y(W)l(eiland,)g(J.)h @F4 @sf(et)i(al.)23 b @F7 @sf(1994,)17 b
(ApJ,)f(L81.)-727 117 y(Whitelo)q(c)o(k,)e(P)l(.)g(A.,)g(Catc)o(hp)q(ole,)h
(R.)f(1992,)j(\\The)e(Cen)o(ter,)f(Bulge,)g(and)i(Disk)e(of)i(the)e(Milky)g
(W)l(a)o(y",)h(ed.)-1805 97 y(b)o(y)h(Leo)h(Blitz.)o -424 227 a
24 @eop1

23 @bop0
/@F2 @newfont
@F2 @sf
[<
FFFF800FF0FFFF803FF807F000FF0C07F000FE0607F001FC0607F001FC0007F001FC0007F0
01FC0007F001FC0007F001FC0007F001FC0007F001FC0007F001FC0007F003F80007F007F0
0007F00FE00007FFFF800007FFFFC00007F007F00007F001F80007F000FC0007F0007E0007
F0007F0007F0007F0007F0007F0007F0007F0007F0007F0007F0007F0007F0007E0007F000
FC0007F001F80007F007F000FFFFFFC000FFFFFE0000>
	 39 34 -2 33 42] 82 @dc
[<
FFFFFFFCFFFFFFFC07F001FC07F0003E07F0001E07F0000E07F0000E07F0000607F0000607
F0000607F0000307F0180307F0180307F0180007F0180007F0380007F0780007FFF80007FF
F80007F0780007F0380007F0180007F0180607F0180607F0180607F0000607F0000E07F000
0E07F0000C07F0001C07F0003C07F000FCFFFFFFFCFFFFFFFC>
	 32 34 -2 33 37] 69 @dc
[<
FFFFE000FFFFE00007F0000007F0000007F0000007F0000007F0000007F0000007F0000007
F0000007F0180007F0180007F0180007F0180007F0380007F0780007FFF80007FFF80007F0
780007F0380007F0180007F0180007F0180C07F0180C07F0000C07F0000C07F0001C07F000
1C07F0001807F0003807F0007807F001F8FFFFFFF8FFFFFFF8>
	 30 34 -2 33 35] 70 @dc
[<
FFF00000C0FFF00001C006000003C006000003C006000007C00600000FC00600001FC00600
003FC00600007FC00600007FC0060000FFC0060001FEC0060003FCC0060007F8C0060007F0
C006000FF0C006001FE0C006003FC0C006007F80C00600FF00C00600FE00C00601FE00C006
03FC00C00607F800C0060FF000C0060FE000C0061FE000C0063FC000C0067F8000C006FF00
00C007FE0000C007FC0000C0FFFC001FFEFFF8001FFE>
	 39 34 -2 33 44] 78 @dc
[<
0003FE0000001FFFC00000FF00F00001F800380003F0000C0007C00006000F800003001F80
0003003F000003803F000001807F000001807E000001807E00000000FE00000000FE000000
00FE00000000FE00000000FE00000000FE00000000FE00000000FE000000007E000001807E
000001807F000001803F000003803F000003801F800007800F8000078007C0000F8003E000
1F8001F8003F8000FF01E380001FFF81800003FE0080>
	 33 34 -3 33 40] 67 @dc
[<
80FF80C7FFE0FF00F8FC003CF0003CE0001EE0001EC0001FC0001FC0001F00003F00003F00
007F0003FF003FFE03FFFE0FFFFC1FFFF83FFFF07FFFC07FFF00FFE000FF0000FC0000FC00
0CF8000CF8000C78001C78001C7C003C3C007C1F03FC07FF8C01FC04>
	 24 34 -3 33 31] 83 @dc
@F7 @sf
[<
000FE00000783C0000E00E0003C00780078003C00F0001E00F0001E01E0000F03E0000F83C
0000787C00007C7C00007C7800003CF800003EF800003EF800003EF800003EF800003EF800
003EF800003EF800003EF800003E7800003C7800003C7C00007C7C00007C3C0000783C0000
781E0000F00E0000E00F0001E0078003C003C0078000E00E0000783C00000FE000>
	 31 36 -3 34 38] 79 @dc
[<
387CFCFCB880808080404020201008>
	 6 15 -3 34 14] 96 @dc

23 @eop0

0 0 23 @bop1 910 45 a @F7 @sf({)17 b(23)g({)-256 147 y @F2 @sf(REFERENCES)
-1166 187 y @F7 @sf(A)o(thanasoula,)g(E.,)e(Biena)o(yme,)o 15 x(O.,)g
(Martinet,)g(L.,)h(&)g(Pfenniger,)g(D.)g(1983,)h(A&A,)e(127,)i(349.)-1794
117 y(Binney)l(,)d(J.J.,)g(Gerhard,)i(O.E.,)f(Stark,)g(A.A.,)f(Bally)l(,)g
(J.,)h(&)g(Uc)o(hida,)g(K.I.)f(1991,)j(MNRAS,)d(252,)i(210.)-1949 117 y
(Binney)l(,)e(J.J.)i(&)g(Sp)q(ergel,)g(D.N.)f(1984,)i(MNRAS,)e(215,)i(59.)
-1129 116 y(Binney)l(,)e(J.J.,)h(&)h(T)l(remaine,)f(S.,)h(1987,)h
(`Galactic)e(Dynamics')g(\(Princeton)h(Univ)o(ersit)o -1 x(y)e(Press,)j
(New)-1786 96 y(Jersey\).)-312 117 y(Blanco,)d(V.M.)g(1988,)i(AJ,)f(95,)g
(1400.)-707 116 y(Blitz,)o 15 x(L.)g(&)h(Sp)q(ergel,)e(D.N.)g(1991a)q(,)i
(ApJ,)f(370,)h(205.)-1009 117 y(Blitz,)o 15 x(L.)f(&)h(Sp)q(ergel,)e(D.N.)g
(1991b,)j(ApJ,)d(379,)i(631.)-1011 117 y(Blum,)e(R.D.,)g(Carr,)h(J.S.,)f
(DeP)o(o)o(y)l(,)g(D.L.,)h(Sellgren,)f(K.,)g(&)h(T)l(erndrup,)g(D.)g(M.)g
(1993,)h(AJ)f(in)g(press.)-1888 116 y(Carney)l(,)j(B.W.,)e(Latham,)j(D.W.)e
(and)h(Laird,)g(J.B.)e(1990,)k(in)d @F3 @sf(Galactic)g(Bulges)p @F7 @sf(,)g
(B.J.)f(Jarvis)i(and)-1748 96 y(D.M.)d(T)l(erndrup,)g(eds.)21 b(p.)g(127.)
-752 117 y(Com)o(b)q(es,)16 b(F.,)f(Debbasc)o(h,)h(F.,)f(F)l(riedli,)f
(D.,)i(&)g(Pfenniger,)g(D.)g(1990,)h(A&)f(A,)f(233,)i(82.)-1632 117 y(Con)o
(top)q(oulos,)h(G.)e(&)g(Grosb)q(ol,)h(P)l(.,)e(Astron)i(Astroph)o(ys)f
(Rev)g(1989,)h(1,)f(261.)-1448 116 y(Dubinski,)f(J.)h(1993,)i(submitted)d
(to)i(ApJ.)-796 117 y(Eggen,O.J,)f(Lynden-Bell,)o 15 x(D.&)g(Sandage,)h
(A.)e(1962)q(,)i(ApJ,)e(136,)i(748.)-1351 116 y(Ev)m(ans,)g(N.W.)e(&)h
(Collett,)f(J.L.)h(1993,)h(submitted)f(to)h(the)f(ApJ)g(Lett.)-1327 117 y
(Gerhard,)g(O.E.)g(&)g(Vietri,)o 15 x(M.)g(1986,)h(MNRAS,)d(223,)j(337.)
-1129 116 y(Grenon,)f(M.)g(1989,)h(Astroph)o(ysics)f(and)h(Space)f
(Science)f(\(Klu)o(w)o(er)g(Academic)o 15 x(Publisher\),)g(156,)i(29.)-1905
117 y(Hasan,)f(H.,)f(Pfenniger,)h(D.,)f(Norman,)h(C.)g(1993,)i(Jan.)j
(preprin)o(t)16 b(of)g(STScI,)g(No.)21 b(403.)-1645 116 y(Ken)o(t,)15 b
(S.E.)h(1992)i(ApJ,)d(387,181.)-642 117 y(Kormendy)l(,)g(J.,)h(1982,)h
(ApJ,)f(257,)h(75.)-729 124 y
23 @eop1

22 @bop0
@F6 @sf
[<
FFE00020000F00002000060000700002000070000200007000020000F000010000F8000100
01F800010001E800010001E800008003C400008003C400008007C400008007840000400782
0000400F020000400F020000400F020000201E010000201E010000203E010000203C010000
103C0080001078008000107800800010F800800008F000400008F000400009E000400009E0
00400005E000200007C000600007C00078007FC003FF>
	 40 34 -2 33 39] 78 @dc
@F5 @sf
[<
004000C000C0018001800180030003000300060006000C000C000C00180018001800300030
003000600060006000C000C000C000C0006000600060003000300030001800180018000C00
0C000C000600060003000300030001800180018000C000C00040>
	 10 50 -5 36 19] 104 @dc
@F5 @sf
[<
C000C000C000600060006000300030003000180018000C000C000C00060006000600030003
00030001800180018000C000C000C000C00180018001800300030003000600060006000C00
0C000C0018001800300030003000600060006000C000C000C000>
	 10 50 -3 36 19] 105 @dc
@F6 @sf
[<
00FE0000000381C0000006003000001C000800001800040000380002000070000100007000
008000F000008000E000004000E000004000E000002000E000002000E000000000F0000000
00F000000000F000000000F000000000F0000000007800000000780000000078000000003C
000008003C000004001E000004000E000004000F000004000700000E000380000E0001C000
0E0000E0000E000070001F000038002700000E006300000380810000007F0080>
	 33 36 -3 34 35] 67 @dc
@F11 @sf
[<
7FF007000700070007000700070007000700070007000700070007000700070007000700FF
0007000300>
	 12 21 -2 20 18] 49 @dc
@F11 @sf
[<
0FE030304018C00CE00EE00E000E000E000C0018003007E0006000380018001C701C701C60
1830300FE0>
	 15 21 -1 20 18] 51 @dc
@F6 @sf
[<
0FE000301800400400E00200F00300F00300700380000380000780003F0001FF0003FE0007
F800070000060000060300060380020180030080008100007E00>
	 17 21 -2 20 23] 115 @dc
/@F9 @newfont
@F9 @sf
[<
FFFFF0FFFFF0>
	 20 2 -3 9 27] 0 @dc
@F5 @sf
[<
FFFFFFFC7FFFFFF80000000000000000000000000000000000000000000000000000000000
0000000000000C0000003C000000F0000003C000000F0000003C000000F0000007C000001F
00000078000001E00000078000001E00000078000000E0000000780000001E000000078000
0001E0000000780000001F00000007C0000000F00000003C0000000F00000003C0000000F0
0000003C0000000C>
	 30 39 -4 31 39] 20 @dc
@F5 @sf
[<
FFFFFFFC7FFFFFF80000000000000000000000000000000000000000000000000000000000
000000C0000000700000003C0000000F00000003C0000000F00000003C0000000F80000003
E0000000780000001E0000000780000001E0000000780000001C00000078000001E0000007
8000001E00000078000003E000000F8000003C000000F0000003C000000F0000003C000000
F0000000C0000000>
	 30 39 -4 31 39] 21 @dc
@F6 @sf
[<
0000003C00FFFC00E30007C001C08003C001C08003C003C04003C003C04003C003C00001E0
01E00001E001E00001E001E00001E001E00000F000F00000F000F00000F000F00000F000F0
00007801E000007801C0000078078000007FFE0000003C01C000003C007000003C003C0000
3C001E00001E000F00001E000780001E000780001E000780000F0003C0000F0003C0000F00
0380000F00078000078007000007800E000007803C00007FFFE000>
	 34 35 -2 33 37] 82 @dc
@F11 @sf
[<
07C01C703018701C600C600CE00EE00EE00EE00EE00EE00EE00EE00EE00E600C600C701C30
18183007C0>
	 15 21 -1 20 18] 48 @dc
@F5 @sf
[<
FFFFFFFF7FFFFFFF0000000000000000000000000000000000000000000000000000000000
000000FFFFFFFFFFFFFFFF0000000000000000000000000000000000000000000000000000
0000000000007FFFFFFFFFFFFFFF>
	 32 22 -3 22 39] 17 @dc
@F7 @sf
[<
4040201010100808080878F8F8F870000000000000000000000070F8F8F870>
	 5 31 -4 20 14] 59 @dc
@F5 @sf
[<
800007E080001FF080003FF8C000781CC001E00E6003C00670078003381E00031FFC00010F
F8000107E00001>
	 32 11 -3 17 39] 24 @dc
@F10 @sf
[<
60F0F060>
	 4 4 -3 3 10] 58 @dc
@F10 @sf
[<
1C0032003200310031001800180018000C008C004600460024001C00000000000000000000
000000030003800300>
	 9 23 -1 22 12] 105 @dc
@F6 @sf
[<
1E0031003080704070403840382038001C001C000E000E000E000700870083808380438043
8023001E0000000000000000000000000000000000000000C001E000E000E0>
	 11 34 -2 33 17] 105 @dc

22 @eop0

0 0 22 @bop1 910 45 a @F7 @sf({)17 b(22)g({)-1040 848 y(T)l(able)f(1:)22 b
(The)16 b(Baade's)g(Windo)o(w)g(\(1)p -18 w @F10 @sf(o)3 18 y @F6 @sf(;)
8 b @F5 @sf(\000)p @F7 @sf(4)p -18 w @F10 @sf(o)2 18 y @F7 @sf(\))16 b(V)l
(elo)q(cit)o(y)f(Ellipsoid)-1244 18 y 2204 -2 z 10 w 2204 -2 z 25 68 a([F)l
(e)p @F6 @sf(=)p @F7 @sf(H])64 b @F6 @sf(N)c @F5 @sf(h)p @F6 @sf(V)p 7 w
@F10 @sf(l)3 -7 y @F5 @sf(i)123 b @F6 @sf(\033)p 7 w @F10 @sf(r)165 -7 y
@F6 @sf(\033)p 7 w @F10 @sf(l)171 -7 y @F6 @sf(\033)p 7 w @F10 @sf(b)159
-7 y @F6 @sf(C)p 7 w @F10 @sf(l)p(r)216 -7 y @F6 @sf(l)p 7 w @F10 @sf(v)
153 -7 y @F6 @sf(\033)p 7 w @F11 @sf(1)164 -7 y @F6 @sf(\033)p 7 w @F11 @sf
(2)156 -7 y @F6 @sf(\033)p 7 w @F11 @sf(3)-1745 89 y @F7 @sf(km)16 b @F6 @sf
(s)p -18 w @F9 @sf(\000)p @F11 @sf(1)52 18 y @F7 @sf(km)f @F6 @sf(s)p -18 w
@F9 @sf(\000)p @F11 @sf(1)59 18 y @F7 @sf(km)h @F6 @sf(s)p -18 w @F9 @sf
(\000)p @F11 @sf(1)59 18 y @F7 @sf(km)g @F6 @sf(s)p -18 w @F9 @sf(\000)p
@F11 @sf(1)329 18 y @F7 @sf(deg)112 b(km)16 b @F6 @sf(s)p -18 w @F9 @sf(\000)p
@F11 @sf(1)59 18 y @F7 @sf(km)g @F6 @sf(s)p -18 w @F9 @sf(\000)p @F11 @sf
(1)51 18 y @F7 @sf(km)g @F6 @sf(s)p -18 w @F9 @sf(\000)p @F11 @sf(1)-2177
49 y 2204 -2 z 25 67 a @F5 @sf(\024)d(\000)p @F7 @sf(0)p @F6 @sf(:)p @F7 @sf
(2)50 b(15)h(-20\(23\))i(142\(25\))e(92\(16\))76 b(83\(13\))68 b
(0.5\(0.21\))92 b(25\(14\))51 b(154\(25)q(\))g(77\(25\))68 b(83\(25\))-2135
96 y @F5 @sf(\025)13 b @F7 @sf(0)127 b(39)51 b(25\(19\))69 b(98\(11\))75 b
(115\(13)q(\))51 b(65\(7\))92 b(-0.26\(0.16)q(\))51 b(-65\(9\))59 b
(126\(13)q(\))51 b(89\(13\))68 b(65\(13\))-2160 31 y 2204 -2 z 86 w 780
-2 z 49 120 a(Note.)21 b(|)16 b @F6 @sf(\033)p 7 w @F10 @sf(r)2 -7 y @F7 @sf
(,)g @F6 @sf(\033)p 7 w @F10 @sf(l)18 -7 y @F7 @sf(and)h @F6 @sf(\033)p
7 w @F10 @sf(b)18 -7 y @F7 @sf(are)f(radial,)g @F6 @sf(l)q @F7 @sf(-)g
(and)h @F6 @sf(b)p @F7 @sf(-)f(prop)q(er)h(motion)f(disp)q(ersions)h(with)f
@F6 @sf(R)p 7 w @F11 @sf(0)16 -7 y @F7 @sf(=)e(8)i(kp)q(c.)-1903 96 y
@F5 @sf(h)p @F6 @sf(V)p 7 w @F10 @sf(l)3 -7 y @F5 @sf(i)26 b @F7 @sf(is)f
(the)h @F6 @sf(l)q @F7 @sf(-comp)q(onen)o(t)f(of)h(the)g(streaming)g(v)o
(elo)q(cit)n(y)e(with)i(resp)q(ect)f(to)h(the)g(whole)f(sample)h(of)-1950
97 y(Spaenhauer)17 b @F4 @sf(et)h(al.)23 b @F7 @sf(.)-408 96 y @F6 @sf(C)p
7 w @F10 @sf(l)p(r)17 -7 y @F5 @sf(\021)13 b @F6 @sf(\033)2 -18 y @F11 @sf
(2)-20 30 y @F10 @sf(l)p(r)3 -12 y @F6 @sf(=)p @F7 @sf(\()p @F6 @sf(\033)p
7 w @F10 @sf(l)2 -7 y @F6 @sf(\033)p 7 w @F10 @sf(r)3 -7 y @F7 @sf(\))j
(is)g(a)h(dimensionless)e(measure)h(of)g(the)g(o\013-diago)q(nal)i(term)d
@F6 @sf(\033)p 7 w @F10 @sf(l)p(r)3 -7 y @F7 @sf(,)g(indep)q(enden)o(t)h
(of)g @F6 @sf(R)p 7 w @F11 @sf(0)2 -7 y @F7 @sf(.)-1949 96 y @F6 @sf(l)p
7 w @F10 @sf(v)19 -7 y @F7 @sf(is)g(the)g(v)o(ertex)f(deviation,)g(giv)o
(en)h(b)o(y)f(equation)h(\(2\);)h(its)f(1)p @F6 @sf(\033)i @F7 @sf(error)e
@F5 @sf(\030)e @F7 @sf(60)p -18 w @F10 @sf(o)2 18 y @F6 @sf(N)5 -18 y
@F9 @sf(\000)p @F11 @sf(0)p @F10 @sf(:)p @F11 @sf(5)2 18 y @F7 @sf(.)-1555
96 y @F6 @sf(\033)p 7 w @F10 @sf(i)18 -7 y @F7 @sf(with)i @F6 @sf(i)e
@F7 @sf(=)f(1)p @F6 @sf(;)8 b @F7 @sf(2)p @F6 @sf(;)g @F7 @sf(3)18 b(are)e
(the)g(eigen)o(v)m(alues)f(of)i(the)f(disp)q(ersion)g(tensor.)-1315 797 y
22 @eop1

21 @bop0
@F3 @sf
[<
FFE00020001F0000200006000060000600007000020000F000020000F000020001F0000200
01F000020003D000010003C800010007C800010007880001000F880001000F080001000F08
0000801E040000801E040000803C040000803C0400008078040000807804000040F0020000
40F002000041E002000041E002000043C002000043C002000027C001000027800100002F80
0100002F000100003F000180003E0003C003FE001FF8>
	 37 34 -2 33 37] 78 @dc
[<
07F0001C1C003807007003807001C0F001E0E000E0E000F0E000F0F00078F00078F00078F0
00787000787800783800701C00700C00E00600C0038380007E00>
	 21 21 -3 20 24] 111 @dc
[<
07C3F81C33C0380F807003C07003C0E001C0E001C0E001C0E001C0F000E0F000E0F000E0F0
00E07000E07800E03800701C00700C00F00700F003837000FC700000380000380000380000
3800003800003800001C00001C00001C00001C00001C00001C0000FE00001E>
	 23 35 -4 34 27] 100 @dc
[<
FFE7FF0F00780E007007003807003807003807003807003807003803801C03801C03801C03
801C03801C03C01C01C00E01E00E01E00E01D00C1FCC1801C3F0>
	 24 21 -1 20 27] 110 @dc
[<
FFE3FF8FFE0F003C00F00E003800E007001C007007001C007007001C007007001C00700700
1C007007001C007003800E003803800E003803800E003803800E003803800E003803C00F00
3801C007001C01E007801C01E007801C01D80E60381FC60C183001C1F807E0>
	 39 21 -1 20 41] 109 @dc
[<
FFF0000F00000E000007000007000007000007000007000007000003800003800003800003
800003800003C00001C00001E0C001E1E001D1E01FC8E001C7C0>
	 19 21 -1 20 19] 114 @dc
[<
FFC000001E0000000E0000000E0000000E0000000E0000000E000000070000000700000007
000000070F80000730E0000740380003801C0003801E0003800F0003800700038007800380
078001C003C001C003C001C003C001C003C001C003C001C003C000E0038000E0038000F007
0000E806000FE60C0000E1F800>
	 26 31 0 20 27] 112 @dc
[<
7078F8783000000000000000000000000E1F1F0F06>
	 8 21 -4 20 14] 58 @dc

21 @eop0

0 0 21 @bop1 910 45 a @F7 @sf({)17 b(21)g({)-942 147 y(While)f(it)h(is)h
(certainly)e(an)i(o)o(v)o(ersimpli\014c)o(ation)f(to)h(imagine)f(that)h
(the)g(bulge)f(is)h(made)f(en)o(tirely)-1917 96 y(of)g(metal)g(ric)o(h)f
(stars)i(on)g(prograde)g(b)q(o)o(xy)f(orbits,)g(it)g(is)g(in)o(triguing)f
(that)i(the)f(surface)g(densit)o(y)f(of)h(the)-1928 96 y(prograde)h(b)q(o)o
(xy)f(orbits)g(resem)o(ble)o(s)f(the)h @F3 @sf(COBE)f @F7 @sf(Bulge)g(in)h
(\015attening)g(and)g(asymmetry)l(.)22 b(In)17 b(a)g(future)-1936 96 y
(pap)q(er)f(\(Zhao)h(1994\),)g(w)o(e)e(plan)h(to)g(construct)g(a)g
(self-consisten)o(t)f(bulge)g(mo)q(del)h(that)g(repro)q(duces)g(b)q(oth)
-1949 97 y(the)g(kinematic)f(and)h(photometric)g(prop)q(erties)g(of)h(the)f
(bulge.)-1075 128 y @F3 @sf(Note)e(added)h(in)g(man)o(uscript:)k @F7 @sf
(T)l(o)d(see)e(ho)o(w)h(a)g(metal)f(ric)o(h)g(bar)h(suc)o(h)g(as)g(the)g
@F3 @sf(COBE)g @F7 @sf(bulge)f(forms,)-1950 96 y(w)o(e)19 b(ha)o(v)o(e)g
(also)h(in)o(v)o(estigated)e(the)h(end)h(state)g(of)f(a)h(N-b)q(o)q(dy)h
(sim)o(ulation,)e(whic)o(h)f(w)o(as)j(generously)-1872 97 y(pro)o(vided)
13 b(b)o(y)h(Jerry)f(Sellw)o(o)q(o)q(d.)21 b(The)14 b(N-b)q(o)q(dy)g(bar)h
(gro)o(ws)g(out)f(of)g(the)g(disc)f(going)i(through)g(instabilit)o(y)l(.)o
-1951 96 a(The)k(bar)g(resem)o(bles)e(the)i @F3 @sf(COBE)f @F7 @sf(Bulge)g
(when)h(view)o(ed)f(at)h @F6 @sf(l)p 7 w @F10 @sf(bar)20 -7 y @F7 @sf(=)f
@F5 @sf(\000)p @F7 @sf(30)p -18 w @F10 @sf(o)22 18 y @F7 @sf(\(Sellw)o(o)q
(o)q(d)h(1993\).)30 b(The)-1890 96 y(sim)o(ulated)16 b(bar)h(retains)g(a)g
(disc)g(kinematic)o(s)f(with)h(98
(prograde)h(sense)e(of)-1934 97 y(rotation.)28 b(Less)19 b(than)f(2
(the)g(particles)g(are)g(on)g(retrograde)h(orbits.)28 b(W)l(e)17 b(ha)o(v)o
(e)h(extracted)f(out)-1902 96 y(the)g(particles)g(in)g(a)h(circle)e(cen)o
(tered)g(on)i(Baade's)g(Windo)o(w)f(with)h(one)g(degree)f(radius,)h(and)g
(plotted)-1918 96 y(their)f(v)o(elo)q(cit)o(y)g(elli)o(psoids)h(in)g
(Figure)f(8.)28 b(The)18 b(striking)g(similarit)n(y)f(with)h(Baade's)g
(Windo)o(w)g(metal)-1903 97 y(ric)o(h)e(sample)h(\(Figure)f(1\))i(and)f
(the)g(prograde)h(stars)g(\(Figure)f(4\))g(implie)o(s)f(that)i(it)e(is)h
(p)q(ossible)g(for)h(the)-1931 96 y(metal)f(ric)o(h)f(Bulge)h(to)h(form)g
(a)g(nearly)f(self-gra)o(vitating)g(bar)h(with)g(most)g(of)g(the)f(stars)i
(on)f(prograde)-1916 96 y(orbits.)24 b(If)16 b(the)h(metal)f(p)q(o)q(or)j
(stars)f(in)e(Figure)h(1)g(are)g(on)h(retrograde)f(orbits,)g(they)g(are)g
(unlik)o(el)o(y)f(from)-1932 97 y(the)i(disc.)26 b(They)18 b(could)f(form)h
(in)g(the)g(Bulge)f(or)h(they)g(could)g(b)q(e)g(accreted)f(systems)h(with)f
(opp)q(osite)-1907 96 y(angular)g(momen)o(tum,)e(a)h(mild)f(form)i(of)f
(NGC7217)i(\(Merri\014eld)d(&)h(Kuijk)o(en)f(1993\).)-1503 193 y(W)l(e)20 b
(w)o(ould)h(lik)o(e)e(to)i(thank)h(James)f(Binney)l(,)o 21 x(Ort)o(win)f
(Gerhard,)i(Rob)q(ert)f(Lupton,)i(Kevin)-1839 96 y(Prendergast)c(and)g
(Neil)d(T)o(yson)j(for)g(helpful)e(discussions.)27 b(HSZ)18 b(also)h
(thanks)g(Jerry)f(Sellw)o(o)q(o)q(d)g(for)-1901 96 y(kindly)c(pro)o
(viding)i(his)f(N-b)q(o)q(dy)h(sim)o(ulation,)f(Linda)h(Spark)o(e)f(and)i
(Donald)f(Lynden-Bell)e(for)i(v)m(aluable)-1950 97 y(commen)o(ts)h(on)h
(the)g(man)o(uscript.)25 b(DNS)18 b(ac)o(kno)o(wledges)f(supp)q(ort)i
(from)f(NSF)f(gran)o(t)i(AST88-5814)q(5)-1910 96 y(\(PYI\),)d(AST91-1738)q
(8)i(and)g(NASA)e(Theory)h(Gran)o(t)g(NA)o(GW-2448.)24 b(RMR)16 b(and)i
(HSZ)e(ac)o(kno)o(wledge)-1936 96 y(supp)q(ort)h(from)g(NASA)e(Long)i(T)l
(erm)f(Space)g(Astroph)o(ysics)g(Program)h(gran)o(t)g(NA)o(GW-2479.)-1742
259 y
21 @eop1

20 @bop0
@F5 @sf
[<
FFFFFFFFFFFFFFFF0001800000018000000180000001800000018000000180000001800000
018000000180000001800000018000000180000001800000018000FFFFFFFFFFFFFFFF0001
80000001800000018000000180000001800000018000000180000001800000018000000180
00000180000001800000018000000180000001800000008000>
	 32 34 -3 32 39] 6 @dc
@F5 @sf
[<
07C018702018201C600C700E700E200E000E000E001C00180038006000C003E00C30181830
1C700CE00EE00EE00EE00EE00E601C7018303018600F8006000C00380030007000E000E000
E008E01CE01C600C700830081C3007C0>
	 15 45 -3 34 22] 120 @dc

20 @eop0

0 0 20 @bop1 910 45 a @F7 @sf({)17 b(20)g({)-1040 147 y(p)q(oin)o(ting)j
(to)o(w)o(ards)g @F6 @sf(l)p 7 w @F10 @sf(v)22 -7 y @F7 @sf(=)f(25)p -18 w
@F10 @sf(o)16 18 y @F5 @sf(\006)13 b @F7 @sf(14)p -18 w @F10 @sf(o)3 18 y
@F7 @sf(,)20 b(while)e(the)h(metal)g(ric)o(h)f(v)o(elo)q(cit)o(y)g(elli)o
(psoid)h(p)q(oin)o(ts)h(to)o(w)o(ards)-1874 96 y @F6 @sf(l)p 7 w @F10 @sf
(v)21 -7 y @F7 @sf(=)d @F5 @sf(\000)p @F7 @sf(65)p -18 w @F10 @sf(o)16 18 y
@F5 @sf(\006)12 b @F7 @sf(9)p -18 w @F10 @sf(o)3 18 y @F7 @sf(.)28 b(In)
18 b(an)h(axisymmetri)o(c)e(Galaxy)l(,)i(the)f(v)o(elo)q(cit)o(y)f(elli)o
(psoid)h(should)h(p)q(oin)o(t)g(to)o(w)o(ards)-1895 96 y @F6 @sf(l)p 7 w
@F10 @sf(v)17 -7 y @F7 @sf(=)14 b(0)p -18 w @F10 @sf(o)18 18 y @F7 @sf(or)j
(90)p -18 w @F10 @sf(o)3 18 y @F7 @sf(,)f(th)o(us,)g(these)f(v)o(ertex)g
(deviations)h(are)g(kinematic)f(signatures)i(of)f(bulge)g(triaxialit)o(y)l
(.)o -1787 129 a(T)l(o)h(b)q(etter)g(understand)h(the)e(origin)h(for)g
(the)g(v)o(ertex)f(deviation)g(in)h(Baade's)g(Windo)o(w,)f(w)o(e)h(ha)o(v)o
(e)-1933 96 y(run)g(n)o(umerical)e(sim)o(ulations)h(of)h(stellar)f(orbits)g
(in)h(a)g(rotating)g(bar)g(p)q(oten)o(tial.)22 b(If)16 b(w)o(e)g(select)g
(only)g(the)-1941 96 y(prograde)i(b)q(o)o(xy)g(orbits)f(\(B-family)f
(orbits\),)h(then)g(this)g(p)q(opulation)i(repro)q(duces)e(the)g
(kinematics)f(of)-1925 96 y(the)h(metal)f(ric)o(h)f(stars)j(in)e(our)h
(sample)g(and)g(the)g(pro)s(jected)f(shap)q(e)h(of)g(the)g(bar.)23 b(On)
17 b(the)f(other)h(hand,)-1937 97 y(the)g(retrograde)g(lo)q(op)h(orbits)g
(\(R-family\))e(ha)o(v)o(e)g(the)h(same)g(kinematics)f(as)h(the)g(metal)g
(p)q(o)q(or)h(stars)g(in)-1929 96 y(our)h(sample.)29 b(While)17 b(the)i
(sample)f(is)h(to)q(o)h(small)e(to)h(de\014nitiv)o(e)o(ly)e(iden)o(tify)g
(di\013eren)o(t)h(metallic)o(iti)o(es)-1891 96 y(with)e(di\013eren)o(t)f
(orbital)h(families,)e(the)i(coincidence)e(is)i(in)o(triguing)g(and)g
(suggestiv)o(e)g(of)g(the)g(promise)g(of)-1950 97 y(larger)g(stellar)g
(samples.)21 b(It)15 b(is)h(also)h(consisten)o(t)f(with)g(Minniti's)e
(\(1993)q(\))j(observ)m(ation)g(that)g(the)f(metal)-1951 96 y(ric)o(h)f
(stars)i(stream)f(ahead)h(of)g(the)f(metal)f(p)q(o)q(or)j(stars.)-919 128 y
(Our)g(n)o(umerical)f(sim)o(ulations)h(sho)o(w)h(the)g(imp)q(ortance)f(of)h
(obtaining)g(prop)q(er)h(motion,)e(radial)-1892 97 y(v)o(elo)q(cit)o(y)o
20 x(and)j(metallici)o(t)o(y)e(data)j(for)f(orbit)g(classi\014cation.)35 b
(While)20 b(w)o(e)g(\014nd)h(that)h(our)f(lac)o(k)f(of)-1838 96 y(kno)o
(wledge)13 b(of)h(line-of-sigh)o(t)g(distances)g(is)f(a)h(ma)s(jor)g
(source)g(of)g(uncertain)o(t)o(y)l(,)f(w)o(e)g(are)h(able)f(to)i(o)o(v)o
(ercome)o -1951 96 a(some)21 b(of)g(this)f(constrain)o(t)h(through)g(Mon)o
(te)g(Carlo)g(realizations.)34 b(This)20 b(suggests)i(that)f(highly)-1844
97 y(accurate)c(prop)q(er)g(line-of-sigh)o(t)f(distance)h(ma)o(y)f(not)h
(b)q(e)g(essen)o(tial)f(for)h(a)h(program)f(of)g(reconstructing)-1936 96 y
(the)f(stellar)f(distribution)g(function.)21 b(Suc)o(h)16 b(a)g(program)h
(w)o(ould)f(allo)o(w)g(us)g(to)g(infer)f(a)i(great)f(deal)g(ab)q(out)-1949
96 y(the)h(formation)g(history)g(of)g(the)g(Galaxy)l(.)24 b(Neglecti)o(ng)
16 b(suc)o(h)h(complications)f(as)i(extinction)e(and)h(disk)-1931 97 y
(con)o(tamination,)f(our)g(sim)o(ulation)g(also)g(\014nds)h(that)g(the)f
(kinematic)o(s)f(of)i(the)f(bar)g(are)g(b)q(est)h(rev)o(ealed)e(at)-1950
96 y(the)h(lo)o(w)g(latitude)g(\014elds.)-379 128 y(The)k(correlations)f
(b)q(et)o(w)o(een)g(metallicit)o -1 x(y)f(and)j(kinematics)d(discussed)i
(here)g(already)g(place)-1861 97 y(signi\014can)o(t)d(constrain)o(ts)g(on)h
(the)f(formation)h(and)g(ev)o(olutionary)e(history)h(of)h(the)f(bulge.)24 b
(It)17 b(suggests)-1924 96 y(that)h(the)f(bulge)h(has)g(not)g(undergone)g
(extremely)e(violen)o(t)g(relaxation)h(and)i(that)f(dissipation)f(lik)o
(ely)o -1918 96 a(pla)o(y)o(ed)e(an)i(imp)q(ortan)o(t)f(role)g(in)g(its)g
(ev)o(olution.)k(Implications)15 b(are)i(discussed)f(in)g(greater)g
(detail)f(in)h @F5 @sf(x)p @F7 @sf(4.)-1947 195 y
20 @eop1

19 @bop0
@F6 @sf
[<
80780040FC0061FE003F830010018008008004004002000001000000C00000200000100000
0800000400000200080100080080060FC007F86003F02000E010>
	 20 21 -2 20 23] 122 @dc
[<
C0000000F00000003C0000000F00000003C0000000F00000003E0000000F80000001E00000
00780000001E0000000780000001E00000007800000078000001E00000078000001E000000
78000001E000000F8000003E000000F0000003C000000F0000003C000000F0000000C00000
00>
	 29 28 -4 25 38] 62 @dc
@F2 @sf
[<
03F8000FFF001C1F803007C07007E0FC03F0FC03F0FE03F8FE03F87C03F83803F80003F800
03F80003F00003F01803E01807E01E0FC01FFF0019FC001800001800001800001800001BC0
001FF8001FFE001FFF001FFF801FFFC01E01E0100020>
	 21 32 -3 31 28] 53 @dc
[<
387CFEFEFE7C38>
	 7 7 -4 6 16] 46 @dc
[<
00FE0007FFC00F83E01F01F03E00F87C007C7C007CFC007EFC007EFC007EFC007EFC007EFC
007EFC007E7C007C7C007C7C007C3E00F81E00F00F83E007FFC000FE00>
	 23 22 -2 21 28] 111 @dc
[<
FFE1FFC0FFE1FFC01F003E001F003E001F003E001F003E001F003E001F003E001F003E001F
003E001F003E001F003E001F003E001F003E001F003E001F003E001F803E001F803E001F40
3C001F307C00FF1FF800FF07E000>
	 26 22 -2 21 31] 110 @dc
[<
00FE0007FF800FC0E01F00603E00307E00007E00007C0000FC0000FC0000FC0000FC0000FC
0000FC00007C00007C01E07E03F03E03F01F03F00F83F007FFE000FF80>
	 20 22 -2 21 25] 99 @dc
[<
FFE0FFE01F001F001F001F001F001F001F001F001F001F001F001F001F001F001F001F001F
001F001F001F001F001F001F001F001F001F001F001F001F001F001F00FF00FF00>
	 11 35 -2 34 16] 108 @dc
[<
01FC3FC007FF3FC00F81BE001F00FE001F007E001F003E001F003E001F003E001F003E001F
003E001F003E001F003E001F003E001F003E001F003E001F003E001F003E001F003E001F00
3E001F003E00FF01FE00FF01FE00>
	 26 22 -2 21 31] 117 @dc
[<
C7F000EFFC00FC0E00F00700E00380E00380C00380C00780000F8001FF000FFF003FFE007F
F8007FE000FC0000F00300E00300E00300600700781F003FFF000FF300>
	 17 22 -2 21 22] 115 @dc
[<
FFE0FFE01F001F001F001F001F001F001F001F001F001F001F001F001F001F001F001F001F
001F00FF00FF0000000000000000000000000000001C003E007F007F007F003E001C00>
	 11 36 -2 35 16] 105 @dc

19 @eop0

0 0 19 @bop1 910 45 a @F7 @sf({)17 b(19)g({)-1040 147 y(mixing)f(pro)q
(cess)i(that)g(w)o(ould)f(erase)g(these)g(correlations.)24 b(On)18 b(the)f
(other)g(hand,)h(the)f(lac)o(k)f(of)i(strong)-1925 96 y(correlations)h(in)g
(the)g(halo)h(fa)o(v)o(ors)f(the)g(notion)h(that)f(the)g(halo)h(did)f
(undergo)h(violen)o(t)e(relaxation,)-1879 96 y(p)q(erhaps)f(as)g(it)f
(formed)g(b)o(y)g(c)o(haotically)e(merging)i(fragmen)o(ts)h(\(Searle)e(&)h
(Zinn,)g(1978\).)-1555 129 y(The)i(strong)i(correlations)e(seen)g(in)g
(the)h(bulge,)f(together)h(with)f(the)h(rapid)f(rotation)h(and)h(the)-1896
96 y(small)c(scale)h(heigh)o(t)f(of)h(the)f(metal)g(ric)o(h)g(bulge,)h(is)f
(consisten)o(t)h(with)f(the)h(notion)g(that)g(the)g(bulge/bar)-1936 96 y
(formed)g(dissipationally)f(out)i(of)f(the)g(galactic)g(disk.)23 b(The)17 b
(high)h(phase)f(space)h(densities)e(seen)h(in)f(the)-1929 96 y(bulge)i
(also)h(imply)e(either)g(a)h(v)o(ery)f(early)h(formation)h(time)e(\()p
@F6 @sf(z)i(>)e @F7 @sf(100\))i(or)g(that)f(dissipation)h(pla)o(y)o(ed)-1904
97 y(an)h(imp)q(ortan)o(t)f(role)g(in)g(its)g(formation.)31 b(It)19 b(is)g
(in)o(triguing)g(to)h(consider)f(the)g(p)q(ossibilit)o(y)f(that)i(the)-1875
96 y(bulge)g(formation)g(is)f(link)o(ed)f(to)j(the)e(ev)o(olution)g(of)h
(the)g(galactic)f(disk.)32 b(Numeric)o(al)18 b(sim)o(ulations)-1865 96 y
(\(Com)o(b)q(es)f @F4 @sf(et)i(al.)24 b @F7 @sf(1990,)18 b(Raha)g @F4 @sf
(et)g(al.)24 b @F7 @sf(1990)q(\))18 b(\014nd)f(that)g(a)g(cold)g(disk)f
(can)h(form)g(a)g(thic)o(k)f(bar)h(through)-1935 97 y(gra)o(vitational)e
(instabilit)o(y)o -4 x(.)20 b(This)14 b(thic)o(k)g(bar)h(app)q(ears)h(to)f
(b)q(e)g(a)g(go)q(o)q(d)h(\014t)f(to)g(the)g(observ)o(ed)f(kinematical)o
-1951 96 a(and)j(photometric)f(prop)q(erties)g(of)g(the)g(Galaxy's)g
(triaxial)g(bulge)g(\(Sellw)o(o)q(o)q(d)g(1993\).)-1492 128 y(The)g
(strong)h(correlations)f(also)h(suggest)g(a)g(p)q(eaceful)f(history)g(for)h
(the)f(ev)o(olution)f(of)i(the)f(bar)g(and)-1948 97 y(the)h(p)q(oten)o
(tial)g(of)h(the)f(inner)f(Galaxy)l(.)25 b(Violen)o(t)16 b(c)o(hanges)h
(in)g(the)g(p)q(oten)o(tial)g(w)o(ould)h(\\mix")f(the)g(stars)-1924 96 y
(in)f(phase)h(space.)22 b(If)16 b(most)g(stars)i(form)e(during)g(the)h
(violen)o(t)e(relaxation)h(phase,)g(one)h(do)q(es)g(not)g(exp)q(ect)-1947
96 y(to)h(observ)o(e)g(strong)g(correlations)g(b)q(et)o(w)o(een)f
(metallici)o(t)o(y)o 17 x(and)h(kinematics.)25 b(On)18 b(the)f(other)h
(hand,)h(if)-1912 97 y(the)g(galactic)f(p)q(oten)o(tial)h(in)f(the)h
(inner)f(Galaxy)h(underw)o(en)o(t)f(only)h(slo)o(w)f(c)o(hanges,)i(then)e
(adiabatic)-1887 96 y(in)o(v)m(ariance)f(implie)o(s)g(that)h(the)f
(orbital)h(structure)f(smo)q(othly)i(ev)o(olv)n(es)e(with)g(the)h(p)q
(oten)o(tial)f(\(Binney)-1916 96 y(and)g(Sp)q(ergel)f(1984,)h(Ev)m(ans)g
(and)g(Collett)f(1993;)h(Dubinski)f(1993\).)-476 204 y @F2 @sf(5.)74 b
(Conclusion)-1070 188 y @F7 @sf(By)15 b(fo)q(cusing)i(on)g(a)g(stellar)f
(sample)g(of)h(62)g(Baade's)g(windo)o(w)f(K)h(gian)o(ts)f(with)h(metallic)o
(it)n(y)l(,)e(radial)-1945 96 y(v)o(elo)q(cit)o(y)o 19 x(and)21 b(prop)q
(er)f(motion)g(measuremen)o(ts,)g(w)o(e)f(are)h(able)g(to)g(\014nd)h
(statistically)e(signi\014can)o(t)-1860 97 y(evidence)g(for)h(a)h(strong)h
(correlation)e(b)q(et)o(w)o(een)f(kinematics)g(and)i(metallic)o(it)o(y)o
-4 x(,)f(extending)f(w)o(ork)-1850 96 y(describ)q(ed)e(in)g(Zhao)i @F4 @sf
(et)g(al.)27 b @F7 @sf(\(1994\).)f(The)18 b(metal)f(p)q(o)q(or)i(stars)g
(app)q(ear)f(to)g(ha)o(v)o(e)f(a)h(v)o(elo)q(cit)o(y)e(ellipsoid)-1918 124 y
19 @eop1

18 @bop0
@F5 @sf
[<
C0C0C0C0C0C0C0C0C0C0C0C0C0C0C0C0C0C0C0C0C0C0C0C0C0C0C0C0C0C0C0C0C0C0C0C0C0
C0C0C0C0C0C0C0C0C0C0C0C0>
	 2 49 -6 36 14] 106 @dc
@F2 @sf
[<
00FFFE00FFFE0007E00007E00007E00007E00007E00007E00007E0FFFFFEFFFFFEE007E070
07E03807E01807E00C07E00E07E00707E00307E00187E001C7E000E7E00077E00037E0001F
E0001FE0000FE00007E00003E00003E00001E00000E0>
	 23 32 -2 31 28] 52 @dc
[<
FFFFE0FFFFE003F80003F80003F80003F80003F80003F80003F80003F80003F80003F80003
F80003F80003F80003F80003F80003F80003F80003F80003F80003F80003F80003F80003F8
0003F80003F80003F80003F80003F80003F80003F800FFFFE0FFFFE0>
	 19 34 -1 33 21] 73 @dc
[<
7FF8007FF8000F80000F80000F80000F80000F80000F80000F80000F80000F80000F80000F
80000F80000F80000F80000F80000F80000F80000F8000FFFC00FFFC000F80000F80000F80
000F80000F80000F80000F83C00F87E007C7E003C7E001E3E000FFC0003F80>
	 19 35 -1 34 17] 102 @dc
[<
00FF0003FFC00FC0701F00303E00187E00007C00007C0000FC0000FC0000FC0000FFFFF8FF
FFF8FC00F8FC00F87C00F87C00F03E01F01E01E00F87C007FF8000FE00>
	 21 22 -2 21 26] 101 @dc
[<
FFF000FFF0001F00001F00001F00001F00001F00001F00001F00001F00001F00001F00001F
00001F00001F00001F83C01E87E01E87E01EC7E01E67E0FE3FC0FE1F00>
	 19 22 -2 21 23] 114 @dc
[<
00F80003FE0007C3000F83000F81800F81800F81800F81800F81800F80000F80000F80000F
80000F80000F80000F80000F80000F80000F80000F8000FFFF00FFFF003F80000F80000780
00078000038000038000018000018000018000018000>
	 17 32 -1 31 22] 116 @dc
[<
FFE1FFC0FFE1FFC01F003E001F003E001F003E001F003E001F003E001F003E001F003E001F
003E001F003E001F003E001F003E001F003E001F003E001F003E001F803E001F803E001F40
3C001F307C001F1FF8001F07E0001F0000001F0000001F0000001F0000001F0000001F0000
001F0000001F0000001F0000001F0000001F000000FF000000FF000000>
	 26 35 -2 34 31] 104 @dc
[<
FFE0FFE0FFE0FFE0FFE0FFE01F001F001F001F001F001F001F001F001F001F001F001F001F
001F001F001F001F001F001F001F001F001F001F001F001F001F001F001F001F001F001F00
1F001F001F001F001F001F001F001F001F001F001F001F801F801F001F801F801F001F403E
403E001F303E303E00FF1FFC1FFC00FF07F007F000>
	 43 22 -2 21 47] 109 @dc
[<
07E03F1FF87F7E0CF87E02F0FC01F0FC01F0FC01F0FC01F07E01F03F01F01FC1F003FDF000
3FF00001F00001F01E01F03F01F03F01E03F03E03F07C01FFF8007FC00>
	 24 22 -2 21 27] 97 @dc
[<
FFFFFFC000FFFFFFF80007F001FE0007F0007F0007F0003F8007F0003F8007F0001FC007F0
001FC007F0001FC007F0001FC007F0001FC007F0001FC007F0001F8007F0003F8007F0003F
0007F0007E0007F001FC0007FFFFF00007FFFFF00007F003F80007F000FC0007F000FE0007
F0007E0007F0007F0007F0007F0007F0007F0007F0007F0007F0007F0007F0007E0007F000
7E0007F000FC0007F001F800FFFFFFE000FFFFFF8000>
	 34 34 -2 33 40] 66 @dc

18 @eop0

0 0 18 @bop1 910 45 a @F7 @sf({)17 b(18)g({)-1040 147 y(latitude)e(\014elds)g
(as)i(most)f(orbits)g(that)g(reac)o(h)f(high)h(latitudes)f(do)h(not)h(ha)o
(v)o(e)e(a)h(\014xed)f(sense)h(of)g(rotation.)-1950 96 y(Observ)m(ations)i
(at)g(o\013-axis)h(lo)o(w)f(latitude)f(\014elds)g @F5 @sf(j)p @F6 @sf(b)p
@F5 @sf(j)f(\024)g @F7 @sf(4)p -18 w @F10 @sf(o)20 18 y @F7 @sf(can)i
(clearly)e(separate)i(the)g(prograde)g(and)-1914 96 y(retrograde)i(orbit)g
(families)e(and)j(will)d(b)q(e)i(v)o(ery)f(imp)q(ortan)o(t)h(in)f
(detecting)g(correlations)g(b)q(et)o(w)o(een)-1866 96 y(metallic)o(it)o(y)o
17 x(and)g(orbit)f(family)l(.)26 b(Observ)m(ations)19 b(along)g(the)f
(minor)f(axis)i(should)f(allo)o(w)g(us)h(to)f(trace)-1903 97 y(the)e
(exten)o(t)g(of)g(the)h(prograde)g(orbit)f(family)l(.)21 b(In)16 b(our)h
(sim)o(ulations,)f(w)o(e)g(\014nd)h(that)g(the)f(prograde)h(stars)-1943
96 y(b)q(egin)i(to)g(disapp)q(ear)h(at)g(1)f(kp)q(c)g(ab)q(o)o(v)o(e)g
(the)f(plane,)h(or)h(at)f @F5 @sf(j)p @F6 @sf(b)p @F5 @sf(j)f @F7 @sf(=)g
(7)p -18 w @F10 @sf(o)3 18 y @F7 @sf(.)29 b(If)18 b(the)h(late)g(\(metal)f
(ric)o(h\))g(M)-1886 96 y(gian)o(ts)g(are)g(asso)q(ciated)h(with)f(this)g
(orbital)f(family)l(,)g(then)h(this)g(w)o(ould)f(pro)q(duce)i(a)f(steep)g
(drop)q(o\013)h(in)-1910 97 y(their)c(n)o(um)o(b)q(er)g(coun)o(ts)i(at)f
(high)g(Galactic)g(latitude.)k(It)15 b(is)h(in)o(triguing)g(that)g(suc)o
(h)g(a)g(drop)q(o\013)i(is)e(seen)f(in)-1950 96 y(observ)m(ations)i(along)g
(the)f(minor)g(axis)g(\(Blanco)g(1988\).)-944 128 y(While)g(the)h
(predictiv)o(e)o 16 x(p)q(o)o(w)o(er)h(of)g(this)f(preliminary)f(mo)q(del)h
(is)g(in)o(triguing,)g(w)o(e)g(note)h(that)g(it)f(is)-1920 97 y(not)g(y)o
(et)f(self-consisten)o(t.)21 b(Rather,)16 b(w)o(e)g(illustrate)g(the)g
(promise)h(of)f(constructing)h(dynamical)f(mo)q(dels)-1942 96 y(of)h
(triaxial)e(systems.)21 b(F)l(urther)16 b(results)g(will)f(b)q(e)h
(discussed)g(in)g(Zhao)h(\(1994\).)-1019 204 y @F2 @sf(4.)74 b(Inferences)
18 b(on)g(the)h(F)-5 b(ormation)18 b(of)h(the)f(Bar)-1417 188 y @F7 @sf
(The)e(correlation)g(b)q(et)o(w)o(een)f(metallic)o(it)n(y)g(and)i
(kinematics)d(seen)i(in)g(the)g(bulge)g(in)g(this)g(study)g(and)-1949 96 y
(in)h(earlier)e(w)o(ork)i(\(Ric)o(h,)f(1990;)i(T)o(yson)g(&)f(Ric)o(h)f
(1991;)i(Minniti)d(1993)q(;)j(Zhao)g @F4 @sf(et)g(al.)25 b @F7 @sf(1994\))
19 b(con)o(trasts)-1932 96 y(with)e(the)h(lac)o(k)e(of)i(correlation)f(b)q
(et)o(w)o(een)g(metalli)o(cit)n(y)f(and)i(kinematics)e(rep)q(orted)i(b)o
(y)f(Carney)g @F4 @sf(et)j(al.)-1920 97 y @F7 @sf(\(1990\))d(for)e(the)g
(Galactic)g(halo.)22 b(This)15 b(con)o(trast)h(suggests)g(t)o(w)o(o)g
(di\013eren)o(t)e(formation)i(histories)f(for)g(the)-1950 96 y(t)o(w)o(o)h
(comp)q(onen)o(ts.)-260 128 y(Our)f(understanding)h(of)g(c)o(hemic)o(al)e
(ev)o(olution)h(implie)o(s)g(that)g(successiv)o(e)f(p)q(opulations)j
(enric)o(h)d(the)-1950 97 y(in)o(terstell)o(ar)h(medium)g(with)h(metals,)f
(so)h(that)h(the)e(y)o(oungest)h(stars)h(are)f(metal)f(ric)o(h)g(and)i(v)m
(ariations)f(in)-1950 96 y(metallic)o(it)o(y)o 17 x(re\015ect)i(v)m
(ariations)g(in)g(either)f(stellar)h(age)h(or)f(formation)h(en)o(vironmen)o
(t)d(\(Eggen)j @F4 @sf(et)h(al.)-1901 96 y @F7 @sf(1962,)e(Binney)e(&)h(T)l
(remaine,)f(1987\).)26 b(The)17 b(existence)e(of)j(strong)g(correlations)f
(b)q(et)o(w)o(een)f(metallici)o(t)o(y)o -1930 97 a(and)i(orbital)f
(families)f(in)i(the)f(bulge)g(suggests)i(that)f(the)f(bulge)g(stars)h
(retain)g(a)f(\\memory")h(of)f(their)-1921 96 y(initial)h(conditions)g
(and)i(ha)o(v)o(e)e(exp)q(erienced)f(neither)h(violen)o(t)g(relaxation)g
(nor)i(an)o(y)e(other)h(c)o(haotic)-1887 124 y
18 @eop1

17 @bop0
@F2 @sf
[<
01FE000FFFC01E07F07801F87E01FCFF00FCFF00FEFF00FEFF00FE7E00FE3C00FE0000FC00
00FC0001F80001F00007C001FF0001FE00001F800007C00003E00003F01F03F01F03F83F81
F83F81F83F81F83F03F01E03F00F07E007FFC000FE00>
	 23 32 -2 31 28] 51 @dc
[<
1F0000007F80000069C00000FC600000FC300000FC3800007818000000180000000C000000
0C0000000E0000000E0000001F0000001F0000003F8000003F8000007FC000007CC000007C
C00000F8600000F8600001F0300001F0300003E0180003E0180007E01C0007C00C000FC00E
000F8006000F800600FFE01FE0FFE01FE0>
	 27 32 -1 21 30] 121 @dc
[<
01FC3FC007FF3FC00F83BE001E00FE003E007E007C003E007C003E00FC003E00FC003E00FC
003E00FC003E00FC003E00FC003E00FC003E00FC003E007C003E007E003E003E003E001F00
7E000F81FE0007FFBE0001FC3E0000003E0000003E0000003E0000003E0000003E0000003E
0000003E0000003E0000003E0000003E0000003E000001FE000001FE00>
	 26 35 -2 34 31] 100 @dc
[<
2070181C0C06060703033B7FFFFFFE7C38>
	 8 17 -4 34 16] 39 @dc
[<
0000E0000E00000000E0000E00000000F0001E00000001F0001F00000001F0001F00000003
F8003F80000003F8003F80000003FC007F80000007FC007FC0000007FC007FC000000FF600
FFE000000FE600FE6000000FE600FE6000001FE301FC3000001FC301FC3000001FC383FC30
00003F8183F81800003F8183F81800007F80C7F81C00007F00C7F00C00007F00C7F00C0000
FF006FE0060000FE006FE0060000FE007FE0060001FC003FC0030001FC003FC0030003FC00
3F80038003F8003F80018003F8007F80018007F0007F0000C007F0007F0000C00FF000FF00
00E0FFFF0FFFF01FFEFFFF0FFFF01FFE>
	 55 34 -1 33 58] 87 @dc
[<
001800C000003800E000003C01E000007C01F000007E03F000007E03F00000FE03D80000FB
07D80001FB079C0001F38F8C0001F18F0C0003E18F060003E0DF060007E0DE070007C0DE03
0007C07E03000F807C01800F807C01800F807801801F007800C0FFE7FF07F8FFE7FF07F8>
	 37 22 -1 21 40] 119 @dc

17 @eop0

0 0 17 @bop1 910 45 a @F7 @sf({)17 b(17)g({)-443 147 y @F2 @sf(3.3.)55 b
(Bey)n(ond)19 b(Baade's)f(Windo)n(w)-1254 203 y @F7 @sf(Our)d(orbit)h(sim)o
(ulations)f(can)h(b)q(e)f(used)h(to)g(b)q(oth)h(understand)f(existing)f
(observ)m(ational)h(anomalies)-1950 97 y(and)h(to)f(suggest)i(promising)e
(directions)f(for)i(future)f(w)o(ork.)-1012 128 y(Grenon)f(\(1989\))h(has)f
(found)h(in)e(the)h(solar)g(neigh)o(b)q(ourho)q(o)q(d)i(high)d(v)o(elo)q
(cit)o(y)f(metal)h(ric)o(h)g(stars)h(with)-1950 96 y(abundance)k(ratios)f
(t)o(ypical)f(of)h(the)g(Bulge)g(stars.)27 b(W)l(e)18 b(\014nd)g(there)g
(are)g(orbits)g(that)h(bring)f(Baade's)-1906 97 y(Windo)o(w)i(stars)h
(either)e(in)o(to)h(the)g(Galactic)f(Cen)o(ter)g(or)i(out)f(of)h(the)f
(corotation)g(radius.)33 b(In)20 b(our)-1858 96 y(sim)o(ulation,)15 b(t)o
(ypically)g(10
(reac)o(h)f(the)h(solar)g(circle.)o -1940 96 a(These)f(orbits)h(ma)o(y)e
(explain)h(the)g(origin)g(of)g(these)g(anomalous)i(stars.)-1215 129 y
(Orbit)e(analysis,)i(together)f(with)g(a)h(careful)e(treatmen)o(t)g(of)i
(the)f(selection)f(e\013ects,)h(ma)o(y)f(explain)-1927 96 y(the)i(the)f
(observ)o(ed)g(hin)o(ts)h(of)g(minor)f(axis)h(rotation)g(\(T)o(yson)g(&)g
(Ric)o(h)f(1991,)i(Blum)e @F4 @sf(et)i(al.)27 b @F7 @sf(1993\).)g(A)-1914
96 y(magnitude-limited)17 b(surv)o(ey)i(will)f(detect)g(more)h(stars)h(on)f
(the)g(near)g(side)g(of)g(the)g(bar,)h(whic)o(h)f(are)-1881 97 y
(streaming)h(to)o(w)o(ards)g(us.)32 b(Other)19 b(explanations)h(for)g
(minor)f(axis)h(rotation)g(in)o(v)o(olv)o(e)e(minor)h(orbit)-1867 96 y
(families)d(of)h(the)g(bar.)24 b(In)16 b(our)i(sim)o(ulation,)e(w)o(e)g
(\014nd)i(10
96 y(minor)f(orbit)g(families)f(of)h(the)g(bar.)22 b(In)16 b(the)g
(classi\014cation)g(sc)o(heme)f(of)h(Pfenniger)g(and)h(F)l(riedli)o 15 x
(\(1991\),)-1949 97 y(these)j(include)e(the)i(anomalous)h(orbits,)g(whic)o
(h)f(are)g(tipp)q(ed)g(retrograde)g(lo)q(op)h(orbits)f(that)h(also)-1859
96 y(rotate)c(around)h(the)e(long-axis,)h(and)g(the)g(Z-orbits,)f(whic)o
(h)g(are)h(prograde)g(b)q(o)o(xy)g(orbits)g(near)g(v)o(ertic)o(al)-1942
96 y(resonances.)k(Since)15 b(the)g(anomalous)i(orbits)f(rotate)g(around)g
(the)f(long)h(axis,)f(they)g(ma)o(y)g(b)q(e)h(imp)q(ortan)o(t)-1950 97 y
(for)h(the)f(minor)f(axis)i(rotation.)-485 128 y(Observ)m(ations)f(of)f
(stellar)g(kinematics)f(and)i(abundances)h(in)e(other)h(bulge)f(\014elds,)g
(com)o(bined)f(with)-1950 96 y(theoretical)k(analysis,)h(should)h(enable)f
(us)g(to)g(determine)f(the)h(stellar)f(distribution)h(function.)29 b(In)
-1883 97 y(Figure)16 b(5,)g(w)o(e)g(sho)o(w)i(the)e(pro)s(jected)f(v)o
(elo)q(cit)o(y)g(ellipsoids)g(of)i(the)f(retrograde)h(and)g(prograde)h
(orbits)e(in)-1945 96 y(di\013eren)o(t)g(bulge)h(\014elds.)24 b(These)17 b
(\014elds)g(ha)o(v)o(e)f(b)q(een)h(selected)f(for)i(their)e(corresp)q
(ondence)h(to)h(on-going)-1926 96 y(observing)f(programs)g(and)g(it)f(w)o
(ould)g(b)q(e)h(straigh)o(tforw)o(ard)g(to)g(compute)f(these)g
(predictions)f(for)i(other)-1947 97 y(\014elds.)29 b(Note)19 b(that)h(the)e
(that)i(lo)o(w)f(latitude)f(\014elds)h(ha)o(v)o(e)f(stronger)i(v)o(ertex)d
(deviations)i(than)h(high)-1882 183 y
17 @eop1

16 @bop0
@F3 @sf
[<
FFE7FF0F00780E007007003807003807003807003807003807003803801C03801C03801C03
801C03801C03C01C01C00E01E00E01E00E01D00C01CC1801C3F000E00000E00000E00000E0
0000E00000E00000700000700000700000700000700000700003F800007800>
	 24 35 -1 34 27] 104 @dc
@F7 @sf
[<
3FC00FF03FC00FF07FC00FF84040080840C00C084080040880800404818006040180060003
80070003000300070003800F0003C00E0001C01E0001E01E0001E03E0001F03E0001F03C00
00F07C0000F87C0000F87C0000F87C0000F87C0000F87C0000F87C0000F83C0000F03E0001
F01E0001E01F0003E00F0003C00780078001C00E0000F03C00001FE000>
	 30 35 -2 34 35] 10 @dc
@F7 @sf
[<
00018000000180000001800000018000000180000001800000018000000180000001800000
018000000180000001800000018000000180000001800000018000FFFFFFFEFFFFFFFE0001
80000001800000018000000180000001800000018000000180000001800000018000000180
00000180000001800000018000000180000001800000018000>
	 31 34 -3 28 38] 43 @dc
@F7 @sf
[<
03FFFC00001F8000000F0000000F0000000F0000000F0000003FC00001EF7800078F1E000E
0F07001E0F07803C0F03C07C0F03E0780F01E0F80F01F0F80F01F0F80F01F0F80F01F0F80F
01F0F80F01F0780F01E07C0F03E03C0F03C01E0F07800E0F0700078F1E0001EF7800003FC0
00000F0000000F0000000F0000000F0000001F800003FFFC00>
	 28 34 -3 33 35] 8 @dc

16 @eop0

0 0 16 @bop1 910 45 a @F7 @sf({)17 b(16)g({)-1040 147 y(densit)o(y)g
(distribution)h(with)f(its)h(ma)s(jor)g(axis)g(p)q(oin)o(ting)g(to)o(w)o
(ards)h(a)f(general)g(angle)g(with)g(resp)q(ect)f(to)-1908 96 y(the)g
(Sun.)25 b(Since)17 b(the)g(ma)s(jor)g(axes)h(of)f(the)g(prograde)i(b)q(o)o
(xy)e(orbits)h(and)g(retrograde)g(lo)q(op)g(orbits)f(are)-1922 96 y(p)q
(erp)q(endicular)f(to)h(eac)o(h)f(other)h(\(see)f(Figure)g(3\),)h(the)f
(asymmetry)g(is)h(opp)q(osite)g(for)g(the)f(t)o(w)o(o)h(families.)o -1942
96 a(While)g(w)o(e)g(ha)o(v)o(e)g(not)i(built)e(a)h(self-consisten)o(t)f
(mo)q(del,)h(it)f(is)h(suggestiv)o(e)f(that)h(assigning)h(the)f(metal)-1912
97 y(ric)o(h)d(stars)i(to)g(predominan)o(tly)e(prograde)j(orbits)e(\014ts)h
(b)q(oth)g(the)f(Baade's)g(Windo)o(w)h(v)o(elo)q(cit)n(y)e(ellipsoid)-1950
96 y(and)j(the)f(rough)h(prop)q(erties)g(of)f(the)h(COBE)f(photometry)l(.)
24 b(In)17 b(fact,)h(one)f(susp)q(ects)h(that)g(the)f(COBE)-1923 96 y
(bulge)e(migh)o(t)g(b)q(e)h(dominated)f(b)o(y)g(stars)h(on)g(prograde)g
(orbits.)21 b(The)16 b(retrograde)g(orbits)f(apparen)o(tly)g(do)-1949 97 y
(not)h(pla)o(y)g(an)g(imp)q(ortan)o(t)g(role)f(in)h(the)g @F3 @sf(COBE)f
@F7 @sf(bulge.)21 b(Also)16 b(see)f @F3 @sf(Note)h(added)g(in)g(the)f(man)o
(uscript)g @F7 @sf(and)-1949 96 y(Figure)h(8)g(on)h(v)o(ertex)e(deviation)h
(seen)f(in)h(the)g(N-b)q(o)q(dy)h(bar)g(sim)o(ulation)f(of)g(Sellw)o(o)q
(o)q(d)h(\(1993\).)-1680 128 y(W)l(e)f(ha)o(v)o(e)f(also)i(lo)q(ok)o(ed)f
(for)h(a)g(correlation)f(b)q(et)o(w)o(een)f(the)h(Jacobi's)h(in)o(tegral)e
@F6 @sf(E)p 7 w @F10 @sf(J)22 -7 y @F7 @sf(and)i(metallici)o(t)o(y)l(.)o
-1948 97 a(The)h(Jacobi's)g(in)o(tegral)f @F6 @sf(E)p 7 w @F10 @sf(J)5 -7 y
@F7 @sf(,)h(equiv)m(alen)o(t)e(of)i(the)g(orbital)f(energy)h @F6 @sf(E)j
@F7 @sf(in)c(a)h(non-rotating)i(bulge,)d(is)h(a)-1913 96 y(constan)o(t)f
(for)f(an)h(orbit)f(and)h(determines)e(its)h(ap)q(ogalactic)h(distance.)k
(On)16 b(the)g(minor)g(axis,)-1267 141 y @F6 @sf(E)p 7 w @F10 @sf(J)47 -7 y
@F5 @sf(\021)41 b @F6 @sf(E)14 b @F5 @sf(\000)d @F7 @sf(\012)p @F6 @sf(L)p
7 w @F10 @sf(z)1053 -7 y @F7 @sf(\(6\))-1379 109 y(=)47 -34 y(1)-24 22 y
24 -2 z 46 w(2)6 -34 y(\()p @F6 @sf(V)11 -21 y @F11 @sf(2)-29 34 y @F10 @sf
(r)26 -13 y @F7 @sf(+)g @F6 @sf(V)11 -21 y @F11 @sf(2)-29 34 y @F10 @sf(l)
31 -13 y @F7 @sf(+)g @F6 @sf(V)11 -21 y @F11 @sf(2)-29 34 y @F10 @sf(b)16
-13 y @F7 @sf(\))g(+)g(\010)g @F5 @sf(\000)g @F7 @sf(\012\()p @F6 @sf(d)p
7 w @F10 @sf(l)p(os)14 -7 y @F5 @sf(\000)g @F6 @sf(R)p 7 w @F11 @sf(0)2
-7 y @F7 @sf(\))p @F6 @sf(V)p 7 w @F10 @sf(l)-1479 135 y @F7 @sf(F)l(or)
18 b(the)h(sample,)f(the)g(Jacobi's)g(in)o(tegral)g @F6 @sf(E)p 7 w @F10 @sf
(J)23 -7 y @F7 @sf(for)h(individual)e(star)i(is)f(p)q(o)q(orly)h
(determined)e(mostly)-1900 96 y(due)g(to)h(the)f(uncertain)o(t)o(y)e(in)i
(the)g(line-of-sigh)o(t)g(distance)g @F6 @sf(d)p 7 w @F10 @sf(l)p(os)3 -7 y
@F7 @sf(,)g(whic)o(h)f(en)o(ters)h(in)o(to)g(the)g(p)q(oten)o(tial)g(\010.)
-1927 96 y(Ho)o(w)o(ev)o(er,)o 16 x(there)g(is)g(a)g @F3 @sf(hin)o(t)g
@F7 @sf(that)h(the)e(metal)h(ric)o(h)f(stars)i(as)g(a)f(whole)h(ha)o(v)o
(e)e(lo)o(w)o(er)g(energy)h(\()p @F6 @sf(E)p 7 w @F10 @sf(J)6 -7 y @F7 @sf
(\))g(than)-1927 97 y(the)h(metal)g(p)q(o)q(or)i(stars.)28 b(See)18 b
(Figure)g(7.)29 b(Suc)o(h)18 b(a)g(correlation)g(app)q(ears)i(only)e
(marginal;)h(applying)-1897 96 y(the)d(P)o(earson)h(correlation)f(test,)g
(w)o(e)g(\014nd)h(a)f(6
(correlation)g(o)q(ccurring)-1946 96 y(randomly)l(.)22 b(This)17 b
(correlation)f(is)g(consisten)o(t)h(with)f(the)g(Zhao)i @F4 @sf(et)g(al.)
24 b @F7 @sf(\(1994\))18 b(\014nding)f(that)g(the)f(metal)-1942 97 y(ric)o
(h)f(stars)i(ha)o(v)o(e)f(lo)o(w)o(er)g(disp)q(ersion)g(and)h(are)g
(kinematic)o(ally)d(b)q(ounded)k(closer)e(to)g(the)g(cen)o(ter)f(and)i
(the)-1947 96 y(plane)d(than)i(the)e(metal)g(p)q(o)q(or)i(stars.)22 b(It)
14 b(is)g(notew)o(orth)o(y)h(that)g(the)f(correlation)h(is)f(almost)h
(indep)q(enden)o(t)-1951 96 y(of)i(the)f(c)o(hoice)f(of)h(p)q(oten)o(tial)g
(and)h(pattern)f(sp)q(eed.)-848 129 y(Although)g(these)g(results)g(are)g
(not)h(de\014nitiv)o(e)e(with)h(the)g(curren)o(t)f(small)h(sample,)g(the)g
(tec)o(hnique)o(s)-1949 96 y(used)g(here)g(can)g(b)q(e)h(useful)f(for)g
(future)g(analysis)h(of)f(observ)m(ational)h(data.)-1376 124 y
16 @eop1

15 @bop0

15 @eop0
0 0 15 @bop1 910 45 a @F7 @sf({)17 b(15)g({)-1040 147 y(p)q(o)q(or)i
(stars.)26 b(By)17 b(selecting)g(only)g(retrograde)h(orbits,)g(w)o(e)f
(can)h(construct)g(a)g(v)o(elo)q(cit)o(y)o 17 x(elli)o(psoid)f(that)-1916
96 y(resem)o(bles)g(the)i(metal)f(p)q(o)q(o)q(r)i(stars.)30 b(Figure)19 b
(4)g(sho)o(ws)h(in)f(gra)o(y)g(scale)g(the)f(distributions)h(in)g(three)
-1884 96 y(v)o(elo)q(cit)o(y)o 18 x(planes)h(with)f(the)g(bar)h(to)o(w)o
(ards)g @F5 @sf(\000)p @F7 @sf(45)p -18 w @F10 @sf(o)3 18 y @F7 @sf(.)30 b
(Comparing)21 b(with)e(Figure)g(1,)h(one)f(notices)g(the)-1874 96 y
(similar)e(v)o(ertex)g(deviations)h(in)g @F6 @sf(V)p 7 w @F10 @sf(l)14 -7 y
@F5 @sf(\000)13 b @F6 @sf(V)p 7 w @F10 @sf(r)21 -7 y @F7 @sf(ellipsoids.)
26 b(The)19 b(results)f(suggest)h(a)f(correlation)g(b)q(et)o(w)o(een)-1903
97 y(metallic)o(it)o(y)o 17 x(and)h(the)f(orbit)h(familie)o(s:)24 b(the)
18 b(metal)g(ric)o(h)f(and)i(metal)f(p)q(o)q(or)i(stars)f(tend)f(to)h(p)q
(opulate)-1900 96 y(B-family)c(and)i(R-family)e(orbits)h(resp)q(ectiv)o
(ely)o -4 x(.)-797 128 y(W)l(e)h(also)h(\014nd)g(the)f(data)h(consisten)o
(t)f(with)h(the)f(bar)h(lying)f(in)g(the)g(plane,)h(b)q(ecause)f(the)h
@F6 @sf(V)p 7 w @F10 @sf(l)14 -7 y @F5 @sf(\000)12 b @F6 @sf(V)p 7 w
@F10 @sf(b)-1917 90 y @F7 @sf(and)20 b @F6 @sf(V)p 7 w @F10 @sf(r)17 -7 y
@F5 @sf(\000)13 b @F6 @sf(V)p 7 w @F10 @sf(b)22 -7 y @F7 @sf
(distributions)19 b(sho)o(w)h(no)g(v)o(ertex)d(deviation)i(at)h(Baade's)f
(Windo)o(w.)31 b(Ho)o(w)o(ev)o(er,)o 19 x(the)-1875 96 y(observ)o(ed)17 b
(anisotrop)o(y)i @F6 @sf(\033)p 7 w @F10 @sf(r)19 -7 y @F6 @sf(>)d(\033)p
7 w @F10 @sf(b)20 -7 y @F7 @sf(is)i(not)g(repro)q(duced)g(in)f(the)h @F6 @sf
(V)p 7 w @F10 @sf(r)16 -7 y @F5 @sf(\000)11 b @F6 @sf(V)p 7 w @F10 @sf(b)
21 -7 y @F7 @sf(diagrams)18 b(for)h(the)e(sim)o(ulated)-1913 96 y(stars.)
26 b(This)18 b(suggests)h(a)f(need)g(to)g(p)q(opulate)g(more)g(lo)o(w)f
@F6 @sf(z)j @F7 @sf(and/or)f @F6 @sf(V)p 7 w @F10 @sf(z)22 -7 y @F7 @sf
(orbits)f(b)o(y)f(adopting)i(a)f(more)-1915 97 y(\015attened)e(densit)o(y)f
(and/or)j(an)f(initially)o 15 x(anisotropic)g(Gaussian)g(v)o(elo)q(cit)o
(y)d(distribution.)-1587 128 y(W)l(e)i(\014nd)g(the)g(observ)o(ed)g(v)o
(elo)q(cit)o(y)f(elli)o(psoid)h(is)g(only)g(consisten)o(t)g(with)g(a)h
(Bar)f(whose)h(near)g(side)f(is)-1948 96 y(in)e(the)g(\014rst)h(Galactic)f
(quadran)o(t)h(as)g(inferred)f(b)o(y)g(Blitz)f(and)i(Sp)q(ergel)f
(\(1991b\),)i(Binney)d @F4 @sf(et)k(al.)22 b @F7 @sf(\(1991\),)-1949 97 y
(Nak)m(ada)c @F4 @sf(et)h(al.)25 b @F7 @sf(\(1992\),)18 b(Whitelo)q(c)o(k)e
(&)h(Catc)o(hp)q(ole)g(\(1992\),)i(and)f(W)l(eiland)e @F4 @sf(et)j(al.)25 b
@F7 @sf(\(1994\).)g(The)18 b(top)-1929 96 y(panel)g(in)g(Figure)g(5)h(sho)o
(ws)h(the)e(v)o(elo)q(cit)o(y)o 17 x(ellipsoids)g(of)h(the)f(retrograde)h
(and)g(prograde)g(orbits)g(for)-1896 96 y(di\013eren)o(t)d(orien)o
(tations)h(of)g(the)g(bar.)24 b(The)17 b(orien)o(tation)g(of)g(the)g
(Baade's)g(Windo)o(w)g(v)o(elo)q(cit)o(y)e(ellipsoid)-1932 97 y(suggests)h
(that)g(the)f(Bar)g(is)g(to)o(w)o(ards)h @F6 @sf(l)p 7 w @F10 @sf(bar)16
-7 y @F6 @sf(<)e @F7 @sf(0)h(but)h(cannot)g(b)q(e)f(used)g(to)h
(accurately)e(constrain)i(the)f(bar)-1950 96 y(angle.)-27 128 y(With)h
(the)g(assumption)i(that)f(the)f(prograde)i(b)q(o)o(xy)e(orbits)h(are)g
(the)f(dominan)o(t)h(orbit)f(families)f(of)-1940 97 y(the)h(metal)f(ric)o
(h)g(bulge)g(based)i(on)f(similariti)o(es)f(in)g(v)o(elo)q(cit)o(y)f
(ellipsoids,)h(w)o(e)g(compute)g(and)i(pro)s(ject)e(the)-1950 96 y(densit)o
(y)h(of)h(the)g(prograde)g(b)q(o)o(xy)g(orbits)g(and)h(\014nd)f(a)g
(Bulge/Bar)f(that)h(resem)o(bles)f(the)g(COBE)h(Bulge)-1937 96 y(in)g
(\015attening)g(and)h(left-righ)o(t)e(asymmetry)l(.)23 b(Figure)17 b(6)g
(sho)o(ws)h(the)f(pro)s(jected)f(view)h(of)g(the)g(prograde)-1927 97 y
(orbits)e(with)g(the)f(bar)h(to)o(w)o(ards)h @F5 @sf(\000)p @F7 @sf(45)p
-18 w @F10 @sf(o)17 18 y @F7 @sf(in)f(gra)o(y)f(scales)h(and)g(con)o
(tours.)22 b(The)14 b(dotted)h(con)o(tours)h(are)e(righ)o(t)-1950 96 y
(side)j(folded)g(o)o(v)o(er)f(to)i(sho)o(w)g(the)f(asymmetry)l(.)24 b(An)
17 b(opp)q(osite)h(left-righ)o(t)e(asymmetry)h(is)g(seen)g(for)g(the)-1923
96 y(retrograde)h(orbits)h(\(not)f(sho)o(wn)h(here\).)26 b(The)18 b
(asymmetry)f(is)g(b)q(ecause)i(the)e(orbits)i(ha)o(v)o(e)e(a)h(triaxial)
-1909 195 y
15 @eop1

14 @bop0
@F3 @sf
[<
81F000C61C00E807007003807001C07001E07000E07000F07000F038007838007838007838
00783800783800781C00701C00701E00601D00C01CC1801C3F000E00000E00000E00000E00
000E00000E00000700000700000700000700000700000700003F8000078000>
	 21 35 -5 34 27] 98 @dc
@F5 @sf
[<
800002C0000660000C3000181800300C00600600C003018001830000C600006C0000380000
3800006C0000C6000183000301800600C00C006018003030001860000CC00006800002>
	 23 24 -8 23 39] 2 @dc

14 @eop0

0 0 14 @bop1 910 45 a @F7 @sf({)17 b(14)g({)-1040 147 y(sampling)h
(function,)f(and)i(hop)q(e)f(to)g(accum)o(ulate)f(some)h(insigh)o(ts)f(to)h
(the)g(imp)q(ortan)o(t)g(orbit)f(families)-1914 96 y(of)h(the)g(bar.)27 b
(The)18 b(phase)h(space)f(are)g(sampled)g(with)g(v)m(arious)h(functions.)
26 b(F)l(or)18 b(one)h(sim)o(ulation,)e(w)o(e)-1907 96 y(only)f(sample)h
(orbits)f(that)h(can)g(reac)o(h)f(Baade's)g(Windo)o(w,)h(and)g(w)o(e)f
(sample)g(the)h(v)o(elo)q(cit)n(y)e(space)i(with)-1943 96 y(a)g(uniform)g
(v)o(elo)q(cit)o(y)e(distribution)i(up)g(to)h(the)f(escap)q(e)g(v)o(elo)q
(cit)n(y)l(.)22 b(Suc)o(h)17 b(a)h(sampling)f(function)g(co)o(v)o(ers)-1931
97 y(the)i(observ)o(ed)g(phase)g(space)g(ev)o(enly)l(.)28 b(Other)19 b
(plausible)g(sampling)g(functions)g(in)o(v)o(olv)n(e)f(launc)o(hing)-1881
96 y(orbits)f(p)q(erp)q(endicularly)e(from)h(one)h(of)g(the)f(axes)g(of)h
(the)f(Bulge.)21 b(Ho)o(w)o(ev)o(er,)o 15 x(an)c(isothermal)f(Gaussian)-1945
96 y(distribution)j(seems)g(attractiv)o(e)g(as)h(a)g(\014rst)g(guess)g(to)g
(the)g(phase)g(space)g(densit)o(y)e(b)q(ecause)i(of)g(its)-1869 97 y
(simplici)o(t)o(y)l(.)o 20 x(It)14 b(also)h(allo)o(ws)g(us)g(to)g(obtain)g
(a)g(clean)f(signature)i(of)f(triaxialit)n(y)l(,)e(b)q(ecause)i(an)o(y)f
(anisotrop)o(y)-1949 96 y(dev)o(elop)q(ed)h(in)h(phase-mixing)g(is)g
(generic)f(of)i(the)f(p)q(oten)o(tial,)f(not)i(due)f(to)h(the)f(sampling)g
(function.)-1773 128 y(Here)f(w)o(e)h(will)g(limit)o 16 x(the)g
(discussion)h(to)g(some)f(seemingly)f(generic)h(results)g(as)i(sho)o(wn)f
(b)o(y)f(one)h(of)-1941 97 y(our)g(sim)o(ulations.)23 b(In)17 b(this)g
(run,)g(20,000)i @F3 @sf(sim)o(ulated)c(orbits)j @F7 @sf(are)f(sampled)g
(according)g(to)g(the)g(densit)o(y)-1932 96 y(of)g(the)f(bar)h(and)g
(launc)o(hed)e(with)h(v)o(elo)q(cities)f(dra)o(wn)i(from)f(an)h(isotropic)f
(Gaussian)i(distribution)d(with)-1947 96 y(100)20 b(km/s)f(disp)q(ersion)h
(and)f(zero)g(streaming)g(v)o(elo)q(cit)o(y)o 18 x(in)g(the)f(bar)i
(frame.)29 b(W)l(e)19 b(group)h(the)e(stars)-1882 97 y(according)g(to)h
(their)e(orbit)h(classes,)g(whic)o(h)g(w)o(e)f(kno)o(w)i(b)o(y)e(c)o(hec)o
(ki)o(ng)g(the)h(angular)h(momen)o(tum)e(and)-1905 96 y(axial)g(ratio)g
(of)h(the)f(orbits)g(for)h(an)f(in)o(tegration)g(time)f(5)i(billi)o(on)f
(y)o(ears.)23 b(The)18 b(inner)e(2)i(kp)q(c)f(is)g(divided)-1929 96 y(in)o
(to)f(20)c @F5 @sf(\002)f @F7 @sf(20)h @F5 @sf(\002)f @F7 @sf(20)18 b
(cubic)d(cells.)21 b(F)l(or)16 b(eac)o(h)h(pass)g(of)g(a)g(family)e(of)i
(orbits)g(through)g(a)g(cell,)e(w)o(e)h(add)h(the)-1944 97 y(densit)o(y)l
(,)e(the)h(v)o(elo)q(cit)o(y)f(\014rst)i(and)g(second)g(momen)o(ts)f(on)h
(that)g(cell)e(\(10)i(records)g(p)q(er)f(orbit)h(family)e(p)q(er)-1942 96 y
(cell\).)o 27 x(In)j(the)h(end,)f(w)o(e)g(obtain)h(the)f(densit)o(y)g(and)h
(the)f(v)o(elo)q(cit)n(y)f(ellipsoid)g(of)i(eac)o(h)f(class)g(of)h(orbits)
-1897 96 y(on)h(a)f(grid,)h(whic)o(h)e(are)i(then)f(pro)s(jected)f(as)i
(mappings)f(in)g(the)g(Galactic)g(co)q(ordinates)h(\()p @F6 @sf(l)q(;)8 b
(b)p @F7 @sf(\);)19 b(the)-1879 96 y(pro)s(jections)e(are)h(done)f(for)h
(12)g(c)o(hoices)f(of)h(the)f(bar)h(angle)g @F6 @sf(l)p 7 w @F10 @sf(bar)
19 -7 y @F7 @sf(ev)o(ery)e(15)p -18 w @F10 @sf(o)21 18 y @F7 @sf(starting)i
(at)g @F6 @sf(l)p 7 w @F10 @sf(bar)18 -7 y @F7 @sf(=)e @F5 @sf(\000)p
@F7 @sf(15)p -18 w @F10 @sf(o)3 18 y @F7 @sf(.)-1921 97 y(T)l(o)j(mak)o(e)e
(a)h(grey)g(scale)g(of)g(v)o(elo)q(cit)o(y)e(distribution,)i(w)o(e)f(also)i
(record)f(the)g(v)o(elo)q(cit)o(y)o 17 x(of)g(eac)o(h)g(star)g(on)-1905
96 y(ev)o(ery)c(pass)j(through)f(the)g(Baade's)f(Windo)o(w)h(p)q(osition.)
22 b(Th)o(us,)15 b(our)h(sim)o(ulated)f(v)o(elo)q(cit)o(y)o 14 x
(distribution)-1950 96 y(is)h(an)h(in)o(tegration)f(through)h @F4 @sf(al)r
(l)h @F7 @sf(of)f(the)f(line)f(of)i(sigh)o(t)f(and)h(is)f(deep)q(er)g
(than)h(most)f(magnitude)h(limi)o(ted)-1949 97 y(observ)m(ations)g(of)g
(the)f(Bulge.)-456 128 y(W)l(e)i(\014nd)h(that)g(b)o(y)f(simply)g
(selecting)f(the)i(prograde)g(stars)h(in)e(the)g(20,000)j(sim)o(ulated)c
(orbits,)-1892 96 y(w)o(e)h(repro)q(duce)g(the)g(orien)o(tation)g(and)h
(shap)q(e)g(of)g(the)f(v)o(elo)q(cit)n(y)f(ellipsoid)g(of)h(metal)g(ric)o
(h)f(and)i(metal)-1902 131 y
14 @eop1

13 @bop0
@F5 @sf
[<
800007E080001FF880007FFCC000F81EC001E00660078003781F00033FFE00011FF8000107
E000010000000000000000800007E080001FF880007FFCC000F81EC001E00660078003781F
00033FFE00011FF8000107E00001>
	 32 22 -3 23 39] 25 @dc
@F7 @sf
[<
7FF1FFCFFE07001C00E007001C00E007001C00E007001C00E007001C00E007001C00E00700
1C00E007001C00E007001C00E007001C00E007001C00E007001C00E007001C00E007001C00
E007001C00E007001C00E007001C00E007001C00E007001C01E0FFFFFFFFE007001C000007
001C000007001C000007001C000007001C000007001C000007001C000007001C00C007003C
01E003803E01E001801E00E000E00B0040007031C080000FC07F00>
	 39 35 0 34 41] 14 @dc

13 @eop0

0 0 13 @bop1 910 45 a @F7 @sf({)17 b(13)g({)-1040 147 y(in)j(absolute)g
(prop)q(er)h(motions)f(and)g(distance)g(on)h(orbit)f(classi\014cation)f
(for)i(one)f(of)g(the)g(stars)h(in)-1861 96 y(the)g(sample,)g(Arp)g
(#1102.)37 b(T)o(ypical)20 b(20)i(km/s)f(v)o(elo)q(cit)o(y)o 20 x
(uncertain)o(t)o(y)f(do)q(es)h(not)h(c)o(hange)f(the)-1834 96 y(orbit)d
(classi\014cation.)26 b(Ho)o(w)o(ev)o(er,)o 17 x(a)18 b(de\014nitiv)o(e)e
(prograde/retrogra)q(de)j(classi\014cation)f(is)f(necessarily)-1911 96 y
(determined)c(b)o(y)g(the)i(sign)f(of)h(the)f(initial)f(angular)i(momen)o
(tum)e @F6 @sf(L)p 7 w @F10 @sf(z)18 -7 y @F5 @sf(\031)g @F7 @sf(\()p
@F6 @sf(d)p 7 w @F10 @sf(l)p(os)10 -7 y @F5 @sf(\000)7 b @F6 @sf(R)p 7 w
@F11 @sf(0)2 -7 y @F7 @sf(\))p @F6 @sf(V)p 7 w @F10 @sf(l)2 -7 y @F7 @sf
(,)14 b(where)g @F6 @sf(d)p 7 w @F10 @sf(l)p(os)17 -7 y @F7 @sf(is)g(the)
-1950 97 y(line-of-sigh)o(t)j(distance,)f @F6 @sf(d)p 7 w @F10 @sf(l)p(os)
19 -7 y @F6 @sf(>)f(R)p 7 w @F11 @sf(0)19 -7 y @F7 @sf(and)j @F6 @sf(d)p
7 w @F10 @sf(l)p(os)18 -7 y @F6 @sf(<)e(R)p 7 w @F11 @sf(0)19 -7 y @F7 @sf
(corresp)q(ond)i(to)f(line-of-sigh)o(t)g(p)q(ositions)h(further)-1927 96 y
(or)e(closer)e(than)i(the)f(tangen)o(t)h(p)q(oin)o(t)f(resp)q(ectiv)o(el)o
(y)l(.)o 20 x(T)l(o)h(in)o(tegrate)f(real)f(stellar)h(orbits,)g(distances)g
(m)o(ust)-1950 96 y(b)q(e)j(determined)f(with)h(accuracy)g(su\016cien)o(t)e
(to)j(place)e(the)h(star)h(on)g(either)e(the)h(near)g(or)g(far)h(side)e
(of)-1904 97 y(the)h(bulge.)27 b(This)18 b(will)f(fa)o(v)o(or)h(studies)g
(of)g(stars)h(whose)g(luminosities)d(ma)o(y)i(b)q(e)g(inferred)f(from)h
(their)-1905 96 y(p)q(ositions)f(on)f(principal)f(sequences)h(in)f(the)h
(HR)g(diagram,)g(or)g(v)m(ariable)g(stars)g(\(Miras)g(or)h(RR)e(Lyraes\))
-1949 96 y(where)j(a)g(luminosit)o(y)e(ma)o(y)i(b)q(e)g(inferred)f(from)h
(pulsation.)27 b(Whitelo)q(c)o(k)16 b(&)i(Catc)o(hp)q(ole)g(\(1992\))i(ha)o
(v)o(e)-1909 97 y(measured)e(the)g(distances)h(of)f(141)i(Miras)e(in)g
(galactic)g(longitude)g(strip)q(es)h @F5 @sf(j)p @F6 @sf(b)p @F5 @sf(j)d
@F7 @sf(=)i(7)p -18 w @F10 @sf(o)15 18 y @F5 @sf(\000)12 b @F7 @sf(8)p -18 w
@F10 @sf(o)21 18 y @F7 @sf(based)19 b(on)-1899 96 y(their)e(p)q(erio)q
(d-luminosit)o(y)h(relation.)26 b(They)18 b(sho)o(w)h(that)g(Miras)e(on)i
(the)f(p)q(ositiv)o(e)f(longitude)h(side)g(of)-1904 96 y(the)f(Bulge)g(is)h
(0)p @F6 @sf(:)p @F7 @sf(5)g(magnitude)f(brigh)o(ter)g(than)h(those)g(on)h
(the)e(negativ)o(e)g(longitude)g(side,)g(indicating)-1917 96 y(a)h
(triaxial)f(geometry)l(.)26 b(Unfortunately)l(,)18 b(Miras)f(or)i(RR)e
(Lyraes)i(are)f(rare)g(and)h(relativ)n(ely)o 17 x(few)f(stars)-1908 97 y
(will)g(b)q(e)g(found)i(on)f(historical)f(plates)g(useful)g(for)h(prop)q
(er)h(motion)e(study)l(.)29 b(The)18 b(red)h(clump)e(migh)o(t)-1893 96 y
(p)q(oten)o(tially)i(presen)o(t)g(another)h(p)q(ossible)g(source)g(of)g
(candidates,)h(but)f(its)f(0.5)i(mag)f(thic)o(kness)f(is)-1865 96 y
(probably)e(to)q(o)i(large.)24 b(W)l(e)17 b(feel)e(that)j(the)f(p)q(oten)o
(tial)g(v)m(alues)g(of)g(this)g(tec)o(hnique)f(is)h(w)o(orth)g(additional)
-1928 97 y(e\013orts)g(to)g(accurately)e(measure)h(reddening)g(and)h
(distances)f(of)g(bulge)g(stars.)-1362 128 y(Although)k(w)o(e)g(still)f
(cannot)i(classify)f(orbits)g(for)h(individual)e(stars,)j(the)e(distinct)f
(v)o(elo)q(cit)o(y)-1853 96 y(ellipsoids)c(for)h(the)g(metal)g(ric)o(h/p)q
(o)q(or)h(ma)o(y)e(b)q(e)h(due)g(to)h(di\013eren)o(t)e(orbit)h(families.)k
(W)l(e)c(ha)o(v)o(e)f(examined)-1951 97 y(this)i(n)o(umericall)o(y)f(b)o
(y)g(p)q(opulating)j(di\013eren)o(t)d(orbit)h(families)f(of)h(the)g(bar)h
(p)q(oten)o(tial.)23 b(Eviden)o(tly)l(,)o 16 x(one)-1928 96 y(has)18 b
(large)g(degrees)f(of)h(freedom)f(in)h(c)o(ho)q(osing)g(a)g(6-dimensional)g
(sampling)f(function)h(for)g(launc)o(hing)-1917 96 y(the)g(orbits.)26 b
(The)18 b(degrees)g(of)g(freedom)g(is)g(only)f(strongly)i(reduced)e(if)h
(one)g(w)o(ere)f(able)h(to)g(solv)o(e)f(the)-1908 97 y(self-consisten)o(t)d
(problem)f(with)i(the)f(Sc)o(h)o(w)o(arzc)o(hi)o(ld)f(tec)o(hnique,)o 14 x
(namely)l(,)g(to)i(select)e(out)i(a)g(set)f(of)h(orbits)-1950 96 y(that)e
(repro)q(duce)h(the)e(densit)o(y)l(.)19 b(Pro)q(ducing)14 b
(self-consisten)o(t)f(bars)g(has)h(b)q(een)f(a)h(long)f(standing)h
(problem,)-1951 96 y(whic)o(h)i(w)o(e)g(will)f(deal)h(with)g(in)g(a)h
(later)f(pap)q(er)h(\(Zhao)g(1994\).)23 b(Rather,)16 b(w)o(e)g(start)h
(with)f(some)h(plausible)-1948 163 y
13 @eop1

12 @bop0
/@F0 @newfont
@F0 @sf
[<
00060000000006000000000E000000000F000000000F000000001F000000001D800000001D
800000001D8000000038C000000038C000000038C000000070600000007060000000706000
0000E030000000E030000000E030000000E018000001C018000001C018000001C00C000003
800C000003800C0000038006000003800600000700060000070003000007000300000E0003
00000E000180008E00018000DC000180007C0000C0003C0000C0001C0000C0000800006000
00000060000000006000000000300000000030000000003000000000180000000018000000
0018000000000C000000000C000000000C0000000006000000000600000000060000000003
000000000300000000030000000001800000000180000000018000000000C000000000C000
000000C0000000006000000000600000000060000000003000000000300000000030000000
001800000000180000000018000000000C000000000C000000000C00000000060000000006
0000000002>
	 39 75 -4 1 42] 113 @dc

12 @eop0

0 0 12 @bop1 910 45 a @F7 @sf({)17 b(12)g({)-1040 147 y(sense)e(of)h
(rotation)g(with)f(resp)q(ect)g(to)h(the)f(rest)g(frame,)g(and)g(ha)o(v)o
(e)g(b)q(o)o(xy)g(shap)q(es)i(aligned)e(with)g(the)g(bar.)-1950 96 y
(R-family)i(orbits)h(ha)o(v)o(e)g(a)g(retrograde)h(sense)f(of)g(rotation)h
(in)f(b)q(oth)h(frames)f(and)h(are)f(lo)q(ops)h(aligned)-1905 96 y(p)q
(erp)q(endicular)f(to)g(the)g(bar.)27 b(F)l(or)18 b(a)h(strong)g(and)f
(rapidly)g(rotating)g(bar,)h(large)f(fractions)g(of)h(phase)-1907 96 y
(space)g(are)g(also)g(tak)o(en)f(b)o(y)g(semi-ergo)q(dic)h(orbits)g(or)g
(higher)f(order)h(bifurcated)f(orbits)h(\(Udry)f(and)-1889 97 y(Pfenniger)e
(1988\).)24 b(T)o(ypically)l(,)o 15 x(these)17 b(orbits)g(tend)f(to)h(o)q
(ccup)o(y)f(man)o(y)h(small)f(islands)g(in)h(phase)g(space.)-1941 96 y
(They)g(neither)e(p)q(ossess)k(three)d(in)o(tegrals)g(of)h(motion)g(nor)g
(do)h(they)e(uniformly)g(\014ll)f(the)i(space)g(allo)o(w)o(ed)-1938 96 y
(b)o(y)e(the)g(Jacobi's)h(in)o(tegral)f @F6 @sf(E)p 7 w @F10 @sf(J)5 -7 y
@F7 @sf(.)21 b(See)16 b(Con)o(top)q(oulos)h(&)e(Grosb)q(o)q(l)h(\(1989\))h
(for)f(a)g(review)f(on)h(planar)g(orbit)-1950 97 y(families,)e(and)j
(Hasan)g @F4 @sf(et)h(al.)23 b @F7 @sf(\(1993\))18 b(on)e(3-dimensional)g
(orbits)h(in)f(a)g(cen)o(trally)f(concen)o(trated)h(bar.)-1805 128 y(F)l
(or)i(the)f(purp)q(ose)i(of)g(classifying)e(the)h(orbits)g(of)g(stars)h
(in)e(Baade's)h(Windo)o(w,)g(w)o(e)g(use)g(a)g(more)-1909 96 y(practical)e
(set)h(of)h(criteri)o(a.)23 b(As)17 b(Baade's)g(Windo)o(w)g(is)g(relativ)o
(e)o(ly)e(far)j(from)f(the)f(Galactic)h(plane)g(with)-1931 97 y(its)e
(closest)g(approac)o(h)h(b)q(eing)f(500)h(p)q(c)g(or)f(1)p @F6 @sf(=)p
@F7 @sf(5)h(of)g(the)f(corotation)h(radius,)f(most)g(of)h(the)f(orbits)g
(through)-1949 96 y(it)i(are)g(semi-ergo)q(dic,)g(or)h(b)q(elong)g(to)g
(high)f(order)h(bifurcation)f(of)g(regular)h(families.)o 24 x(Regular)f
(orbits)-1922 96 y(t)o(ypically)o 16 x(exist)g(close)g(to)h(the)f(bar)h
(plane.)25 b(By)17 b(running)h(orbit)g(pairs)f(that)h(are)g(initially)o
16 x(o\013set)h(b)o(y)e(a)-1919 97 y(small)g(amoun)o(t,)h(0)p @F6 @sf(:)p
@F7 @sf(001)h(p)q(c,)f(and)g(determining)e(the)i(Ly)o(apuno)o(v)g(exp)q
(onen)o(ts)g(for)g(the)f(orbits,)h(w)o(e)f(\014nd)-1914 96 y(that)i(most)g
(orbits)f(are)h(semi-ergo)q(dic.)27 b(Ho)o(w)o(ev)o(er,)17 b(most)i
(orbits)g(do)f(ha)o(v)o(e)g(\014xed)g(sense)h(of)f(rotation)-1895 96 y
(for)h(the)f(in)o(tegration)h(times,)e(here)h(5)h(billion)f(y)o(ears)g
@F5 @sf(\030)g @F7 @sf(50)h(rotation)h(p)q(erio)q(ds,)f(th)o(us)g(the)f
(orbits)h(are)-1893 97 y(readily)e(classi\014ed)g(in)o(to)g(broad)h
(families)e(based)j(on)f(their)e(sense)i(of)g(rotation)g(and)g(morphology)l
(.)26 b(A)o(t)-1919 96 y(the)16 b(end)h(of)g(the)f(n)o(umerical)f(in)o
(tegration)h(of)h(an)g(orbit,)f(w)o(e)g(obtain)h(its)f(range)h(of)g
(angular)g(momen)o(tum)-1944 96 y(z-comp)q(onen)o(t)g @F6 @sf(L)p 7 w
@F10 @sf(z)20 -7 y @F7 @sf(and)h(its)e(r.m.s.)22 b(shap)q(e)17 b
(parameter)g @F6 @sf(Q)d @F7 @sf(=)14 -50 y @F0 @sf(q)p 195 -2 z 50 w
@F5 @sf(h)p @F6 @sf(x)p -14 w @F11 @sf(2)2 14 y @F5 @sf(i)p @F6 @sf(=)p
@F5 @sf(h)p @F6 @sf(y)2 -14 y @F11 @sf(2)2 14 y @F5 @sf(i)q @F7 @sf(,)i
(where)g @F6 @sf(x)h @F7 @sf(and)g @F6 @sf(y)h @F7 @sf(de\014ne)f(the)-1940
96 y(long)g(and)h(in)o(termedi)o(ate)e(axes)h(of)g(the)f(bar)i(plane)e
(resp)q(ectiv)o(ely)o -4 x(.)22 b(W)l(e)17 b(\014nd)g(that)g(the)g
(prograde)h(orbits)-1935 97 y(ha)o(v)o(e)d(a)i(b)q(o)o(xy)f(shap)q(e,)h
(and)g(the)f(retrograde)h(orbits)f(are)h(lo)q(ops.)22 b(In)16 b(this)g
(pap)q(er,)g(w)o(e)g(classify)g(stars)h(in)o(to)-1950 96 y(three)g(broad)i
(groups:)26 b(the)18 b(prograde)h(b)q(o)o(xy)e(orbits,)i(the)e(retrograde)i
(lo)q(op)f(orbits)h(and)f(orbits)g(that)-1909 96 y(c)o(hange)e(their)g
(sense)g(of)g(rotation.)-544 129 y(It)h(w)o(ould)h(b)q(e)f(desirable)g(to)i
(classify)e(the)g(orbit)h(of)g(eac)o(h)f(individual)f(star)j(in)e(the)h
(sample.)25 b(But)-1915 96 y(w)o(e)18 b(\014nd)h(that)g(the)f
(line-of-sigh)o(t)h(distances)f(for)h(these)f(62)i(stars)f(are)g(so)g
(uncertain)f(that)h(ev)o(en)f(the)-1894 96 y(sense)f(of)f(rotation)i(of)f
(a)f(star)i(cannot)f(b)q(e)f(determined.)21 b(Figure)16 b(3)h(sho)o(ws)h
(the)e(e\013ect)g(of)h(uncertain)o(ties)-1944 131 y
12 @eop1

11 @bop0
@F6 @sf
[<
FFC0001C00001C00001C00000E00000E00000E00000E0000070000070000071E0007218003
C0E003C07003803803803801C03C01C01E01C01E01C01E00E00F00E00F00E00F00E00F0870
0F08700708700708780E04740E04631C03C0F0>
	 24 31 1 20 24] 112 @dc
@F6 @sf
[<
07E00F001C1C1880380338403000B8007000780070003800F0003800F0003800F0003C00F0
003A0078003A0078003900780039003C0038803C0070801C0070400E007040070060400380
C02000E18000003F0000>
	 27 21 -2 20 31] 11 @dc
@F6 @sf
[<
3000007000003800003800003800003800001C00001C00001C00001C00000E00000E00000E
00000E00008700008701808783C08783C04741C04630C03C0F00>
	 18 21 -2 20 22] 114 @dc
@F6 @sf
[<
C00000E00000E00000E00000E0000070000070000070000070000038000038000038F00039
0C001E07001E03801C01C01C01C00E01E00E00F00E00F00E00F00700780700780700780380
7803807801803800C03800E0700070700018E0000F80>
	 21 32 -2 20 25] 26 @dc
@F6 @sf
[<
FFE0203FFF000F003001F00006003800F00002003800F00002007C00F00002007C00F00001
007A007800010079007800010079007800010078807800008078803C00008078403C000080
78203C000080F0203C000040F0101E000040F0081E000040F0081E000040F0041E000020F0
040F000020F0020F000020F0010F000021E0010F000011E00087800011E00047800011E000
47800011E00027800009E00023C00009E00013C00009E0000BC00009E0000BC00007C00005
E00007C00005E00007C00003E0007FC00001FF>
	 48 34 -2 33 47] 77 @dc
@F9 @sf
[<
00FC0007038008004010002020001020001040000840000880000480780480FC0480FC0480
FC0480FC0480780480000440000840000820001020001010002008004007038000FC00>
	 22 24 -2 19 27] 12 @dc
@F6 @sf
[<
07E0001C1C00380200300180700040700040F00000F00000F00000F0000078000078000078
00003C00003C00001C03C00E03C00701C00380C000E080003F00>
	 18 21 -2 20 21] 99 @dc
@F11 @sf
[<
1F8020C0603070187018001C000C000E0F8E186E301E601EE00EE00EE00EE00CE00C601830
18183007C0>
	 15 21 -1 20 18] 57 @dc
@F6 @sf
[<
0F01C018C620302620701E10700E10F00E10F00708F00700F00700F0070078038078038078
03803803803C01C01C01C00E01C00601C00302E001C4E0007860>
	 21 21 -2 20 26] 97 @dc
@F10 @sf
[<
3F0060C0C020C01080008000C000C000C0006000207010700C2007C0>
	 12 14 -2 13 15] 99 @dc
@F4 @sf
[<
3F800060E000F03000F01800701C00000E00000E00000E0000070000070001E70006170006
0B800E07800E03801E03801E01C01E01C01E01C01E01C00F00E00F00E00F00E00700E00780
7003807001C07001C07000E0B80030B8001F18>
	 21 31 -2 20 22] 103 @dc
@F6 @sf
[<
C00000E00000E00000E00000E0000070000070000070000070000038000038000039F0383F
08C41E04C41E03C21E01C21E01C20E00E10E00E00E00E00E00E00700700700700700700700
7003803803803803803803803801C01C01C01C00C00C>
	 24 32 -2 20 29] 22 @dc
@F2 @sf
[<
FFFFF0FFFFF07FFFF03FFFF01FFFF00FFFF00E003807001803801801C01800E01800700000
3800003E00001F00000F800007C00007E00003F00003F00003F83803F87C01F8FE01F8FE03
F8FE03F8FE03F07C07F07007E03C1FC00FFF0003FC00>
	 21 32 -3 31 28] 50 @dc
[<
03FFFFF80003FFFFF8000003F800000003F800000003F800000003F800000003F800000003
F800000003F800000003F800000003F800000003F800000003F800000003F800000003F800
000003F800000003F800000003F800000003F800000003F800000003F800000003F80000C0
03F800C0C003F800C0C003F800C0C003F800C0E003F801C0E003F801C06003F801807003F8
03807803F807807E03F80F807FFFFFFF807FFFFFFF80>
	 34 34 -2 33 39] 84 @dc
[<
7FF0FFE07FF0FFE00F801F000F801F000F801F000F801F000F801F000F801F000F801F000F
801F000F801F000F801F000F801F000F801F000F801F000F801F000F801F000F801F000F80
1F000F801F00FFFFFF00FFFFFF000F8000000F8000000F8000000F8000000F800C000F801E
000F803F000F803F0007C03F0003E01F0001F80E00007FFC00000FF000>
	 27 35 -1 34 31] 12 @dc
[<
0007FC0000003FFF800000FC07E00003F001F80007E000FC000FC0007E001F80003F003F80
003F803F00001F807F00001FC07F00001FC07E00000FC0FE00000FE0FE00000FE0FE00000F
E0FE00000FE0FE00000FE0FE00000FE0FE00000FE0FE00000FE0FE00000FE07E00000FC07E
00000FC07F00001FC03F00001F803F00001F801F80003F001F80003F000FC0007E0007E000
FC0003F001F80000FC07E000003FFF80000007FC0000>
	 35 34 -3 33 42] 79 @dc
[<
180FC0001C3FF8001EE07C001FC03E001F801F001F001F801F000F801F000FC01F000FC01F
000FC01F000FC01F000FC01F000FC01F000FC01F000FC01F000F801F000F801F001F001F80
1E001FF07C001F3FF8001F0FE0001F0000001F0000001F0000001F0000001F0000001F0000
001F0000001F0000001F0000001F0000001F000000FF000000FF000000>
	 26 35 -2 34 31] 98 @dc

11 @eop0

0 0 11 @bop1 910 45 a @F7 @sf({)17 b(11)g({)-1040 147 y(logarithmic)g
(slop)q(e)h @F5 @sf(\000)p @F6 @sf(p)e @F7 @sf(=)g @F5 @sf(\000)p @F7 @sf
(1)p @F6 @sf(:)p @F7 @sf(8)i(or)g @F6 @sf(\013)f @F5 @sf(\021)f @F7 @sf(2)c
@F5 @sf(\000)g @F6 @sf(p)17 b @F7 @sf(=)f(0)p @F6 @sf(:)p @F7 @sf(2)i(and)g
(normalization)f(constan)o(ts)i @F6 @sf(r)p 7 w @F11 @sf(0)18 -7 y @F7 @sf
(=)d(1)i(kp)q(c)-1916 96 y(and)g @F6 @sf(\032)p 7 w @F11 @sf(0)19 -7 y
@F7 @sf(=)e(0)p @F6 @sf(:)p @F7 @sf(7)p @F6 @sf(M)p 7 w @F9 @sf(\014)3 -7 y
@F6 @sf(pc)p -18 w @F9 @sf(\000)p @F11 @sf(3)2 18 y @F7 @sf(,)i(whic)o(h)f
(corresp)q(onds)i(to)f(7)p @F6 @sf(:)p @F7 @sf(3)13 b @F5 @sf(\002)e @F7 @sf
(10)p -18 w @F11 @sf(9)3 18 y @F6 @sf(M)p 7 w @F9 @sf(\014)20 -7 y @F7 @sf
(within)18 b(1)g(kp)q(c.)25 b(W)l(e)18 b(set)f(the)h(shap)q(e)-1914 96 y
(parameters)f @F6 @sf(b)p 7 w @F11 @sf(20)16 -7 y @F7 @sf(=)e(0)p @F6 @sf
(:)p @F7 @sf(8)j(and)f @F6 @sf(b)p 7 w @F11 @sf(22)16 -7 y @F7 @sf(=)f(0)p
@F6 @sf(:)p @F7 @sf(4)h(so)g(that)h(the)f(bar)g(has)h(axis)f(ratios)g
@F6 @sf(a)e @F7 @sf(:)g @F6 @sf(b)g @F7 @sf(:)f @F6 @sf(c)h @F7 @sf(=)g(1)h
(:)f(0)p @F6 @sf(:)p @F7 @sf(88)g(:)g(0)p @F6 @sf(:)p @F7 @sf(76)-1931 96 y
(in)g(p)q(oten)o(tial)h(con)o(tour)g(and)g(1)e(:)g(0)p @F6 @sf(:)p @F7 @sf
(8)g(:)f(0)p @F6 @sf(:)p @F7 @sf(5)j(in)f(densit)o(y)g(con)o(tour,)h(and)g
(a)g(pro)s(jected)f(\015attening)h(0.3|0.4,)-1950 97 y(insensitiv)o(e)g
(to)j(the)e(orien)o(tation)h(of)g(the)g(bar's)h(long)f(axis)g(in)g(the)g
(Galactic)f(plane.)27 b(The)18 b(densit)o(y)f(is)-1907 96 y(ev)o(erywhere)e
(p)q(ositiv)o(e)h(de\014nite.)22 b(W)l(e)16 b(also)i(adopt)f(the)g(bar)g
(pattern)g(sp)q(eed)g(of)g(Binney)e @F4 @sf(et)k(al.)24 b @F7 @sf(\012)15 b
(=)g(81)-1937 96 y(km/s/kp)q(c,)20 b(whic)o(h)e(corresp)q(onds)i(to)g(a)f
(corotation)h(radius)f @F6 @sf(R)p 7 w @F10 @sf(cor)22 -7 y @F7 @sf(=)g(2)p
@F6 @sf(:)p @F7 @sf(4)g(kp)q(c;)h(the)f(pattern)g(sp)q(eed)-1883 97 y
(comes)f(in)o(to)g(the)h(equations)f(of)h(motion)g(for)g(orbits)g(through)g
(the)f(Coriolis)h(force)f(and)h(cen)o(trifugal)-1895 96 y(force)d(terms)g
(\(see,)f @F4 @sf(e.g.)23 b @F7 @sf(,)16 b(Binney)f(and)i(T)l(remaine)e
(1987\).)-999 128 y(The)h(analytic)g(bar)h(p)q(oten)o(tial)f(used)h(here)f
(shares)h(man)o(y)f(adv)m(an)o(tages)i(of)f(the)f(prolate)h(ellipsoidal)
-1944 97 y(bar)i(used)g(in)f(Binney)g @F4 @sf(et)i(al.)30 b @F7 @sf
(\(1991\).)g(It)18 b(can)h(repro)q(duce)g(not)g(only)g(the)f(cusp)q(ed)h
(orbit)g(to)g(\014t)g(the)-1891 96 y(CO)f(gas)i @F6 @sf(l)q @F7 @sf(-v)e
(diagram,)h(but)f(also)h(the)f(observ)o(ed)g(radial)g(pro\014le)g(and)h
(\015attening)g(of)f(the)g(Bulge.)27 b(In)-1901 96 y(addition,)15 b(its)f
(simple)g(analytical)g(form)h(is)f(e\016cien)o(t)f(for)i(orbit)g(in)o
(tegration,)f(and)h(it)g(is)f(straigh)o(tforw)o(ard)-1949 97 y(to)19 b
(extend)f(the)h(densit)o(y/p)q(oten)o(tial)f(pair)g(to)i(mo)q(del)e(the)h
@F3 @sf(COBE)f @F7 @sf(Bulge)g(b)o(y)h(adding)g(higher)g(order)-1889 96 y
(expansion)d(terms.)21 b(The)16 b(simple)f(analytical)g(p)q(oten)o(tial)h
(is)g(also)g(e\016cien)o(t)e(for)j(orbit)f(in)o(tegration.)k(Suc)o(h)-1950
96 y(a)d(cusp)q(ed)h(p)q(o)o(w)o(er)f(la)o(w)g(pro\014le)g(is)g(seen)f(in)h
(the)g(2.2)g @F6 @sf(\026)h @F7 @sf(data)g(from)f(1)g(p)q(c)h(out)f(to)h
(ab)q(out)g(400)g(p)q(c)f(on)h(the)-1930 97 y(minor)g(axis)h(and)h(1)f(kp)q
(c)f(on)i(the)e(ma)s(jor)h(axis)g(\(Ken)o(t)f(1992\).)30 b(Inside)18 b
(10p)q(c,)i(the)f(putativ)o(e)f(cen)o(tral)-1890 96 y(blac)o(k)e(hole)h
(ma)o(y)f(in\015uence)g(the)h(v)o(elo)q(cit)n(y)e(distribution.)23 b(Bey)o
(ond)16 b(1)i(kp)q(c,)e(although)i(the)f(bulge)g(ligh)o(t)-1935 96 y
(steep)q(ens)f(considerably)g(with)g(a)h(spatial)f(densit)o(y)g(p)q(o)o(w)o
(er)g(la)o(w)g(index)g @F5 @sf(\030)g @F7 @sf(-3.7,)g(the)g(rest)g(of)h
(the)f(Galaxy)-1949 96 y(starts)h(con)o(tributing)e(to)h(the)f(p)q(oten)o
(tial.)21 b(Giv)o(en)15 b(these,)g(our)h(p)q(oten)o(tial)g(mo)q(del)f
(migh)o(t)g(not)i(b)q(e)e(to)q(o)i(bad)-1949 97 y(as)g(an)g(appro)o
(ximation)f(of)g(p)q(oten)o(tial)g(through)i(the)e(Bulge)f(region,)h
(namely)l(,)f(from)h(10)h(p)q(c)f(to)h(2)g(kp)q(c.)-1330 203 y @F2 @sf
(3.2.)55 b(The)18 b(Classi\014cation)h(of)g(Orbits)-1287 204 y @F7 @sf(A)o
(thanassoula)14 b @F4 @sf(et)h(al.)22 b @F7 @sf(\(1983\))14 b(classify)f
(planar)g(orbits)h(in)e(a)i(bar)g(p)q(oten)o(tial)e(to)i(three)e(main)h
(groups:)-1949 96 y(the)k(B-family)l(,)e(the)i(R-family)g(and)g(the)g
(semi-ergo)q(dic)g(families.)23 b(B-family)15 b(orbits)j(ha)o(v)o(e)e(a)i
(prograde)-1926 141 y
11 @eop1

10 @bop0
@F2 @sf
[<
7FFFE07FFFE001F80001F80001F80001F80001F80001F80001F80001F80001F80001F80001
F80001F80001F80001F80001F80001F80001F80001F80001F80001F80001F80001F80001F8
0001F80001F800FFF800FFF80001F800007800001800>
	 19 32 -4 31 28] 49 @dc
[<
FFFF8000FFFF800007F0000007F0000007F0000007F0000007F0000007F0000007F0000007
F0000007F0000007F0000007F0000007F0000007F0000007FFFF0007FFFFE007F007F007F0
01FC07F000FC07F0007E07F0007E07F0007F07F0007F07F0007F07F0007F07F0007F07F000
7E07F0007E07F000FC07F001FC07F007F0FFFFFFE0FFFFFF00>
	 32 34 -2 33 38] 80 @dc
@F6 @sf
[<
0F000011800030C000706000703000603800601C00E01C00F00E00F00E00F00F00F00700F0
07807007807803C07803C07803C07FFFC03C01E03C01E03C01E01C01E01E00E01E00E00E00
E00F00E00700E00700E00380E00180E001C0E000E0C00060C0003180000F00>
	 19 35 -2 34 23] 18 @dc
[<
00400000004000000040000000200000002000000020000000200000001000000010000000
100000007F00000388E000060810000E0808001C0804001C0402001C0401001C0400801C04
00401C0200400E0200200E0200200E02002007010020870100308381003083810070438080
F0438081F0230081E01E0080C0000040000000400000004000000040000000200000002000
0000200000002000000010000000100000001000000010000000080000000800>
	 28 45 -2 34 32] 32 @dc
@F10 @sf
[<
FC003000300030001800180019E01A300C080C0C0C060C0606030603060346032303238222
461C3C>
	 16 20 0 13 18] 112 @dc
@F6 @sf
[<
3FFF00000001F000000000F000000000F000000000F000000000F000000000780000000078
00000000780000000078000000003C000000003C000000003C000000003C000000003E0000
00003D000000007C800000007CC00000007840000000F820000000F010000000F008000001
F004000001E002000001E003000003E001800003C000800007C000400007C0002000078000
10000F800008000F80000C000F80000F00FFF8003FE0>
	 35 34 -2 33 28] 89 @dc
@F6 @sf
[<
1803001C03801C03800E07000E070006070007070003070003060001060001820001820000
820000820080C3008041004041003041003FFFFF1FFFFF07FFFF>
	 24 21 -2 20 28] 25 @dc
[<
007F0000000380E0800006001180000C0009C000180007C000380003C000700003C0007000
01E000F00001E000F00001E000E00001E000E00000F000E00000F000E00000F000F0003FFF
00F000000000F000000000F000000000F0000000007800000000780000000078000000003C
000004003C000002001E000002000E000002000F000002000700000700038000070001C000
070000E00007000070000F80001C001380000E0021800003C0C08000007F0040>
	 34 36 -3 34 38] 71 @dc
@F10 @sf
[<
3F030060C480C02C00C01C00800C00800E00C00E00C00D00C00C806018802018801010400C
304007C000>
	 18 14 -2 13 23] 11 @dc
@F6 @sf
[<
FFFC00000007C000000003C000000003C000000003C000000003C000000001E000000001E0
00000001E000000001E000000000F000000000F000000000F000000000F000000000780000
000078000000007FFF0000007801E000003C007000003C003C00003C001E00003C000F0000
1E000F00001E000780001E000780001E000780000F000380000F000380000F000380000F00
038000078007000007800F000007801C00007FFFF000>
	 33 34 -2 33 31] 80 @dc
@F6 @sf
[<
07E0001C3800381C00700600700700700380F003C0F001C0F001E0F001E07800F07800F078
00F03C00F03C00F01C00F00E00E00700E00380C000C180003E00>
	 20 21 -2 20 23] 111 @dc
@F6 @sf
[<
3000C003C07001C006203800E006103800E00E083800E00E083800E007081C007007041C00
7007001C007003801C007003800E003801C00E003801C00E003801C00E003801C087001C00
E087801C00E087801E00E087401D00E047201880C046183061803C07E01F00>
	 38 21 -2 20 43] 109 @dc

10 @eop0

0 0 10 @bop1 910 45 a @F7 @sf({)17 b(10)g({)-942 147 y(Stellar)i
(kinematic)o(s)g(and)i(metallic)o(it)n(y)e(observ)m(ations)i(can)f(do)g
(far)g(more)g(than)h(con\014rm)e(the)-1861 96 y(accum)o(ulating)d
(evidence)g(for)h(bar)h(triaxialit)o(y)o -4 x(.)23 b(By)16 b(com)o(bining)h
(observ)m(ations)h(of)f(stellar)g(kinematics)-1929 96 y(and)k(metalli)o
(cit)n(y)d(in)i(m)o(ultiple)o 19 x(bulge)g(windo)o(ws,)g(it)g(is)g(p)q
(ossible)g(to)g(signi\014can)o(tly)f(constrain)h(the)-1861 96 y
(distribution)c(function)g(of)h(di\013eren)o(t)e(stellar)h(p)q
(opulations.)22 b(These)16 b(distribution)g(functions)h(con)o(tain)f(a)-1949
97 y(tremendous)g(amoun)o(t)h(of)f(information)g(ab)q(out)i(the)e
(formation)h(history)f(of)g(the)g(bulge.)22 b(Ken)o(t's)15 b(\(1992\))-1948
96 y(oblate)h(rotator)h(mo)q(del)f(has)h(b)q(een)f(a)g(v)o(ery)f(useful)h
(to)q(ol)h(for)f(analyzing)g(and)h(syn)o(thesizing)e(kinematical)o -1951
96 a(data.)27 b(In)17 b(the)h(rest)f(of)i(this)e(section,)h(w)o(e)f(tak)o
(e)g(the)h(\014rst)g(steps)g(to)o(w)o(ards)g(a)h(dynamical)d(mo)q(del)i
(for)g(a)-1912 97 y(triaxial)f(bulge)g(b)o(y)h(in)o(tegrating)f(orbits)h
(in)f(the)h(p)q(oten)o(tial)f(of)h(a)h(triaxial)d(bulge)i(and)g(examining)f
(the)-1914 96 y(kinematical)f(and)i(photometric)f(prop)q(erties)h(of)g
(di\013eren)o(t)f(orbit)h(familie)o(s.)25 b(These)18 b(sim)o(ulations)f
(will)-1916 96 y(also)h(enable)e(us)i(to)f(determine)f(whic)o(h)g(\014elds)h
(app)q(ear)h(most)f(promising)g(for)h(constraining)f(the)g(bulge)-1930 97 y
(stellar)f(distribution)f(functions.)115 203 y @F2 @sf(3.1.)55 b(The)19 b
(P)n(oten)n(tial)-1117 194 y @F7 @sf(W)l(e)d(use)h(a)g(densit)o(y/p)q
(oten)o(tial)f(pair)h(similar)e(to)i(that)g(of)g(the)g(prolate)g(bar)g
(prop)q(osed)h(b)o(y)e(Binney)-1939 96 y @F4 @sf(et)i(al.)23 b @F7 @sf
(\(1991\).)340 96 y @F6 @sf(\032)p @F7 @sf(\()p @F6 @sf(r)o(;)8 b(\022)q
(;)g( )r @F7 @sf(\))13 b(=)h @F6 @sf(\032)p 7 w @F11 @sf(0)2 -7 y @F7 @sf
(\()14 -34 y @F6 @sf(r)-31 22 y 42 -2 z 46 w(r)p 7 w @F11 @sf(0)6 -41 y
@F7 @sf(\))p -21 w @F9 @sf(\000)p @F10 @sf(p)3 21 y @F7 @sf(\(1)d(+)g
@F6 @sf(Y)g @F7 @sf(\()p @F6 @sf(\022)q(;)d( )r @F7 @sf(\)\))557 b(\(4\))
-1330 124 y(\010\()p @F6 @sf(r)o(;)8 b(\022)q(;)g( )r @F7 @sf(\))k(=)i(4)p
@F6 @sf(\031)r(G\032)p 7 w @F11 @sf(0)2 -7 y @F6 @sf(r)1 -21 y @F11 @sf(2)
-19 34 y(0)3 -13 y @F7 @sf(\()15 -34 y @F6 @sf(r)-32 22 y 42 -2 z 46 w(r)p
7 w @F11 @sf(0)7 -41 y @F7 @sf(\))p -21 w @F10 @sf(\013)2 21 y @F6 @sf(P)
7 b @F7 @sf(\()p @F6 @sf(\022)q(;)h( )r @F7 @sf(\))557 b(\(5\))-1950 120 y
(where)17 b(\()p @F6 @sf(r)o(;)8 b(\022)q(;)g( )r @F7 @sf(\))17 b(are)h
(the)f(spherical)g(co)q(ordinates)i(\014xed)e(on)h(the)f(bar)i(so)f(that)g
(\()p @F6 @sf(\022)q(;)8 b( )r @F7 @sf(\))16 b(=)g(\(0)p -18 w @F10 @sf(o)
3 18 y @F6 @sf(;)8 b @F7 @sf(0)p -18 w @F10 @sf(o)2 18 y @F7 @sf(\))18 b
(and)-1915 96 y(\(90)p -18 w @F10 @sf(o)3 18 y @F6 @sf(;)8 b @F7 @sf(0)p
-18 w @F10 @sf(o)2 18 y @F7 @sf(\))16 b(p)q(oin)o(ts)h(to)g(the)f
(rotation)h(axis)f(and)h(the)f(long)g(axis)g(resp)q(ectiv)o(ely)-5 b(,)-831
137 y @F6 @sf(P)7 b @F7 @sf(\()p @F6 @sf(\022)q(;)h( )r @F7 @sf(\))14 b(=)
99 -34 y(1)-105 22 y 185 -2 z 46 w @F6 @sf(\013)p @F7 @sf(\(1)e(+)f @F6 @sf
(\013)p @F7 @sf(\))17 -34 y @F5 @sf(\000)91 -34 y @F6 @sf(Y)g @F7 @sf(\()p
@F6 @sf(\022)q(;)d( )r @F7 @sf(\))-233 22 y 308 -2 z 46 w(\(2)k @F5 @sf(\000)f
@F6 @sf(\013)p @F7 @sf(\)\(3)g(+)g @F6 @sf(\013)p @F7 @sf(\))-1367 103 y
(and)418 96 y @F6 @sf(Y)g @F7 @sf(\()p @F6 @sf(\022)q(;)d( )r @F7 @sf(\))
13 b(=)h @F5 @sf(\000)p @F6 @sf(b)p 7 w @F11 @sf(20)p -7 w @F6 @sf(P)p 7 w
@F11 @sf(20)2 -7 y @F7 @sf(\()p @F6 @sf(cos\022)q @F7 @sf(\))e(+)f @F6 @sf
(b)p 7 w @F11 @sf(22)1 -7 y @F6 @sf(P)p 7 w @F11 @sf(22)2 -7 y @F7 @sf(\()p
@F6 @sf(cos\022)q @F7 @sf(\))p @F6 @sf(cos)p @F7 @sf(2)p @F6 @sf( )-1452
120 y @F7 @sf(are)19 b(linear)g(com)o(binations)g(of)h(spherical)f
(harmonic)g(functions)h(of)f(the)h @F6 @sf(l)f @F7 @sf(=)h(2,)g @F6 @sf(m)f
@F7 @sf(=)g(0)p @F6 @sf(;)8 b @F7 @sf(2)20 b(mo)q(des.)-1873 96 y(W)l(e)f
(adopt)i(the)e(same)h(densit)o(y)e(radial)i(pro\014le)f(as)h(Binney)e
@F4 @sf(et)j(al.)32 b @F7 @sf(\(1991\),)22 b(with)d(the)g(p)q(o)o(w)o(er)h
(la)o(w)-1871 124 y
10 @eop1

9 @bop0
@F0 @sf
[<
FFFFFFFFFFFFC07FFFFFFFFFFFE03FFFFFFFFFFFE03FFFFFFFFFFFE01C00000001FFF00E00
0000000FF0070000000001F0038000000000F801C0000000003801E0000000001800E00000
00000C0070000000000400380000000004001C0000000002000E0000000000000E00000000
0000070000000000000380000000000001C0000000000000E0000000000000700000000000
0070000000000000380000000000001C0000000000000E0000000000000700000000000003
800000000000078000000000000FC000000000001FC000000000001FC000000000003F8000
000000007F0000000000007E000000000000FE000000000001FC000000000003F800000000
0003F8000000000007F000000000000FE000000000000FC000000000001FC000000000003F
8000000000007F0000000000007F000000000000FE000000000001FC000000000203F80000
00000403F8000000000407F0000000000C0FE000000000180FC000000000381FC000000000
F03F8000000003F07F000000001FE07F00000001FFE0FFFFFFFFFFFFE0FFFFFFFFFFFFC0>
	 55 58 -2 -1 60] 88 @dc
@F10 @sf
[<
40402020101070F0F060>
	 4 10 -3 3 10] 59 @dc
[<
7800C600E3006300018001800180018000C000C000C000C000600060006008600430043002
2001C0000000000000000000000000001000380018>
	 13 29 0 22 14] 106 @dc
@F11 @sf
[<
FFFFFCFFFFFC000000000000000000000000000000000000FFFFFCFFFFFC>
	 22 10 -2 12 27] 61 @dc
/@F8 @newfont
@F8 @sf
[<
FFFF00FFFF007FFF003FFF001003800C018006018003018001800000E00000700000780000
3C00001E00001F00001F80780F80FC0F80FC0F80FC1F80FC1F00787E003FFC000FF000>
	 17 24 -2 23 22] 50 @dc
@F11 @sf
[<
FF801C001C001C001C001C001C001C001C001C001C101E381D38FCF0>
	 13 14 0 13 14] 114 @dc
@F11 @sf
[<
FF80001C00001C00001C00001C00001C00001CF8001F0E001C07001C03801C01801C01C01C
01C01C01C01C01C01C01801C03801C03001F0E00FCF800>
	 18 20 0 13 20] 112 @dc
@F2 @sf
[<
FFE00000FFE000001F0000001F0000001F0000001F0000001F0000001F0000001F0000001F
0000001F0FC0001F3FF8001FE0FC001FC03E001F803F001F001F801F001F801F000FC01F00
0FC01F000FC01F000FC01F000FC01F000FC01F000FC01F000FC01F001F801F001F801F001F
001F803E001FF07C00FF3FF800FF0FE000>
	 26 32 -2 21 31] 112 @dc
@F4 @sf
[<
FFFF000003E0000001E0000001E0000001E0000001E0000000F0000000F0000000F0000000
F0000000780000007800000078000000780000003C0000003C0000003C0000003C0000001E
0000001E0000001E0000001E0000000F0000800F0010800F0010400F001040078010400780
1020078018200780183003C0181803C0181E03C0381FFFFFF8>
	 29 34 -9 33 35] 84 @dc
[<
6003C0E00620700610700E10700E087007083807043803803803803803801C01C01C01C01C
01C01C01C00E00E00E00E00F00E00F00E007C0C0072180071F000700000380000380000380
0003800001C00001C00001C00001C00000E00000E00000E0000FE00000F000>
	 22 35 -3 34 25] 104 @dc
[<
07C000187000301800700C00700E00700700F00780F00380F003C0F003C07801E07801E078
01E03801E03C01E01C01E00E01C00701C003818001C300007E00>
	 19 21 -5 20 25] 111 @dc
[<
1F0031C060E06070E038E038E03CE01EE01EE01E700F700F700F700F380F380738073C061E
0E1D0C1CF81C000E000E000E000E0007000700070007000380038003803F8003C0>
	 16 35 -5 34 22] 98 @dc
[<
1FC000203000400800E00400F00600F00600700700000700000F00003E0003FE0007FC000F
F0000F00000C00000C03000C038004018002008001830000FC00>
	 17 21 -3 20 20] 115 @dc
[<
3000007000003800003800003800003800001C00001C00001C00001C00000E00000E00000E
00000E00008700008701808703C08783C04741C02620801C1F00>
	 18 21 -5 20 21] 114 @dc
[<
03C0000C30001C08001C08001C04001C02001C02001C01001C01001C01000E00800E00800E
00800700808700C08700C08381C04383C04387C02307C01E0380>
	 18 21 -5 20 22] 118 @dc
[<
0F0780308C40305C40703C20701C20F01C20F00E10F00E00F00E00F00E0078070078070078
07003807003C03801C03800E03800E03800705C00185C000F9C00001C00000E00000E00000
E00000E00000700000700000700000700000380000380000380003F800003C>
	 22 35 -5 34 25] 100 @dc
[<
383C00446200E2C100F1C080F1C04061C04000E02000E00000E00000E00000700000700000
70000070002038002038602038F0103CF008347004621003C1E0>
	 20 21 -3 20 22] 120 @dc
[<
1C00320071007080708070803840380038001C001C001C000E000E00870087008700430043
0023001C000000000000000000000000000000000001C001C001E000C0>
	 11 33 -5 32 15] 105 @dc
[<
3003C0700620380610380E10380E083807081C07041C03801C03801C03800E01C00E01C00E
01C00E01C08700E08700E08780E08780E04740C02631C01C0F00>
	 22 21 -5 20 27] 110 @dc
[<
3E000043800080C000E06000F03000F03800601C00001C00000E00000E0003CE000C3E001C
0F001C07001C07001C07001C03801C03801C03801C03800E01C00E01C00E01C00701C08700
E08700E08380E04380E04380702300701E0030>
	 20 31 -5 20 24] 121 @dc
[<
3003001E00700700310038038030803803807080380380704038038038401C01C038201C01
C01C001C01C01C001C01C01C000E00E00E000E00E00E000E00E00E000E00E00E0087007007
0087007007008780780700878078070047606606002610C10C001C0F80F800>
	 35 21 -5 20 40] 109 @dc
[<
0FFE0000E00000E00000E0000070000070000070000070000038000038000F380030B80030
5C00703C00701C00F01C00F00E00F00E00F00E00F00E007807007807007807003807003C03
801C03800E03800E03800705C00184C000F840>
	 18 31 -5 20 22] 113 @dc
[<
07C3C00C26201C1E201C0E10180E101C0E101C07081C07001C07001C07000E03800E03800E
03800703808701C08701C08381C04381C04380E02300E01E0060>
	 21 21 -5 20 26] 117 @dc
[<
FFC0001C00001C00001C00000E00000E00000E00000E0000070000070000071E0007238003
C1C003C0E003807003807001C07801C03C01C03C01C03C00E01E00E01E00E01E00E01E0870
1E08700E08700E08780C04741C02621801C1F0>
	 23 31 -1 20 25] 112 @dc
[<
FFE0FFF0FFF0>
	 12 3 -4 11 17] 45 @dc
[<
0FC000183000300C00700200700100F00100F00000F00000F00000F0000078000078000078
00003800003C00001C07800E078007038003018001C100007E00>
	 17 21 -5 20 22] 99 @dc

9 @eop0

0 0 9 @bop1 922 45 a @F7 @sf({)17 b(9)f({)-929 147 y(W)l(e)i(w)o(ould)h
(lik)o(e)f(to)h(estimate)f(whether)h(the)g(metal)f(ric)o(h)g(and)i(metal)f
(p)q(o)q(or)h(p)q(opulations)g(are)-1881 96 y(statistically)d(di\013eren)o
(t.)26 b(T)l(o)19 b(test)f(this,)g(w)o(e)g(de\014ne)f(a)i(statistic)f
(that)g(quan)o(ti\014es)g(the)g(di\013erences)f(in)-1904 96 y(their)f(v)o
(elo)q(cit)n(y)f(momen)o(ts)g(in)h(the)g @F6 @sf(V)p 7 w @F10 @sf(l)14 -7 y
@F5 @sf(\000)10 b @F6 @sf(V)p 7 w @F10 @sf(r)20 -7 y @F7 @sf(plane:)-491
127 y @F6 @sf(T)20 b @F7 @sf(=)56 -34 y(1)-61 22 y 98 -2 z 46 w @F6 @sf(\033)
2 -17 y @F11 @sf(2)-20 29 y @F10 @sf(l)10 -12 y @F6 @sf(\033)2 -14 y
@F11 @sf(2)-20 27 y @F10 @sf(r)15 -81 y @F7 @sf(\()p @F6 @sf(N)p 7 w
@F10 @sf(p)13 -7 y @F5 @sf(\000)11 b @F7 @sf(1\)\()p @F6 @sf(N)p 7 w
@F10 @sf(r)14 -7 y @F5 @sf(\000)g @F7 @sf(1\))-363 22 y 363 -2 z 19 46 a
(6\()p @F6 @sf(N)p 7 w @F10 @sf(r)15 -7 y @F7 @sf(+)g @F6 @sf(N)p 7 w
@F10 @sf(p)13 -7 y @F5 @sf(\000)g @F7 @sf(2\))58 -76 y @F0 @sf(X)-85 91 y
@F10 @sf(i;j)r @F11 @sf(=1)p @F10 @sf(;)p @F11 @sf(2)-1 -49 y @F7 @sf(\()p
@F6 @sf(\033)2 -21 y @F8 @sf(2)2 28 y @F11 @sf(r)p @F10 @sf(;ij)15 -7 y
@F5 @sf(\000)g @F6 @sf(\033)2 -21 y @F8 @sf(2)1 28 y @F11 @sf(p)p @F10 @sf
(;ij)4 -7 y @F7 @sf(\))p -21 w @F11 @sf(2)2 21 y @F6 @sf(;)385 b @F7 @sf
(\(3\))-1950 137 y(where)17 b(the)f(subscripts)h(1)h(and)f(2)g(stand)h
(for)f @F6 @sf(r)i @F7 @sf(and)e @F6 @sf(l)h @F7 @sf(directions)e(resp)q
(ectiv)o(el)o(y)l(,)o 16 x @F6 @sf(\033)2 -18 y @F8 @sf(2)2 25 y @F11 @sf
(r)18 -7 y @F7 @sf(and)i @F6 @sf(\033)2 -18 y @F8 @sf(2)2 25 y @F11 @sf(p)
18 -7 y @F7 @sf(are)f(the)-1934 96 y(v)o(elo)q(cit)o(y)o 15 x(disp)q
(ersion)e(tensors)i(of)f(the)f(metal)g(ric)o(h)f(and)j(metal)d(p)q(o)q(or)k
(p)q(opulations)e(and)h @F6 @sf(N)p 7 w @F10 @sf(r)18 -7 y @F7 @sf(and)f
@F6 @sf(N)p 7 w @F10 @sf(p)18 -7 y @F7 @sf(are)-1950 96 y(the)h(n)o(um)o
(b)q(er)f(of)h(stars)h(in)f(the)g(metal)f(ric)o(h)g(and)i(the)e(metal)h(p)q
(o)q(or)h(sample.)23 b(The)18 b(normalization)e(of)i @F6 @sf(T)-1925 97 y
@F7 @sf(w)o(as)f(c)o(hosen)g(so)g(that)g(if)f(the)h(t)o(w)o(o)f(samples)h
(had)g(b)q(een)g(dra)o(wn)g(from)f(the)h(sample)f(paren)o(t)g(p)q
(opulation,)-1939 96 y(then)h(the)h(exp)q(ectation)f(v)m(alue)g(of)h @F6 @sf
(T)23 b @F7 @sf(=)16 b(1)p @F6 @sf(:)i @F7 @sf(F)l(or)f(our)h(Baade's)g
(Windo)o(w)g(sample,)f(w)o(e)g(\014nd)h @F6 @sf(T)k @F7 @sf(=)17 b(3)p
@F6 @sf(:)p @F7 @sf(12.)-1917 96 y(Figure)f(2)h(sho)o(ws)h(the)e(cum)o
(ulativ)o(e)o 15 x(distribution)h(of)g(the)f @F6 @sf(T)23 b @F7 @sf
(statistic,)16 b @F6 @sf(P)7 b @F7 @sf(\()p @F6 @sf(T)g @F7 @sf(\),)16 b
(computed)g(with)h(10,000)-1939 96 y(Mon)o(te)h(Carlo)h(realizations)f(in)g
(whic)o(h)g @F6 @sf(N)p 7 w @F10 @sf(r)21 -7 y @F7 @sf(=)f(39)j(and)f
@F6 @sf(N)p 7 w @F10 @sf(p)19 -7 y @F7 @sf(=)f(15)h(stars)h(are)e(dra)o
(wn)h(from)g(a)g(paren)o(t)-1896 97 y(p)q(opulation)g(whose)f(v)o(elo)q
(cit)o(y)o 16 x(momen)o(ts)f(are)h(the)f(mean)h(of)g(our)g(full)e(62)j
(star)f(sample.)25 b(Th)o(us,)18 b(if)f(the)-1918 96 y(t)o(w)o(o)e
(samples)f(w)o(ere)g(dra)o(wn)h(from)g(the)f(same)h(paren)o(t)g(p)q
(opulation,)g(then)g(the)f(probabilit)o(y)g(of)h(randomly)-1950 96 y
(measuring)j(a)g(T)g(v)m(alue)f(as)i(large)e(as)i(the)e(observ)o(ed)h(v)m
(alue)f(is)h(less)f(than)h(0.03.)27 b(In)17 b(other)h(w)o(ords,)g(the)-1914
97 y(data)f(argues)h(for)f(a)g(gen)o(uine)f(kinematic)o(s)g(di\013erence)f
(b)q(et)o(w)o(een)h(the)g(metal)g(ric)o(h)g(stars)h(and)g(the)g(metal)-1942
96 y(p)q(o)q(or)h(stars.)538 203 y @F2 @sf(3.)74 b(Implications)-1088 182 y
@F7 @sf(Baade's)17 b(Windo)o(w)g(\()p @F6 @sf(l)q(;)8 b(b)p @F7 @sf(\))14 b
(=)i(\(1)p @F6 @sf(;)8 b @F5 @sf(\000)p @F7 @sf(4\))p -18 w @F10 @sf(o)19
18 y @F7 @sf(is)17 b(a)h(line)e(of)i(sigh)o(t)f(nearly)f(through)i(the)f
(rotation)h(axis)f(of)-1927 96 y(the)d(Bulge,)g(so)h(if)e(the)i(Bulge)e
(is)h(axisymmetric,)o 13 x(the)h(v)o(elo)q(cit)n(y)e(ellipsoid)g(should)i
(align)f(with)g(the)g(line)g(of)-1950 97 y(sigh)o(t,)i(in)h(other)f(w)o
(ords,)h(the)g(v)o(ertex)o 16 x(angle)f(should)i(b)q(e)e(either)g @F6 @sf
(l)p 7 w @F10 @sf(v)17 -7 y @F7 @sf(=)f(0)p -18 w @F10 @sf(o)19 18 y @F7 @sf
(or)i(90)p -18 w @F10 @sf(o)3 18 y @F7 @sf(.)22 b @F4 @sf(The)c(observe)n
(d)h(vertex)-1939 96 y(deviations)g(violate)h(the)f(axisymmetry)e(and)i(r)n
(e)n(quir)n(e)e(the)i(p)n(otential)g(to)g(b)n(e)f(non-axisymmetric.)26 b
@F7 @sf(Also)-1927 96 y(the)17 b(di\013eren)o(t)f(v)o(elo)q(cit)n(y)f
(ellipsoids)h(of)h(the)g(metal)f(ric)o(h)g(sample)g(and)i(the)e(metal)h(p)q
(o)q(or)h(sample)e(argues)-1934 97 y(for)h(stars)g(of)g(di\013eren)o(t)e
(metallici)o(t)o(y)f(o)q(ccup)o(ying)j(di\013eren)o(t)e(regions)i(of)g
(phase)f(space)h(or)g(di\013eren)o(t)e(orbit)-1946 96 y(families)g(of)h
(the)g(triaxial)g(p)q(oten)o(tial.)-691 124 y
9 @eop1

8 @bop0
@F6 @sf
[<
07C0001C3000380800300C00700400700600F00200F00300F00100F0010078018078018078
01803C01803C01801C01800E038007038003870000EF00003E00003C0000F80001F00001E0
0001C00001800001800001800001800000C000004010006020001860000F80>
	 20 35 -2 34 22] 14 @dc

8 @eop0

0 0 8 @bop1 922 45 a @F7 @sf({)17 b(8)f({)-929 147 y(Although)g(the)g
(sample)h(is)f(small,)f(the)i(v)o(ertex)e(deviations)h(are)g(already)h
(signi\014can)o(t)f(at)h(the)f(2|3)-1943 96 y @F6 @sf(\033)i @F7 @sf(lev)o
(el.)o 22 x(W)l(e)e(estimate)g(that)h(the)g(con\014dence)f(lev)o(el)o 15 x
(w)o(ould)h(go)h(up)f(to)g(5)p @F6 @sf(\033)h @F7 @sf(with)f(a)g(sample)f
(four)h(times)-1939 96 y(as)g(large,)g(the)g(1)g @F6 @sf(\033)h @F7 @sf
(error)f(in)g(the)f(v)o(ertex)g(angle)g(b)q(eing)i(60)p @F6 @sf(N)5 -18 y
@F9 @sf(\000)p @F11 @sf(0)p @F10 @sf(:)p @F11 @sf(5)19 18 y @F7 @sf
(degrees.)23 b(There)16 b(is)h(also)g(a)h(hin)o(t)e(that)-1935 96 y(the)f
(metal)f(ric)o(h)g(stars)h(stream)g(ahead)h(of)f(the)f(metal)h(p)q(o)q(or)h
(ones)f(with)g(a)g(small)g(systematic)e(v)o(elo)q(cit)o(y)g(in)-1950 97 y
(the)i(p)q(ositiv)o(e)f @F6 @sf(l)h @F7 @sf(direction.)20 b(The)15 b
(relativ)o(e)o 13 x(streaming)g(v)o(elo)q(cit)o(y)e @F6 @sf(l)q @F7 @sf
(-comp)q(onen)o(t)i @F6 @sf(\016)r @F5 @sf(h)p @F6 @sf(V)p 7 w @F10 @sf(l)
2 -7 y @F5 @sf(i)f @F7 @sf(=)g(\(45)8 b @F5 @sf(\006)g @F7 @sf(30\)km/s.)
-1949 96 y(The)19 b(small)f(size)f(of)i(our)g(sample)g(brings)g(the)f(w)o
(orry)h(ab)q(out)g(the)g(selection)e(e\013ect.)28 b(The)19 b(existence)o
-1894 96 a(of)f(the)f(v)o(ertex)f(deviation)g(do)q(es)j(dep)q(end)e(up)q
(on)h(the)g(distinction)e(of)i(the)f(w)o(eak-line)f(sample)h(and)h(the)-1923
97 y(strong-line)d(sample.)21 b(The)15 b(total)g(65)h(star)f(sample,)g
(e.g.,)f(sho)o(ws)i(no)f(v)o(ertex)f(deviation.)20 b(Generally)14 b(one)
-1950 96 y(exp)q(ects)i(that)i(random)g(noise)f(and/or)h(the)f(canceling)f
(v)o(ertex)f(deviations)i(of)h(v)o(elo)q(cit)n(y)e(elli)o(psoids)g(for)-1930
96 y(the)g(metal)f(ric)o(h)h(sample)f(and)i(the)f(metal)f(p)q(o)q(o)q(r)i
(sample)f(tend)g(to)h(symmetriz)o(e)e(the)h(v)o(elo)q(cit)n(y)f
(ellipsoid.)o -1951 97 a(Ho)o(w)o(ev)o(er,)o 15 x(the)h(v)o(ertex)f
(deviation)h(for)h(the)f(metal)g(ric)o(h)f(and)i(metal)f(p)q(o)q(or)i
(sample)e(is)g(stable)h(when)f(one)-1945 96 y(mo)o(v)o(es)f(the)h
(metallici)o(t)o(y)e(cuto\013)j(b)o(y)f(0)p @F6 @sf(:)p @F7 @sf(2)h(dex.)j
(It)c(app)q(ears)i(ev)o(en)d(less)h(sensitiv)o(e)e(to)j(magnitude)f
(cuto\013.)-1830 128 y(The)h(observ)o(ed)f(v)o(elo)q(cit)o(y)o 16 x
(distribution)g(also)i(app)q(ears)g(to)f(ha)o(v)o(e)g(wider)f(wings)h
(than)h(a)f(Gaussian.)-1932 97 y(Since)g(the)h(v)o(elo)q(cit)o(y)f(disp)q
(ersion)h(tensor)h(is)f(only)g(su\016cien)o(t)f(to)i(describ)q(e)e(a)i
(Gaussian)g(distribution)-1901 96 y(function,)c(one)g(migh)o(t)f(w)o
(onder)h(whether)g(the)g(v)o(ertex)e(deviation)i(biases)g(to)g(a)h(few)e
(high)i(v)o(elo)q(cit)n(y)e(stars.)-1950 96 y(W)l(e)h(test)g(this)g(b)o(y)g
(w)o(eigh)o(ting)f(eac)o(h)h(data)h(p)q(oin)o(t)f(in)g(Figure)g(1)g(with)g
(the)g(in)o(v)o(erse)f(of)h(the)g(sp)q(eed,)g(namely)l(,)-1951 97 y(the)j
(length)h(of)f(the)h(v)o(elo)q(cit)n(y)e(v)o(ector)h(\(Lynden-Bell)f
(1994\).)30 b(W)l(e)18 b(\014nd)h(that)g(the)f(v)o(ertex)f(deviation)-1895
96 y(remains)f(after)g(w)o(e)g(apply)g(less)g(w)o(eigh)o(t)f(on)i(the)f
(high)g(v)o(elo)q(cit)o(y)f(stars.)-1204 128 y(The)i(goal)h(of)g(this)g(w)o
(ork)f(is)h(to)g(gain)g(insigh)o(t)f(in)o(to)g(the)h(kinematic)o(s)f(of)h
(the)f(bulge.)25 b(Tw)o(o)18 b(biases)-1917 97 y(p)q(oten)o(tially)d
(a\013ect)i(the)f(data.)23 b(The)16 b(prop)q(er)h(motion)g(sample)f
(excludes)f(stars)i(to)q(o)h(blue)e(to)g(b)q(e)h(gian)o(ts;)-1946 96 y
(some)h(metal)g(p)q(o)q(or)i(stars)f(ma)o(y)f(ha)o(v)o(e)g(b)q(een)g(lost)h
(from)f(the)h(prop)q(er)g(motion)f(sample.)28 b(Those)19 b(stars)-1896 96 y
(with)f(radial)g(v)o(elo)q(citi)o(es)e(are)i(originally)f(from)h(the)g
(Ric)o(h)f(\(1988\))i(sample;)f(these)f(w)o(ere)g(c)o(hosen)h(from)-1911
97 y(a)h(w)o(ell)f(de\014ned)g(gian)o(t)h(branc)o(h/clump)f(in)h(the)f
(R,I)g(color-magnitude)h(diagram.)29 b(Galaxy)19 b(mo)q(dels)-1889 96 y
(sho)o(w)d(that)g(at)g(the)f(p)q(osition)h(of)g(Baade's)f(Windo)o(w,)h
(stars)g(c)o(hosen)f(from)h(the)f(principal)f(sequences)h(are)-1950 96 y
(o)o(v)o(erwhelmi)o(ngly)g(bulge)h(mem)o(b)q(ers.)-682 227 y
8 @eop1

7 @bop0
@F8 @sf
[<
FF80FF801E001E001E001E001E001E001E001E001E001E001E001E001E007E007E00000000
00000000003C007E007E007E007E003C00>
	 9 27 -1 26 12] 105 @dc
[<
3F8071E0F8F0F878F878707800780078007800780078007800780078007800780078007800
7800780078007803F803F8000000000000000000F001F801F801F801F800F0>
	 13 34 3 26 14] 106 @dc
@F6 @sf
[<
3E0000630000F18000F0C00070E00000700000700000700000380000380000380000380000
1C00001C00001C00001C00000E00000E00000E00000E000007000007000007000007000103
80010380010380008380004380002300001E00000000000000000000000000000000000000
0000000000000000000000C00001E00001E00000E0>
	 19 44 1 33 20] 106 @dc

7 @eop0

0 0 7 @bop1 922 45 a @F7 @sf({)17 b(7)f({)-1027 147 y(v)o(elo)q(cit)o(y)o
16 x(disp)q(ersion)i(tensor)g @F6 @sf(\033)2 -18 y @F8 @sf(2)1 18 y @F7 @sf
(,)f(whose)h(6)g(indep)q(enden)o(t)e(comp)q(onen)o(ts)i @F6 @sf(\033)2 -18 y
@F8 @sf(2)2 27 y(ij)19 -9 y @F7 @sf(are)f(de\014ned)g(to)h(b)q(e)g(the)-1925
96 y(follo)o(wing)e(a)o(v)o(erages)g(of)g(v)o(elo)q(cit)o(y)f(o)o(v)o(er)g
(the)h(sample,)-351 147 y @F6 @sf(\033)2 -21 y @F8 @sf(2)1 28 y @F10 @sf
(ij)19 -7 y @F7 @sf(=)61 -34 y @F6 @sf(N)-82 22 y 130 -2 z 46 w(N)h @F5 @sf
(\000)10 b @F7 @sf(1)6 -34 y(\()p @F5 @sf(h)p @F6 @sf(V)p 7 w @F10 @sf(i)
3 -7 y @F6 @sf(V)p 7 w @F10 @sf(j)4 -7 y @F5 @sf(i)i(\000)f(h)p @F6 @sf(V)p
7 w @F10 @sf(i)2 -7 y @F5 @sf(ih)p @F6 @sf(V)p 7 w @F10 @sf(j)6 -7 y @F5 @sf
(i)p @F7 @sf(\))572 b(\(1\))-1950 146 y(where)20 b(the)g(subscripts)g
@F6 @sf(i)g @F7 @sf(and)h @F6 @sf(j)i @F7 @sf(stand)e(for)g(an)o(y)f(of)h
(the)f(directions)f @F6 @sf(l)q @F7 @sf(,)h @F6 @sf(r)i @F7 @sf(or)f @F6 @sf
(b)f @F7 @sf(in)g(the)g(Galactic)-1855 96 y(co)q(ordinate,)d(the)f(brac)o
(k)o(eted)f(quan)o(tities)g(are)i(the)f(v)o(elo)q(cit)n(y)f(momen)o(ts)h
(for)g(the)h(sample,)e(and)i @F6 @sf(N)22 b @F7 @sf(is)16 b(the)-1946 96 y
(sample)h(size.)24 b(The)17 b(cross)h(terms)f @F6 @sf(\033)2 -18 y @F11 @sf
(2)-20 31 y @F10 @sf(l)p(r)3 -13 y @F6 @sf(;)8 b(\033)2 -18 y @F11 @sf(2)
-20 31 y @F10 @sf(br)2 -13 y @F6 @sf(;)g(\033)2 -18 y @F11 @sf(2)-20 31 y
@F10 @sf(l)p(b)19 -13 y @F7 @sf(determine)16 b(the)h(directions)g(of)h
(the)f(v)o(elo)q(cit)n(y)f(ellipsoid.)-1924 97 y(Diagonalizing)j(the)f(3)h
@F5 @sf(\002)f @F7 @sf(3)h(tensor,)g(w)o(e)g(obtain)g(the)f(three)g(eigen)o
(v)m(alues)g(and)h(eigen)o(v)o(e)o(ctors)f(of)g(the)-1894 96 y(disp)q
(ersion)g(tensor,)f(whic)o(h)g(giv)o(es)g(the)g(axis)g(ratio)h(and)g
(orien)o(tation)f(of)h(the)f(v)o(elo)q(cit)o(y)o 16 x(ellipsoid.)24 b(The)
-1922 96 y(cen)o(troid)15 b(of)i(the)f(ellipsoid)f(is)h(giv)o(en)f(b)o(y)h
(the)g(v)o(elo)q(cit)o(y)o 15 x(\014rst)h(momen)o(ts)e(of)i(the)f(sample.)k
(The)d(results)f(of)-1950 97 y(the)h(diagonalization)g(are)g(summarized)f
(in)h(T)l(able)f(1,)i(and)f(are)g(illustrated)f(more)h(directly)o 16 x(in)g
(Figure)-1934 96 y(1.)60 128 y(Figure)h(1)i(sho)o(ws)g(the)f
(distributions)g(of)g(stars)h(pro)s(jected)e(in)h(three)g(v)o(elo)q(cit)n
(y)f(planes)h(for)g(the)-1880 97 y(samples)e(with)f(di\013eren)o(t)g
(metallici)o(t)o(y)f(ranges.)24 b(Eac)o(h)17 b(distribution)f(is)h(\014tted)g
(b)o(y)f(a)h(v)o(elo)q(cit)o(y)e(ellipsoid)-1936 96 y(with)k(a)g(Gaussian)h
(pro\014le,)e(whose)i(1.5)f @F6 @sf(\033)h @F7 @sf(and)g(2.5)f @F6 @sf(\033)h
@F7 @sf(lev)o(el)o(s)e(are)h(indicated)e(b)o(y)i(dotted)g(and)g(solid)-1890
96 y(ellipses,)13 b(whic)o(h)g(enclose)g(67
(stars.)22 b(In)13 b(the)h @F6 @sf(V)p 7 w @F10 @sf(l)9 -7 y @F5 @sf(\000)
6 b @F6 @sf(V)p 7 w @F10 @sf(r)17 -7 y @F7 @sf(plane,)14 b(w)o(e)g(note)g
(that)g(although)-1949 96 y(the)i(v)o(elo)q(cit)o(y)f(distribution)h(of)g
(the)h(whole)f(sample)g(is)g(nearly)g(symmetric)f(with)h(resp)q(ect)g(to)h
@F6 @sf(V)p 7 w @F10 @sf(r)20 -7 y @F7 @sf(and)g @F6 @sf(V)p 7 w @F10 @sf
(l)-1943 90 y @F7 @sf(axes,)g(the)f(long)i(axes)f(of)g(the)g(v)o(elo)q
(cit)n(y)e(distributions)i(of)g(the)g(metal)f(ric)o(h)g(and)i(the)e(metal)g
(p)q(o)q(o)q(r)i(stars)-1933 96 y(p)q(oin)o(t)g(to)o(w)o(ards)h(opp)q
(osite)f(sides)g(of)g(the)g(line)f(of)h(sigh)o(t,)g(and)g(are)g(curiously)g
(p)q(erp)q(endicular)f(to)i(eac)o(h)-1910 96 y(other.)27 b(W)l(e)18 b
(quan)o(tify)g(this)g(b)o(y)g(computing)g(the)g(v)o(ertex)f(angle)h @F6 @sf
(l)p 7 w @F10 @sf(v)3 -7 y @F7 @sf(,)h(the)f(angle)g(b)q(et)o(w)o(een)f
(the)h(line)g(of)-1901 97 y(sigh)o(t)e(and)h(the)f(long)h(axis)f(of)g(the)g
(v)o(elo)q(cit)o(y)f(elli)o(psoid:)-236 146 y(tan)p(\(2)p @F6 @sf(l)p 7 w
@F10 @sf(v)3 -7 y @F7 @sf(\))f(=)58 -34 y(2)p @F6 @sf(\033)2 -18 y @F11 @sf
(2)-20 30 y @F10 @sf(l)p(r)-118 10 y 159 -2 z 46 w @F6 @sf(\033)2 -14 y
@F11 @sf(2)-20 27 y @F10 @sf(r)16 -13 y @F5 @sf(\000)d @F6 @sf(\033)2 -17 y
@F11 @sf(2)-20 30 y @F10 @sf(l)727 -47 y @F7 @sf(\(2\))-1950 146 y(F)l(or)
16 b(the)f(metal)g(ric)o(h)g(and)h(metal)f(p)q(o)q(or)i(samples,)f @F6 @sf
(l)p 7 w @F10 @sf(v)18 -7 y @F7 @sf(is)g(\()p @F5 @sf(\000)p @F7 @sf(65)
10 b @F5 @sf(\006)g @F7 @sf(9\))p -18 w @F10 @sf(o)18 18 y @F7 @sf(and)16 b
(\(25)11 b @F5 @sf(\006)f @F7 @sf(14\))p -18 w @F10 @sf(o)18 18 y @F7 @sf
(resp)q(ectiv)o(ely)o -4 x(.)20 b(On)-1950 96 y(the)15 b(other)g(hand,)h
(plotting)f(the)h(data)g(on)f(the)g @F6 @sf(V)p 7 w @F10 @sf(b)12 -7 y
@F5 @sf(\000)9 b @F6 @sf(V)p 7 w @F10 @sf(r)19 -7 y @F7 @sf(and)16 b @F6 @sf
(V)p 7 w @F10 @sf(l)11 -7 y @F5 @sf(\000)9 b @F6 @sf(V)p 7 w @F10 @sf(b)
18 -7 y @F7 @sf(diagrams)16 b(in)f(the)g(lo)o(w)o(er)f(t)o(w)o(o)h(ro)o
(ws)-1949 97 y(of)i(Figure)e(1,)i(w)o(e)e(do)i(not)g(see)f(signi\014can)o
(t)g(v)o(ertex)o 15 x(deviations.)-1178 155 y
7 @eop1

6 @bop0
@F2 @sf
[<
FFFFFF8000FFFFFFF00007F003FC0007F0007E0007F0003F0007F0001F8007F0000FC007F0
0007E007F00007E007F00003F007F00003F007F00003F007F00003F807F00003F807F00003
F807F00003F807F00003F807F00003F807F00003F807F00003F807F00003F807F00003F007
F00003F007F00003F007F00007F007F00007E007F00007E007F0000FC007F0001F8007F000
3F0007F0007E0007F003FC00FFFFFFF000FFFFFF8000>
	 37 34 -2 33 43] 68 @dc
[<
FFE00FFFF8FFE00FFFF80600007F000600007F00030000FE00030000FE00038001FE000180
01FC00018001FC0000FFFFF80000FFFFF80000E007F800006007F00000600FF00000300FE0
0000300FE00000381FE00000181FC00000183FC000000C3F8000000C3F8000000E7F800000
067F00000006FF00000003FE00000003FE00000003FE00000001FC00000001FC00000000F8
00000000F800000000F8000000007000000000700000>
	 37 34 -2 33 42] 65 @dc

6 @eop0

0 0 6 @bop1 922 45 a @F7 @sf({)17 b(6)f({)-1027 147 y(gian)o(ts,)h(has)h
(b)q(een)g(studied)f(in)f(detail)h(for)h(the)f(metalli)o(cit)n(y-disp)q
(ersion)f(correlations)h(b)o(y)g(Zhao)h @F4 @sf(et)h(al.)-1925 96 y @F7 @sf
(\(1994\).)26 b(They)17 b(sho)o(w)i(that)e(the)h(Bulge)e(is)h(\015attened)h
(b)o(y)f(rotation)h(with)f(the)g(more)h(metal)e(ric)o(h)h(stars)-1921 96 y
(rotating)h(faster)f(and)h(ha)o(ving)f(lo)o(w)o(er)f(disp)q(ersion.)25 b
(They)17 b(further)f(explain)h(the)g(observ)o(ed)f(anisotrop)o(y)-1927 96 y
(\()p @F6 @sf(\033)p 7 w @F10 @sf(l)16 -7 y @F6 @sf(>)d(\033)p 7 w @F10 @sf
(b)2 -7 y @F7 @sf(\))i(is)f(due)h(to)g(rotation)g(broadening)g(and)h(\()p
@F6 @sf(\033)p 7 w @F10 @sf(r)16 -7 y @F6 @sf(>)e(\033)p 7 w @F10 @sf(b)
2 -7 y @F7 @sf(\))g(is)h(due)f(to)h(in)o(trinsic)e(anisotrop)o(y)i(at)g(a)g
(lev)o(el)o -1951 97 a(consisten)o(t)j(with)g(extragalactic)f(bulges)h
(\(Kormendy)g(1982\))h(and)g(n)o(umerical)d(sim)o(ulations)i(of)g(bars)-1905
96 y(\(Pfenniger)e(and)h(F)l(riedli)e(1991\).)23 b(Ho)o(w)o(ev)o(er,)o 15 x
(the)17 b(Zhao)g @F4 @sf(et)h(al.)23 b @F7 @sf(\(1994\))18 b(analysis)f
(did)f(not)h(mak)o(e)f(use)g(of)-1944 96 y(the)h(cross)h(terms)f(of)g(the)g
(v)o(elo)q(cit)o(y)f(disp)q(ersion)h(tensor.)25 b(W)l(e)17 b(study)g(the)g
(same)h(sample)f(of)g(stars,)h(and)-1924 97 y(dev)o(elop)d(a)i(new)f(tec)o
(hnique)f(to)h(infer)g(the)g(triaxialit)n(y)f(of)h(the)g(bulge)g(from)h
(the)f(v)o(elo)q(cit)n(y)f(data)i(alone.)-1806 128 y(In)f @F5 @sf(x)p
@F7 @sf(2,)g(w)o(e)g(rep)q(ort)h(v)o(ertex)e(deviations)h(for)h(b)q(oth)g
(the)g(metal)e(p)q(o)q(or)j(and)f(metal)f(ric)o(h)g(stars.)22 b(In)16 b
@F5 @sf(x)q @F7 @sf(3,)-1945 96 y(w)o(e)i(presen)o(t)g(a)h(realistic)e(p)q
(oten)o(tial)h(for)g(the)g(bulge.)28 b(W)l(e)18 b(sim)o(ulate)f(the)h
(bulge)h(stellar)e(distribution)-1898 97 y(b)o(y)f(running)i(orbits)f(in)f
(the)h(p)q(oten)o(tial)f(and)i(use)f(these)f(sim)o(ulations)g(to)i
(compute)e(the)h(pro)s(jections)f(of)-1935 96 y(di\013eren)o(t)g(orbital)g
(families)f(in)h(v)m(arious)h(bulge)f(\014elds.)21 b(These)c(sim)o
(ulations)e(are)i(used)f(to)h(in)o(terpret)e(the)-1947 96 y(observ)m
(ational)i(results)f(presen)o(ted)f(in)h(the)f(previous)h(section:)21 b(w)o
(e)15 b(suggest)j(that)e(the)g(metal)f(ric)o(h)g(stars)-1949 97 y(are)j
(on)g(primarily)e(prograde)j(orbits,)e(while)g(the)g(metal)g(p)q(o)q(or)j
(stars)e(also)g(p)q(opulate)g(the)g(retrograde)-1916 96 y(orbits.)k(These)
16 b(sim)o(ulations)g(also)h(suggest)g(that)g(com)o(bining)e(observ)m
(ations)j(of)e(sev)o(eral)g(\014elds)g(can)g(put)-1947 96 y(imp)q(ortan)o
(t)i(constrain)o(ts)g(on)g(the)f(bulge)g(stellar)g(distribution)h
(function.)25 b(The)17 b(implications)g(for)g(the)-1917 97 y(formation)g
(of)f(the)g(Bulge/Bar)g(are)g(discussed)g(in)g @F5 @sf(x)p @F7 @sf(4)h
(and)g(conclusions)f(are)g(giv)o(en)g(in)g @F5 @sf(x)p @F7 @sf(5.)-939 203 y
@F2 @sf(2.)56 b(Data)19 b(Analysis)-1099 188 y @F7 @sf(F)l(or)h(the)g(62)h
(K)f(gian)o(t)g(sample)f(de\014ned)h(b)o(y)g(Zhao)h @F4 @sf(et)g(al.)34 b
@F7 @sf(\(1994\),)22 b(Ric)o(h)d(\(1988,)j(1990\))g(has)-1857 97 y
(measured)16 b(its)h(radial)f(v)o(elo)q(cit)o(y)f(and)i(metallici)o(t)o(y)e
(and)i(Spaenhauer)g @F4 @sf(et)i(al.)k @F7 @sf(\(1992\))18 b(has)g
(measured)e(its)-1941 96 y(prop)q(er)h(motions.)k(The)c(prop)q(er)g
(motion)f(v)o(elo)q(citi)o(es)f(are)i(computed)e(assuming)i @F6 @sf(R)p
7 w @F11 @sf(0)16 -7 y @F7 @sf(=)d(8)j(kp)q(c)f(and)h(that)-1949 96 y(the)g
(whole)h(sample)f(of)h(Spaenhauer)g @F4 @sf(et)h(al.)27 b @F7 @sf(\(1992\))
19 b(is)e(stationary)l(.)26 b(The)18 b(stars)g(are)g(group)q(ed)h(in)o(to)e
(a)-1917 96 y(metal)e(ric)o(h)h(sample)g(with)g([F)l(e)p @F6 @sf(=)p @F7 @sf
(H])c @F5 @sf(\025)i @F7 @sf(0)i(and)h(a)g(metal)e(p)q(o)q(or)j(sample)e
(with)g([F)l(e)p @F6 @sf(=)p @F7 @sf(H])c @F5 @sf(\024)i(\000)p @F7 @sf(0)p
@F6 @sf(:)p @F7 @sf(2.)-1630 129 y(T)l(o)20 b(illustrate)e(the)h(tec)o
(hnique)f(of)h(constructing)h(the)f(v)o(elo)q(cit)o(y)f(elli)o(psoid,)h(w)o
(e)g(in)o(tro)q(duce)g(the)-1874 124 y
6 @eop1

5 @bop0

5 @eop0
0 0 5 @bop1 922 45 a @F7 @sf({)17 b(5)f({)-256 147 y @F2 @sf(1.)56 b(In)n
(tro)r(duction)-1082 199 y @F7 @sf(While)17 b(starcoun)o(ts)j(\(Nak)m(ada)g
@F4 @sf(et)g(al.)30 b @F7 @sf(1992,)21 b(Whitelo)q(c)o(k)d(&)g(Catc)o(hp)q
(ole)h(1992)q(\),)h(observ)m(ations)-1885 96 y(of)e(the)g(in)o(tegrated)f
(ligh)o(t)h(of)g(the)g(Bulge)f(at)h(near)g(infrared)g(w)o(a)o(v)o(ele)o
(ngths)g(\(Blitz)o 17 x(&)g(Sp)q(ergel)g(1991b;)-1908 96 y(W)l(eiland)e
@F4 @sf(et)j(al.)24 b @F7 @sf(1994\))18 b(and)f(measuremen)o(ts)f(of)h
(the)f(kinematics)g(of)h(the)f(atomic)h(and)g(molecular)e(gas)-1937 97 y
(\(de)j(V)l(aucouleurs)h(1964;)i(Gerhard)e(&)g(Vietri)e(1986;)k(Binney)l
(,)o 18 x(Gerhard,)f(Stark,)f(Bally)e(&)i(Uc)o(hida)-1892 96 y(1991\))h
(all)d(suggest)i(that)g(the)f(Bulge)f(is)h(triaxial,)f(stellar)g
(kinematical)g(observ)m(ations)i(ha)o(v)o(e)e(not)i(y)o(et)-1907 96 y(rev)o
(ealed)d(Bulge)h(triaxialit)o(y)o 16 x(\(de)h(Zeeu)o(w)e(1993)q(\).)26 b
(Detection)17 b(of)h(either)e(v)o(ertex)g(deviation)h(or)h(minor)-1918 96 y
(axis)g(rotation)i(w)o(ould)e(b)q(e)h(c)o(haracteristic)d(signatures)k(of)e
(triaxialit)o(y)l(.)o 27 x(While)f(Blum)h @F4 @sf(et)i(al.)29 b @F7 @sf
(\(1993\))-1896 97 y(rep)q(ort)19 b(large)f(\(negativ)o(e\))g(radial)g(v)o
(elo)q(cit)o(y)e(in)i(their)g(minor)g(axis)h(\014eld)e(\()p @F6 @sf(l)q(;)
8 b(b)p @F7 @sf(\))17 b(=)h(\()p @F5 @sf(\000)p @F7 @sf(1)p @F6 @sf(;)8 b
@F7 @sf(2\),)18 b(Sharples,)-1898 96 y(W)l(alk)o(er)g(&)g(Cropp)q(er)i
(\(1990\))g(do)f(not)g(see)f(these)g(large)h(v)o(elo)q(citi)o(es)e(at)i
(Baade's)g(Windo)o(w)g(\(1)p @F6 @sf(;)8 b @F5 @sf(\000)p @F7 @sf(4\))p
-18 w @F10 @sf(o)2 18 y @F7 @sf(.)-1894 96 y(Th)o(us,)16 b(it)g(is)g(not)h
(clear)e(if)h(minor)g(axis)g(rotation)h(has)g(b)q(een)f(detected.)k(In)c
(this)g(pap)q(er,)h(w)o(e)e(will)h(presen)o(t)-1951 97 y(the)g(\014rst)h
(clear)e(evidence)f(for)j(v)o(ertex)e(deviation,)g(a)i(\\smoking)f(gun")h
(of)g(bulge)f(triaxialit)o(y)l(.)o -1630 128 a(In)k(the)h(last)g(few)f(y)o
(ears,)h(the)g(\014rst)g(bulge)g(prop)q(er)g(motion)g(data)h(ha)o(v)o(e)e
(b)q(ecome)g(a)o(v)m(ailable)-1841 96 y(\(Spaenhauer,)j(Jones)g(&)f
(Whitford)g(1992\))h(and)g(there)e(will)g(so)q(on)i(b)q(e)f(radial)g(v)o
(elo)q(citi)o(es)f(and)-1814 97 y(abundances)15 b(for)f(these)f(stars)h
(\(Sadler)g(et.)20 b(al)13 b(1994\).)22 b(These)14 b(data)g(and)h(the)e
(published)g(radial)h(v)o(elo)q(cit)o(y)o -1951 96 a(data,)h @F4 @sf(e.g.)
23 b @F7 @sf(,)14 b(those)h(collected)e(in)h(Ken)o(t)g(\(1992\),)h(ma)o(y)f
(p)q(oten)o(tially)g(pro)o(vide)f(insigh)o(t)h(in)o(to)g(the)h(principle)o
-1951 96 a(orbit)h(families)e(presen)o(t)h(in)h(the)f(bulge,)g(if)h(they)f
(can)h(b)q(e)g(compared)g(with)f(theoretical)g(mo)q(dels.)21 b(Ken)o(t's)
-1951 97 y(\(1992\))c(oblate)e(rotator)h(mo)q(del)f(has)h(b)q(een)f(v)o
(ery)f(useful)h(in)g(analyzing)g(existing)f(kinematical)f(data)j(and)-1949
96 y(the)g(next)f(step)h(is)g(the)f(dev)o(elopmen)o(t)f(of)i
(self-consisten)o(t)g(triaxial)f(mo)q(dels)h(for)g(the)f(bulge.)21 b(This)
16 b(pap)q(er)-1949 96 y(rep)q(orts)j(in)o(termedi)o(ate)e(results)h(from)g
(suc)o(h)g(a)g(program)h(\(Zhao)g(1994\),)h(whose)f(goal)g(is)f(to)g
(deriv)o(e)f(a)-1904 97 y(p)q(oten)o(tial)g(from)h(the)f(bulge)g(ligh)o(t)g
(distribution)g(giv)o(en)g(b)o(y)g(the)g @F3 @sf(COBE)h @F7 @sf(data,)g
(and)g(to)g(subsequen)o(tly)-1920 96 y(build)e(a)g(self-consisten)o(t)g
(dynamical)f(mo)q(del)h(for)h(the)f(Galactic)f(bulge)h(and)h(inner)f
(disk.)-1560 128 y(In)h(this)h(pap)q(er,)g(w)o(e)g(study)g(the)f(o)o(v)o
(erlap)g(sample)h(of)g(stars)h(at)f(Baade's)g(Windo)o(w)g(b)q(et)o(w)o
(een)e(the)-1911 97 y(prop)q(er)h(motion)g(data)g(of)g(427)g(Bulge)f(K)g
(&)h(M)f(gian)o(ts)h(b)o(y)f(Spaenhauer)h @F4 @sf(et)h(al.)23 b @F7 @sf
(\(1992\))18 b(and)f(the)g(radial)-1945 96 y(v)o(elo)q(cit)o(y)o 18 x(and)j
(metallici)o(t)o(y)e(data)i(of)f(88)i(K)e(gian)o(ts)g(b)o(y)g(Ric)o(h)g
(\(1988,)i(1990\).)32 b(This)19 b(o)o(v)o(erlap,)g(62)h(K)-1875 124 y
5 @eop1

4 @bop0
@F4 @sf
[<
703800F87C00FC7E00FC7E00B85C0080400040200040200020100020100010080008040004
0200020100010080>
	 17 15 -13 34 25] 92 @dc
[<
81FC0000C6070000C80180007000C0007000600060003000600038002000180020001C0020
001C0020001C0000001E0000001E0000001E0000003C0000007C000003FC00001FF800007F
F00000FFE00001FF000001F8000003E0000003C0000003C0000003C0010003C0010001C001
8001C0018000C0018000E00180006001C0003001C0001802C0000E0C400001F020>
	 27 36 -3 34 27] 83 @dc
[<
804020100808040402021E1E1E1E0E>
	 7 15 -3 4 15] 44 @dc
[<
3C0000620000F30000F1800071800001C00001C00000C00000E00000E00000E00000E00000
7000007000007000007000007000003800003800003800003800003800001C00001C00001C
00001C00001C00000E00000E00000E0000FFF0000E00000700000700000700000700000700
00038000038000038000018600018F0000CF00004700003E>
	 24 45 2 34 15] 102 @dc
[<
600F00E01880701880703840703840703840381C20381C003838003838001C70001FC0001E
00001D80000E40000E20600E10F00E10F0070C700702100701E00700000380000380000380
0003800001C00001C00001C00001C00000E00000E00000E0000FE00000F000>
	 20 35 -3 34 22] 107 @dc
[<
FFFFF0000F801E0007800700078003C0078001E0078000F003C000F003C0007803C0007803
C0007801E0003C01E0003C01E0003801E0007800F0007800F000F000F001E000FFFF800078
0780007801E0007800F000780078003C003C003C001E003C001E003C000E001E000F001E00
0F001E000F001E000E000F001E000F001C000F003800FFFFE0>
	 32 34 -3 33 34] 66 @dc
[<
E000F000F800F8007000000000000000000000000000000000000000000000000E000F000F
800F800700>
	 9 21 -6 20 15] 58 @dc
[<
03E0F00006138C000C0F02001C0F01001C0701001C0700801C0700801C0700401C0700401C
0700400E0380200E0380200E038020070380208701C0308701C0308381C0704381C0F04380
E1F02300E1F01E0060E0>
	 28 21 -5 20 32] 119 @dc
[<
FFF80FFE000F8001F000078001E000078001E000078001E000078003C00003C003C00003C0
03C00003C007800003C007800001E00F800001E00F000001E00F000001F01F000000F81E00
0000F41E000000F23C000000F0BC000000787C000000783800000078180000007808000000
3C040000003C010000003C008000003C004000001E002000001E001000001E000400001E00
0200000F000100000F0001C0000F0001E000FFF807FC>
	 38 34 -3 33 37] 75 @dc
[<
804020100804040202021D3F3F1E0C>
	 8 15 -11 34 15] 39 @dc
[<
080000100000300000200000600000600000400000C00000C00000C00000C00000C00000C0
0000C00000C00000C00000C00000C00000C00000C00000E00000E00000E000006000006000
007000007000003000003000003800001800001800001C00000C00000E0000060000060000
03000003000001800000C00000C00000600000300000100000080000040000020000010000
0080>
	 17 50 -8 36 20] 40 @dc
[<
FFFC0780038003800380038001C001C001C001C000E000E000E000E0007000700070007000
38003800380038001C001C001C001C03CE002E001E0006000300030001>
	 16 33 -6 32 25] 49 @dc
[<
3F000041C000806000E03000F01800F00C00600E000007000007000003800003C00003C001
F1E00309E00605E00E02F00E01F01C00F01E00F01E00F81E00781E00781E00780F00380F00
380F003807003807803803803801C03800E030007030003860000FC0>
	 21 34 -5 32 25] 57 @dc
[<
801E00807F0040FF8063FFC05F80C020006010002008003004001002000001000000C00000
20000018000006000003000001C00700E00880F008407808403808203C08203C04101E0410
1E04101E02101E02201E01001C00801C008018006038001060000FC0>
	 23 34 -4 32 25] 50 @dc
[<
80000040000020000010000008000004000006000003000001800001800000C00000600000
6000003000003000003800001800001C00000C00000C00000E000006000006000007000007
00000300000300000380000380000380000180000180000180000180000180000180000180
00018000018000018000018000018000018000010000030000030000020000060000040000
0800>
	 17 50 0 36 20] 41 @dc
[<
8040004020002010001008000804000402000402000201000201000201001D0E803F1F803F
1F801E0F000C0600>
	 17 15 -8 34 25] 34 @dc

4 @eop0

0 0 4 @bop1 922 45 a @F7 @sf({)17 b(4)f({)-929 147 y @F4 @sf(\\Star)n(c)n
(ounts,)h(me)n(asur)n(ements)g(of)g(the)g(inte)n(gr)n(ate)n(d)g(light)h
(and)g(the)f(kinematics)i(of)e(the)g(atomic)h(and)-1950 96 y(mole)n(cular)i
(gas)f(al)r(l)i(pr)n(ovide)e(str)n(ong)g(evidenc)n(e)i(that)e(the)h(Bulge)h
(is)e(triaxial,)h(and)f(is)g(r)n(otating)g(fairly)-1904 96 y(r)n(apid)r
(ly.)i(T)l(o)16 b(date,)g(ther)n(e)f(is)h(little)h(evidenc)n(e)g(for)e
(triaxiality)h(in)f(the)h(stel)r(l)q(ar)g(kinematics:)23 b(the)16 b
(available)-1948 96 y(stel)r(lar)23 b(velo)n(cities)g(ar)n(e)e(c)n
(onsistent)i(with)f(Kent's)g(\(1992\))f(oblate)i(mo)n(del.)35 b(This)21 b
(unsatisfactory)-1843 97 y(situation)d(is)f(exp)n(e)n(cte)n(d)h(to)g(impr)n
(ove)f(r)n(apid)r(ly.")-776 128 y @F7 @sf(Tim)i(de)h(Zeeu)o(w)f(in)h @F3 @sf
(Galactic)g(Bulges)g @F7 @sf(H.)f(Dejonghe)i(and)f(H.J.)f(Habing)h(\(eds.\))
33 b(\(Klu)o(w)o(er:)-1857 96 y(Dordrec)o(h)o(t\))16 b(p.)21 b(191.)-397
1993 y
4 @eop1

3 @bop0

3 @eop0
0 0 3 @bop1 922 45 a @F7 @sf({)17 b(3)f({)-905 147 y(sense)i(in)g(the)f
(rest)h(frame.)27 b(They)17 b(ha)o(v)o(e)h(nearly)f(round)i(lo)q(op)g
(shap)q(es)g(and)g(are)f(aligned)-1663 96 y(p)q(erp)q(endicularly)d(to)i
(the)f(bar,)g(hence)f(limit)g(the)h(triaxialit)o(y)o 15 x(of)h(the)f(bar)h
(p)q(oten)o(tial.)-1512 96 y(The)g(correlations)g(b)q(et)o(w)o(een)g(the)f
(metallici)o(t)o(y)f(and)j(the)f(orbit)g(families)f(can)h(dev)o(elop)-1685
96 y(if)h(the)g(Bulge)f(forms)i(dissipativ)o(ely)o 17 x(on)g(a)g(su\016cien)o
(tl)o(y)e(long)i(time)e(scale.)27 b(Ho)o(w)o(ev)o(er,)o 18 x(it)-1657 97 y
(is)18 b(di\016cult)e(to)i(explain)f(suc)o(h)h(correlations)g(if)f(most)i
(stars)f(in)g(the)g(inner)f(Galaxy)h(form)-1666 96 y(during)e(the)g
(violen)o(t)f(relaxation)h(phase.)-882 2121 y
3 @eop1

2 @bop0
@F9 @sf
[<
0F801FC0306060306030C018C018C018C0186030603030601FC00F80>
	 13 14 -2 14 18] 14 @dc
@F2 @sf
[<
01E00003F00003F00003F00003F00003F00003F00003F00001F00001F00001F00000F00000
F00000F000007800007800003800001800001C00000C00000600C00300C00300C00180E000
C0E000607FFFF07FFFF87FFFF87FFFFC7FFFFE7FFFFE780000600000>
	 23 34 -3 33 28] 55 @dc

2 @eop0

0 0 2 @bop1 922 45 a @F7 @sf({)17 b(2)f({)-209 147 y @F2 @sf(ABSTRA)n(CT)
-936 167 y @F7 @sf(W)l(e)k(study)h(a)g(sample)f(of)h(62)h(Baade's)e(Windo)o
(w,)i(\()p @F6 @sf(l)q(;)8 b(b)p @F7 @sf(\))20 b(=)h(\(1)p @F6 @sf(;)8 b
@F5 @sf(\000)p @F7 @sf(4\))p -18 w @F10 @sf(o)3 18 y @F7 @sf(,)21 b(K)f
(gian)o(ts)-1601 96 y(that)h(ha)o(v)o(e)e(published)h(prop)q(er)h
(motions,)h(radial)e(v)o(elo)q(cit)o(y)o 19 x(and)h(metallici)o(t)o(y)e
(Using)-1606 96 y @F6 @sf(R)p 7 w @F11 @sf(0)21 -7 y @F7 @sf(=)g(8)h(kp)q
(c,)f(w)o(e)g(construct)g(the)h(v)o(elo)q(cit)n(y)e(ellipsoids,)g(namely)h
(the)g(3)13 b @F5 @sf(\002)g @F7 @sf(3)20 b(v)o(elo)q(cit)o(y)-1634 97 y
(disp)q(ersion)e(tensors,)g(for)h(the)e(metal)g(ric)o(h)g(stars)i(\()p([F)l
(e)p @F6 @sf(=)p @F7 @sf(H])c @F5 @sf(\025)h @F7 @sf(0\))i(and)h(metal)e
(p)q(o)q(or)i(stars)-1666 96 y(\()p([F)l(e)p @F6 @sf(=)p @F7 @sf(H])e
@F5 @sf(\024)h(\000)p @F7 @sf(0)p @F6 @sf(:)p @F7 @sf(2\).)30 b(After)18 b
(diagonalizing)h(the)g(tensor,)g(w)o(e)g(\014nd)g(a)h(v)o(ertex)d
(deviation)-1640 96 y(c)o(haracteristic)f(of)i(a)h(non-axisymmetric)d
(system.)25 b(Eigen)o(v)m(alues)18 b(for)g(the)g(t)o(w)o(o)f(v)o(elo)q
(cit)o(y)-1670 96 y(ellipsoids)h(\()p @F6 @sf(\033)p 7 w @F11 @sf(1)2 -7 y
@F7 @sf(,)i @F6 @sf(\033)p 7 w @F11 @sf(2)1 -7 y @F7 @sf(,)g @F6 @sf(\033)p
7 w @F11 @sf(3)1 -7 y @F7 @sf(\))g(are)g(\(126,)h(89,)f(65\))p @F5 @sf(\006)p
@F7 @sf(13)h(km/s)f(for)f(the)h(metal)e(ric)o(h)h(sample)-1628 97 y(and)h
(\(154,)h(77,)f(83\))p @F5 @sf(\006)p @F7 @sf(25)h(km/s)e(for)h(the)f
(metal)g(p)q(o)q(or)i(sample)e(with)g(their)g(long)g(axes)-1630 96 y(p)q
(oin)o(ting)e(to)h(t)o(w)o(o)f(nearly)g(p)q(erp)q(endicular)g(directions)g
(\()p @F6 @sf(l)p 7 w @F10 @sf(v)2 -7 y @F6 @sf(;)8 b(b)p 7 w @F10 @sf(v)
3 -7 y @F7 @sf(\))16 b(=)g(\()p @F5 @sf(\000)p @F7 @sf(65)c @F5 @sf(\006)g
@F7 @sf(9)p -18 w @F10 @sf(o)2 18 y @F6 @sf(;)c @F7 @sf(+14)k @F5 @sf(\006)g
@F7 @sf(9)p -18 w @F10 @sf(o)2 18 y @F7 @sf(\))-1680 96 y(and)18 b(\()p
@F6 @sf(l)p 7 w @F10 @sf(v)3 -7 y @F6 @sf(;)8 b(b)p 7 w @F10 @sf(v)3 -7 y
@F7 @sf(\))16 b(=)h(\(25)12 b @F5 @sf(\006)g @F7 @sf(14)p -18 w @F10 @sf
(o)3 18 y @F6 @sf(;)c @F5 @sf(\000)p @F7 @sf(11)k @F5 @sf(\006)g @F7 @sf
(14)p -18 w @F10 @sf(o)3 18 y @F7 @sf(\))18 b(resp)q(ectiv)o(el)o(y)l(.)o
25 x(The)g(v)o(ertex)o 17 x(deviations)f(of)h(the)-1672 97 y(v)o(elo)q
(cit)o(y)o 14 x(ellipsoids)c(cannot)h(b)q(e)h(consisten)o(tly)d(explained)h
(b)o(y)h @F4 @sf(any)g @F7 @sf(oblate)g(mo)q(del.)20 b(W)l(e)15 b(are)-1706
96 y(able)g(to)h(reject)e(the)i(h)o(yp)q(othesis)f(that)h(the)f(metal)g(p)q
(o)q(or)i(and)f(metal)f(ric)o(h)f(p)q(opulations)j(are)-1706 96 y(dra)o
(wn)g(from)f(the)g(same)g(distribution)g(at)g(b)q(etter)g(than)h(the)f
(97
(orbits)g(in)f(a)h(realistic)e(bar)i(p)q(oten)o(tial)f(with)g(a)h
(Gaussian)h(v)o(elo)q(cit)n(y)-1631 96 y(distribution,)c(allo)o(wing)h(us)g
(to)g(sim)o(ulate)f(and)i(in)o(terpret)d(observ)m(ations.)27 b(W)l(e)18 b
(conclude)-1668 96 y(that)g(the)g(data)h(are)f(consisten)o(t)g(with)g(a)g
(triaxial)f(bulge)h(p)q(oin)o(ting)g(to)o(w)o(ards)h(\()p @F6 @sf(l)q(;)
8 b(b)p @F7 @sf(\))17 b(with)-1663 97 y @F6 @sf(l)d(<)g @F7 @sf(0)p -18 w
@F10 @sf(o)17 18 y @F7 @sf(and)h @F6 @sf(b)f @F7 @sf(=)g(0)p -18 w @F10 @sf
(o)17 18 y @F7 @sf(as)h(suggested)g(b)o(y)f(earlier)f(w)o(ork)i(on)g(gas)g
(dynamics)f(and)h(the)f(observ)o(ed)-1706 96 y(ligh)o(t)19 b
(distribution.)31 b(W)l(e)19 b(also)h(predict)f(that)h(lo)o(w)g(latitude)f
(\()p @F5 @sf(j)p @F6 @sf(b)p @F5 @sf(j)g(\024)g @F7 @sf(4)p -18 w @F9 @sf
(\016)2 18 y @F7 @sf(\))h(bulge)f(\014elds)-1624 96 y(should)h(sho)o(w)f
(the)g(v)o(ertex)f(deviation)g(more)h(strongly)g(and)h(w)o(ould)f
(therefore)g(b)q(e)g(the)-1638 96 y(b)q(est)g(lo)q(cations)h(for)f(future)f
(prop)q(er)i(motion)f(studies.)28 b(In)19 b(the)g(classi\014cation)f(sc)o
(heme)-1645 97 y(of)i(A)o(thanassoula,)i(Biena)o(yme)o(,)d(Martinet)g(&)h
(Pfenniger)g(\(A&A)f(1983,)j @F2 @sf(127)p @F7 @sf(,)f(349\),)-1616 96 y
(the)d(metal)f(ric)o(h)f(stars)j(app)q(ear)g(to)f(o)q(ccup)o(y)g(the)f
(B-family)f(orbits)j(whic)o(h)e(rotate)h(in)f(the)-1668 96 y(prograde)f
(sense)e(in)g(the)h(rest)f(frame)g(and)i(ha)o(v)o(e)e(b)q(o)o(xy)g(shap)q
(es)i(that)f(are)g(aligned)f(with)g(and)-1705 97 y(supp)q(orting)19 b(the)e
(bar.)26 b(The)18 b(metal)f(p)q(o)q(or)i(stars)g(in)e(the)h(sample)f(lag)h
(b)q(ehind)g(the)f(metal)-1671 96 y(ric)o(h)e(Bulge)h(and)h(app)q(ear)h
(to)f(o)q(ccup)o(y)f(R-family)f(orbits)i(whic)o(h)f(rotate)h(in)f(the)g
(retrograde)-1821 124 y
2 @eop1

1 @bop0
/@F1 @newfont
@F1 @sf
[<
C01FF000E0FFFE00F3FFFF00FFE01F80FF0007C0FC0003E0F80003F0F00001F0F00001F0E0
0001F8E00001F8E00001F8600001F8000003F8000007F800000FF800007FF00007FFF000FF
FFE003FFFFE007FFFFC00FFFFF801FFFFE003FFFFC007FFFE0007FFE0000FFC00000FF0000
00FE0000E0FE0000E0FC0000E0FC0001E0FC0001E07C0001E07C0003E03E0007E01F001FE0
0F807FE007FFF9E003FFF0E0007F8060>
	 29 41 -4 40 38] 83 @dc
[<
FFFEFFFEFFFE0FE00FE00FE00FE00FE00FE00FE00FE00FE00FE00FE00FE00FE00FE00FE00F
E00FE00FE00FE00FE00FE0FFE0FFE0FFE0000000000000000000000000000007000F801FC0
3FE03FE03FE01FC00F800700>
	 15 43 -3 42 19] 105 @dc
[<
007FF00003FFFE000FC01F801F0007C03C0001E07C0001F0F80000F8F80000F8F80000F8F8
0000F87C0001F83E0007F01FFFFFF007FFFFE00FFFFFC01FFFFF801FFFFF003FFFF8003E00
00003C000000380000003800000018FF80001FFFE0000FC1F8001F80FC001F007C003F007E
007F007F007F007F007F007F007F007F007F007F007F007F003F007E101F007C381F80FC7C
0FC1FE7C03FFE7F800FF81F0>
	 30 40 -2 26 34] 103 @dc
[<
FFFE3FFF80FFFE3FFF80FFFE3FFF800FE003F8000FE003F8000FE003F8000FE003F8000FE0
03F8000FE003F8000FE003F8000FE003F8000FE003F8000FE003F8000FE003F8000FE003F8
000FE003F8000FE003F8000FE003F8000FE003F8000FF003F8000FF003F8000FD803F8000F
D803F0000FCE03F000FFC787E000FFC1FFC000FFC07F0000>
	 33 27 -3 26 38] 110 @dc
[<
01FC03FC0FFF0FFC3F839FFC7F00DF807E007F80FE003F80FE003F80FE003F80FE003F807F
003F803F003F803F803F800FE03F8007FC3F8000FFFF80000FFF8000003F8000003F800000
3F8007003F800F803F801FC03F001FC07E001FC07E000F81F80007FFF00001FF8000>
	 30 27 -2 26 33] 97 @dc
[<
001F8000FFC001F86003F87003F03807F03807F03807F03807F03807F03807F03807F00007
F00007F00007F00007F00007F00007F00007F00007F00007F00007F00007F00007F000FFFF
F0FFFFF01FFFF007F00003F00003F00001F00000F00000F00000F000007000007000007000
007000>
	 21 38 -1 37 27] 116 @dc
[<
003FC3FF8001FFF3FF8003F03BFF8007E00FF80007E007F8000FE007F8000FE003F8000FE0
03F8000FE003F8000FE003F8000FE003F8000FE003F8000FE003F8000FE003F8000FE003F8
000FE003F8000FE003F8000FE003F8000FE003F8000FE003F8000FE003F8000FE003F8000F
E003F8000FE003F800FFE03FF800FFE03FF800FFE03FF800>
	 33 27 -3 26 38] 117 @dc
[<
FFFF00FFFF00FFFF000FE0000FE0000FE0000FE0000FE0000FE0000FE0000FE0000FE0000F
E0000FE0000FE0000FE0000FE0000FE0000FE0000FF01C0FF03E0FF07F0FD87F0FD87FFFCE
3EFFC7FCFFC1F0>
	 24 27 -2 26 28] 114 @dc
[<
001FF00000FFFE0003F81F0007E003800FC001C01F8000E03F8000E07F0000007F0000007F
000000FF000000FF000000FF000000FFFFFFE0FFFFFFE0FF0007E0FF0007E07F0007E07F00
07C07F000FC03F800FC01F800F800F801F8007C01F0003F07E0001FFF800003FE000>
	 27 27 -2 26 32] 101 @dc
[<
C1FF00F7FFC0FF01E0FC0070F80038F00038F0003CE0003C60007C0000FC0003FC00FFF807
FFF81FFFF03FFFE07FFF807FFE00FFE000FC0000F80070F00070F000707000703800F01E03
F00FFFF003FE30>
	 22 27 -2 26 27] 115 @dc
[<
003FE00001FFFC0007F07F000FC01F801F800FC03F800FE03F800FE07F0007F07F0007F0FF
0007F8FF0007F8FF0007F8FF0007F8FF0007F8FF0007F8FF0007F8FF0007F87F0007F07F00
07F07F0007F03F0007E03F800FE01F800FC00FC01F8003F07E0001FFFC00003FE000>
	 29 27 -2 26 34] 111 @dc
[<
7FFF807FFF807FFF8007F00007F00007F00007F00007F00007F00007F00007F00007F00007
F00007F00007F00007F00007F00007F00007F00007F00007F00007F00007F00007F000FFFF
C0FFFFC0FFFFC007F00007F00007F00007F00007F00007F00007F00007F03E07F07F03F07F
03F87F01F87F00FE3E003FFC0007F0>
	 24 42 -2 41 21] 102 @dc
[<
FFFFFFFE0000FFFFFFFFC000FFFFFFFFF00003F8001FF80003F80007FC0003F80003FE0003
F80001FF0003F80000FF0003F80000FF8003F80000FF8003F80000FF8003F80000FF8003F8
0000FF8003F80000FF8003F80000FF0003F80000FF0003F80001FE0003F80001FE0003F800
03FC0003F80007F00003FFFFFFE00003FFFFFE000003F800FFC00003F8001FE00003F8000F
F00003F80007F80003F80003FC0003F80003FC0003F80003FE0003F80001FE0003F80001FE
0003F80001FE0003F80003FE0003F80003FC0003F80003FC0003F80007FC0003F8000FF800
03F8001FF000FFFFFFFFC000FFFFFFFF8000FFFFFFF80000>
	 41 41 -3 40 49] 66 @dc
[<
FFFEFFFEFFFE0FE00FE00FE00FE00FE00FE00FE00FE00FE00FE00FE00FE00FE00FE00FE00F
E00FE00FE00FE00FE00FE00FE00FE00FE00FE00FE00FE00FE00FE00FE00FE00FE00FE00FE0
0FE00FE0FFE0FFE0FFE0>
	 15 42 -3 41 19] 108 @dc
[<
007FFFFFE000007FFFFFE000007FFFFFE00000003FC0000000003FC0000000003FC0000000
003FC0000000003FC0000000003FC0000000003FC0000000003FC0000000003FC000000000
3FC0000000003FC0000000003FC0000000003FC0000000003FC0000000003FC0000000003F
C0000000003FC0000000003FC0000000003FC0000000003FC0000000003FC0000000003FC0
0000E0003FC000E0E0003FC000E0E0003FC000E0E0003FC000E0E0003FC000E0F0003FC001
E0F0003FC001E070003FC001C078003FC003C078003FC003C07E003FC007C07F803FC03FC0
7FFFFFFFFFC07FFFFFFFFFC07FFFFFFFFFC0>
	 43 40 -2 39 48] 84 @dc
[<
FFF01FFF80FFF01FFF80FFF01FFF8003C007F00001C00FE00001E01FE00000F01FC0000078
3F8000003C7F0000001EFE0000001FFE0000000FFC00000007F800000007F00000000FF000
00000FF00000001FF00000003FF80000007F3C000000FF1E000000FE0F000001FC07800003
F807800007F003C000FFFC0FFF00FFFC0FFF00FFFC0FFF00>
	 33 27 -1 26 36] 120 @dc
[<
0FC00000003FE00000007C78000000FE3C000000FE1E000000FE0E000000FE0F0000007C07
0000003807800000000380000000038000000001C000000001C000000003E000000003E000
000007F000000007F00000000FF80000000FF80000000FF80000001FDC0000001FDC000000
3FDE0000003F8E0000007F8F0000007F070000007F07000000FE03800000FE03800001FC01
C00001FC01C00003FC01E00003F800E00007F800F00007F000700007F0007000FFFE03FF80
FFFE03FF80FFFE03FF80>
	 33 39 -1 26 36] 121 @dc
[<
FFFE1FFFC3FFF8FFFE1FFFC3FFF8FFFE1FFFC3FFF80FE001FC003F800FE001FC003F800FE0
01FC003F800FE001FC003F800FE001FC003F800FE001FC003F800FE001FC003F800FE001FC
003F800FE001FC003F800FE001FC003F800FE001FC003F800FE001FC003F800FE001FC003F
800FE001FC003F800FE001FC003F800FE001FC003F800FF001FE003F800FF001FE003F800F
D801FF003F800FCC01F9803F000FC603F8C07F00FFC383F0707E00FFC1FFE03FFC00FFC07F
800FF000>
	 53 27 -3 26 57] 109 @dc
[<
FFFFF00FFFFEFFFFF00FFFFEFFFFF00FFFFE03FC00007F8003FC0000FF0003FC0001FF0003
FC0001FE0003FC0003FC0003FC0007F80003FC000FF80003FC000FF00003FC001FE00003FC
003FC00003FC003FC00003FC007F800003FC00FF000003FE01FF000003FF01FE000003FF83
FC000003FDE7F8000003FCFFF8000003FC7FF0000003FC3FE0000003FC1FC0000003FC07C0
000003FC0380000003FC01E0000003FC00F0000003FC0078000003FC003C000003FC001E00
0003FC0007000003FC0003C00003FC0001E00003FC0000F00003FC0000780003FC00003C00
03FC00001E00FFFFF001FFFCFFFFF001FFFCFFFFF001FFFC>
	 47 41 -3 40 54] 75 @dc
[<
001FE00000FFFC0003F01E0007E007000FC003801F8001C03F8001C07F8000007F0000007F
000000FF000000FF000000FF000000FF000000FF000000FF000000FF0000007F0000007F00
00007F800E003F801F001F803F800FC03F8007E03F8003F01F0000FFFE00001FF800>
	 26 27 -2 26 31] 99 @dc
[<
003FC3FF8000FFF3FF8003F03BFF8007C00FF8000F8007F8001F8003F8003F8003F8007F00
03F8007F0003F8007F0003F800FF0003F800FF0003F800FF0003F800FF0003F800FF0003F8
00FF0003F800FF0003F8007F0003F8007F0003F8007F0003F8003F8003F8001F8003F8000F
C007F80007E00FF80003F03FF80000FFFBF800001FE3F800000003F800000003F800000003
F800000003F800000003F800000003F800000003F800000003F800000003F800000003F800
000003F800000003F80000003FF80000003FF80000003FF800>
	 33 42 -2 41 38] 100 @dc
[<
2000300018000C0006000300030001800180018000C000C01CC07FC07FC0FFC0FFC0FF807F
007F001C00>
	 10 21 -5 41 19] 39 @dc
[<
00000E00000700000000001F00000F80000000001F00000F80000000001F80001F80000000
003F80001FC0000000003F80001FC0000000003FC0003FC0000000007FC0003FE000000000
7FC0003FE000000000FFE0007FF000000000FFE0007FF000000000FFF000FFF000000001FE
7000FF3800000001FE7000FF3800000001FE7801FF3800000003FC3801FE1C00000003FC38
01FE1C00000007FC1C03FC1E00000007F81C03FC0E00000007F81E07FC0E0000000FF00E07
F8070000000FF00E07F8070000001FF00F0FF0078000001FE0070FF0038000001FE0070FF0
038000003FE0039FE003C000003FC0039FE001C000003FC003FFE001C000007F8001FFC000
E000007F8001FFC000E00000FF8000FF8000F00000FF0000FF8000700000FF0000FF800070
0001FF0000FF0000780001FE0000FF0000380001FE0001FF0000380003FC0001FE00001C00
03FC0001FE00001C00FFFFE07FFFF007FFF0FFFFE07FFFF007FFF0FFFFE07FFFF007FFF0>
	 68 41 -1 40 71] 87 @dc
[<
00078003C00000078003C000000FC007E000000FC007E000000FC007E000001FE00FF00000
1FE00FF000003FF01FF800003FF01FB800003FF01FB800007F783F3C00007F383F1C0000FF
383F1E0000FE1C7E0E0000FE1C7E0E0001FE1EFC0F0001FC0EFC070001FC0EFC070003F807
F8038003F807F8038007F807F803C007F003F001C007F003F001C00FE007E000E0FFFE7FFC
0FFEFFFE7FFC0FFEFFFE7FFC0FFE>
	 47 27 -1 26 50] 119 @dc

1 @eop0

0 0 1 @bop1 46 192 a @F1 @sf(Signatures)24 b(of)g(Bulge)f(T)-6 b
(riaxialit)n(y)25 b(from)e(Kinematics)g(in)-1014 82 y(Baade's)g(Windo)n(w)
-431 114 y @F7 @sf(HongSheng)17 b(Zhao)-699 63 y(Departmen)o(t)f(of)g
(Astronom)o(y)l(,)g(Colum)o(bia)g(Univ)o(e)o(rsit)o(y)l(,)o -975 61 a(538)i
(W)l(est)e(120)h(Street,)e(New)h(Y)l(ork,)f(NY)h(10027)-636 105 y(Da)o
(vid)g(N.)f(Sp)q(ergel)-826 63 y(Departmen)o(t)h(of)g(Astroph)o(ysical)g
(Sciences,)o 15 x(Princeton)g(Univ)o(ersi)o(t)o(y)l(,)o -1005 60 a(P)o
(eyton)g(Hall,)f(Princeton,)h(NJ)g(08544)-549 105 y(R.)g(Mic)o(hael)f(Ric)o
(h)15 -18 y @F11 @sf(1)-707 82 y @F7 @sf(Departmen)o(t)h(of)g(Astronom)o
(y)l(,)g(Colum)o(bia)g(Univ)o(e)o(rsit)o(y)l(,)o -975 60 a(538)i(W)l(est)e
(120)h(Street,)e(New)h(Y)l(ork,)f(NY)h(10027)-1059 195 y(Receiv)o(e)o(d)f
525 -2 z 525 x(;)73 b(accepted)16 b 525 -2 z -1221 1429 a 780 -2 z 68 57 a
@F11 @sf(1)2 18 y @F7 @sf(Alfred)f(P)l(.)h(Sloan)h(F)l(oundation)g(F)l
(ello)o(w)-823 126 y
1 @eop1